\documentclass[aps,prl,twocolumn,showkeys,superscriptaddress]{revtex4-1}

\usepackage{placeins}
\usepackage[artemisia]{textgreek}
\usepackage{amssymb}
\usepackage{epsfig}
\usepackage{amsmath}
\usepackage{graphicx}
\usepackage{dcolumn}
\usepackage{latexsym}
\usepackage{lipsum}  
\usepackage{color}
\usepackage{epstopdf}
\usepackage{subfigure}
\usepackage{nicefrac}
\usepackage{blindtext}
\usepackage[ulem=normalem]{changes}
\usepackage{float}
\usepackage{soul}
\usepackage[T1]{fontenc}
\usepackage[hidelinks]{hyperref}

\hypersetup{
    colorlinks,
    linkcolor={red!50!black},
    citecolor={blue!50!black},
    urlcolor={blue!80!black}
}

\begin{document}

\title{Control over epitaxy and the role of the InAs/Al interface in hybrid two-dimensional electron gas systems}

\author{Erik Cheah}
\email{echeah@phys.ethz.ch}
\affiliation{Solid State Physics Laboratory, ETH Zurich, 8093 Zurich, Switzerland}

\author{Daniel Z. Haxell}
\affiliation{IBM Research Europe - Zurich, 8803 Rüschlikon, Switzerland}

\author{R\"udiger Schott}
\affiliation{Solid State Physics Laboratory, ETH Zurich, 8093 Zurich, Switzerland}

\author{Peng Zeng}
\affiliation{ScopeM, ETH Zurich, 8093 Zurich, Switzerland}

\author{Ekaterina Paysen}
\affiliation{Paul-Drude-Institut für Festkörperelektronik, Leibniz-Institut im Forschungsverbund Berlin e. V., 10117 Berlin, Germany}

\author{Sofieke C. ten Kate}
\affiliation{IBM Research Europe - Zurich, 8803 Rüschlikon, Switzerland}

\author{Marco Coraiola}
\affiliation{IBM Research Europe - Zurich, 8803 Rüschlikon, Switzerland}

\author{Max Landstetter}
\affiliation{Solid State Physics Laboratory, ETH Zurich, 8093 Zurich, Switzerland}

\author{Ali B. Zadeh}
\affiliation{ScopeM, ETH Zurich, 8093 Zurich, Switzerland}

\author{Achim Trampert}
\affiliation{Paul-Drude-Institut für Festkörperelektronik, Leibniz-Institut im Forschungsverbund Berlin e. V., 10117 Berlin, Germany}

\author{Marilyne Sousa}
\affiliation{IBM Research Europe - Zurich, 8803 Rüschlikon, Switzerland}

\author{Heike Riel}
\affiliation{IBM Research Europe - Zurich, 8803 Rüschlikon, Switzerland}

\author{Fabrizio Nichele}
\affiliation{IBM Research Europe - Zurich, 8803 Rüschlikon, Switzerland}

\author{Werner Wegscheider}
\affiliation{Solid State Physics Laboratory, ETH Zurich, 8093 Zurich, Switzerland}
\affiliation{Quantum Center, ETH Zurich, 8093 Zurich, Switzerland}

\author{Filip Krizek}
\affiliation{Solid State Physics Laboratory, ETH Zurich, 8093 Zurich, Switzerland}
\affiliation{IBM Research Europe - Zurich, 8803 Rüschlikon, Switzerland}
\affiliation{Institute of Physics, Czech Academy of Sciences, 162 00 Prague, Czech Republic}

\date{\today}

\begin{abstract}

In-situ synthesised semiconductor/superconductor hybrid structures became an important material platform in condensed matter physics. Their development enabled a plethora of novel quantum transport experiments with focus on Andreev and Majorana physics. The combination of InAs and Al has become the workhorse material and has been successfully implemented in the form of one-dimensional structures and two-dimensional electron gases. In contrast to the well-developed semiconductor parts of the hybrid materials, the direct effect of the crystal nanotexture of Al films on the electron transport still remains unclear. This is mainly due to the complex epitaxial relation between Al and the semiconductor. We present a study of Al films on shallow InAs two-dimensional electron gas systems grown by molecular beam epitaxy, with focus on control of the Al crystal structure. We identify the dominant grain types present in our Al films and show that the formation of grain boundaries can be significantly reduced by controlled roughening of the epitaxial interface. Finally, we demonstrate that the implemented roughening does not negatively impact either the electron mobility of the two-dimensional electron gas or the basic superconducting properties of the proximitized system.

\end{abstract}
\pacs{}
\maketitle

\section{Introduction}

Material systems that combine semiconductors (SEs) and superconductors (SCs) have recently prompted novel research directions in condensed matter physics. The main motivation to study these systems are different approaches to quantum computing, e.g. Andreev \cite{hays2021coherent} and Transmon qubits \cite{larsen2015semiconductor,casparis2018superconducting} or topological systems hosting Majorana bound states \cite{sarma2015majorana}. 

The combination of InAs and in-situ deposited Al has become an established material platform, either in the form of proximitized quasi one-dimensional hybrid nanowires or shallow two-dimensional electron gas (2DEG) systems \cite{krogstrup2015epitaxy, shabani2016two}. The choice of InAs as the semiconducting part is due to its relatively high spin-orbit coupling and advantageous band alignment at the shallow SE/SC interface \cite{mikkelsen2018hybridization}. Furthermore, it exhibits good etching selectivity and chemical stability in device fabrication processes. The choice of Al is motivated by the possibility to achieve a high degree of epitaxial order on InAs \cite{krogstrup2015epitaxy}, as well as by the fact that Al and InAs are mutually compatible with in-situ deposition in typical III-V Molecular Beam Epitaxy (MBE) systems, which often contain both the In and Al sources. 

It was shown that the crystal structure of epitaxial Al films differs, depending on the implemented SE system. Krogstrup, et al. have shown in Ref. \cite{krogstrup2015epitaxy} that in-situ Al deposition on specific nanowire facets results in the formation of a single crystal Al film with thicknesses down to a few nanometers. Improving the interface between the materials resulted in a break-through, where a so-called "hard" superconducting gap was induced into the hybrid system, meaning that no sub-gap states were observed in tunneling spectroscopy measurements. This is in contrast to materials with ex-situ deposited Al, where sup-gap states are typically present \cite{chang2015hard}. 

Nanowire-based systems have shown great promise as a platform for investigating transport phenomena \cite{mourik2012signatures,vaitiekenas2020flux,kurtossy2021andreev}, however the lack of reliable schemes for scaling up to large device arrays limits their perspective for industrial applications. To address the scaling issue, growth of larger scale nanowire networks via selective area growth was recently developed \cite{friedl2018template,krizek2018field}. Yet, the complementary 2DEG systems show more promise for large-scale applications, due to their compatibility with top-down fabrication techniques \cite{aghaee2022inas}.

In comparison to nanowires, the 2D systems reduce the spatial degrees of freedom for relaxation of the Al layer. This promotes the formation and co-existence of Al grains with various orientations after in-situ Al deposition, as previously reported in literature \cite{sarney2018reactivity, sarney2020aluminum, wang2020dependence}.  
Such grains in the Al film determine the local epitaxial relations to the underlying SE crystal and induce sharp grain boundaries into the system. There are strong indications that the formation of grain boundaries has a significant effect on the properties of superconducting thin films. For example, it was shown that the superconducting properties (e.g. critical temperature and critical magnetic field) of thin Al films can significantly exceed typical bulk values \cite{ferguson2007energy}, where extreme cases are found in granular aluminum films consisting of nano-sized Al grains \cite{pracht2016enhanced}. A different study has shown that the presence of grain boundaries modifies the oxidation dynamics of Al films \cite{nguyen2018atomic}, which can cause oxidation beyond the thickness of native oxide (around 3 nm). This effect is not restricted to thin Al layers only, as the presence of grain boundaries in combination with strain have reportedly altered the SC properties of Nb thin films \cite{david2014stress}. Despite the presence of grain boundaries, a hard superconducting gap was also reported for Al in-situ deposited on shallow InAs 2DEG structures \cite{kjaergaard2016quantized}. 

In terms of hybrid materials, the direct effect of these local crystallographic changes in thin Al remains vastly unexplored, but recent studies have suggested that local disorder in superconducting films could play a key role in the functionality of superconducting qubits \cite{richardson2016fabrication, de2021materials} and proximitized SE/SC structures \cite{cole2016proximity, thomas2022disorder}. Therefore, the nanotexturing of the Al thin films should be investigated in more detail, with particular emphasis on the growth of grain-free materials.

The direct way to improve the crystalline quality of the Al is to control the Al orientation on top of the SE 2DEG. This can be done either by the modification of the surface chemistry by capping the SE surface with different materials, where Sb-based interlayers have shown most promising results in promoting the growth of single crystalline Al films in recent studies \cite{sarney2018reactivity,sarney2020aluminum}. A disadvantage in this case is that presence of Sb often causes problems with the stability of the interface and device fabrication \cite{thomas2019toward, moehle2021insbas}. Another approach is to control the orientation of the Al by modifying the lattice constant of the underlying SE \cite{wang2020dependence}. In that case, the engineered modification of the lattice constant within the quantum well region is necessarily related to a change in the electronic properties of the 2DEG, which may affect the desired balance between electron mobility and proximitized superconductivity.

\begin{figure*}[hbt!]
\vspace{0.2cm}
\includegraphics[scale=0.25]{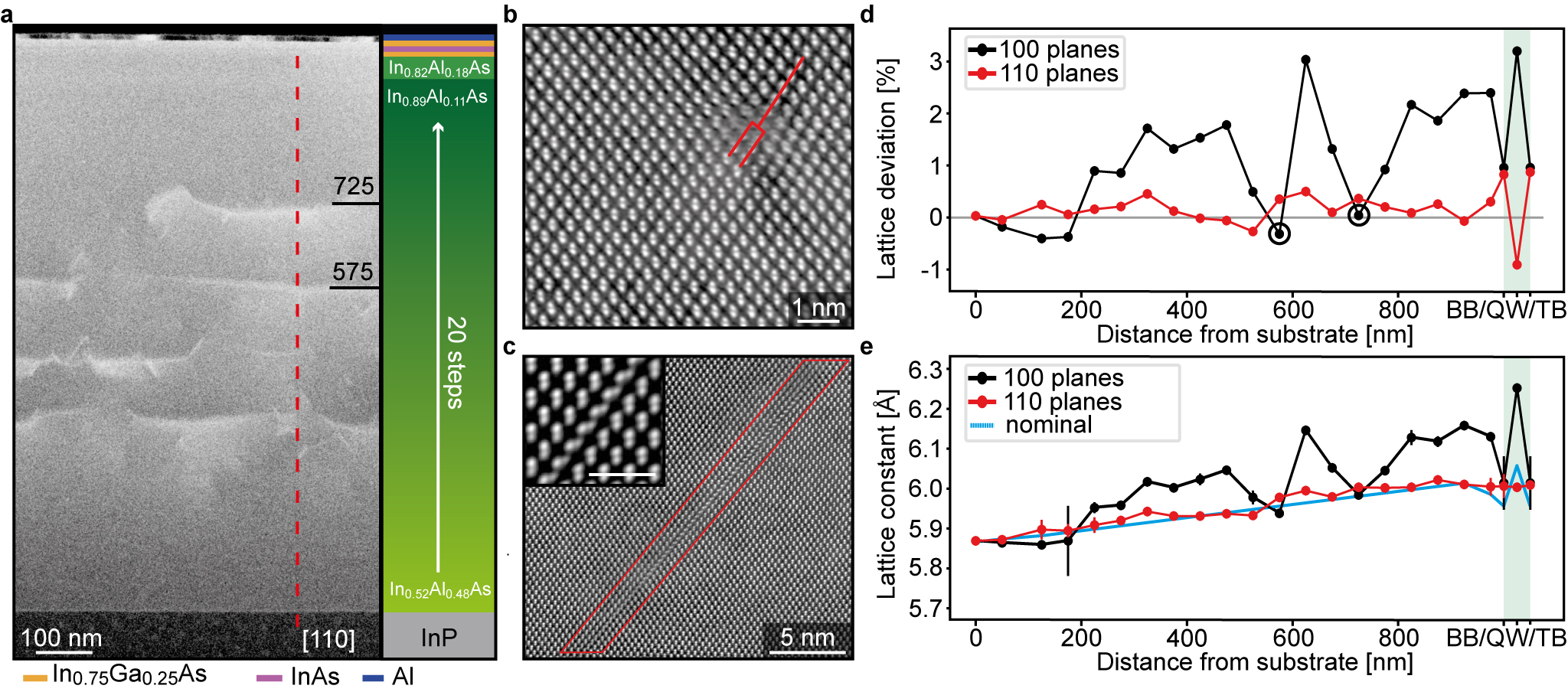}
\vspace{-0.4cm}
\caption{(a)   ADF-STEM image of the shallow InAs 2DEG structure. The red line highlights locations of the images used for the lattice constant evaluation. An illustration of the structure is shown in the right panel. (b) HAADF-STEM image of a single misfit dislocation. (c) ADF-STEM image of an extended twin defect, with a higher magnification image in the inset (scalebar is 1 nm). (d) The deviation of the measured lattice constant as a function of the distance from the substrate (extracted from HAADF-STEM images taken in the center of each grown layer) from the bulk lattice constant for the individual layers. The lattice constant is evaluated from the spacings of both the (001) and (110) planes. BB stands for bottom barrier, QW for quantum well and TB for top barrier, which are plotted in the green shaded area. (e) Same dependency, but for the lattice constant of each individual layer in the structure. The blue line shows the bulk lattice constant of each layer.}
\label{fig1}
\end{figure*}

Here, we introduce a novel approach to control thin Al film crystallinity. First, we give a detailed description of the growth of shallow InAs 2DEGs and analyse the strain evolution in the structure. This allows us to identify the most abundant defects responsible for relaxation within the metamorphic buffer layer and show that the lateral lattice spacing remains stable within the QW region of the 2DEG. Next, we identify the two dominant grain orientations responsible for grain boundaries in our in-situ grown epitaxial Al films. Furthermore, we demonstrate that controlled roughening of the SE surface can be implemented to grow grain-boundary-free Al films over scales of at least 5$\mu$m. Finally, we find that the roughening does not negatively affect the mobility and electron density in the 2DEG, while preserving the functionality of the material in hybrid Josephson junctions (JJs).

\begin{figure}[hb!]
\vspace{0.2cm}
\includegraphics[scale=0.125]{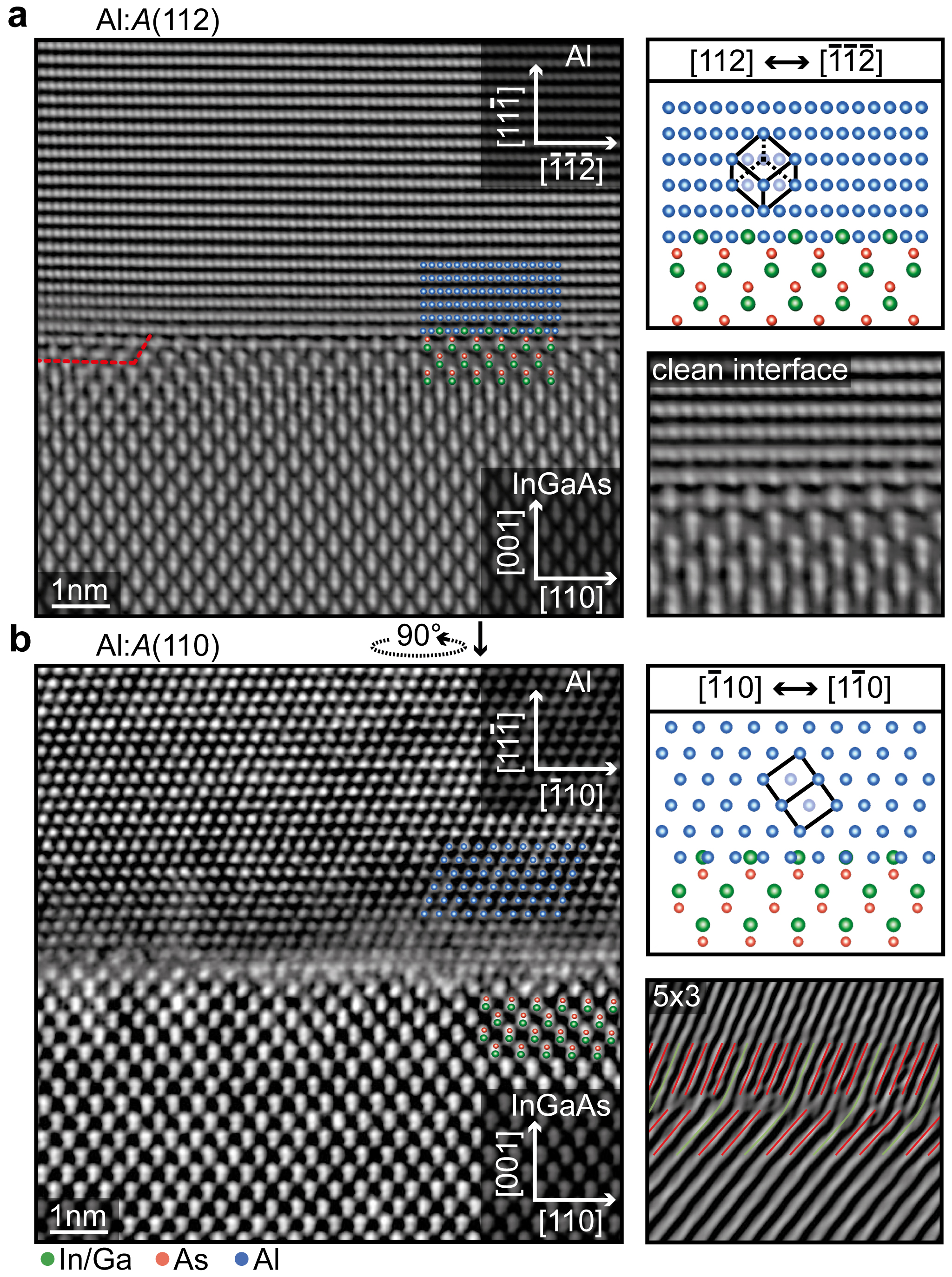}
\vspace{-0.4cm}
\caption{(a) HAADF-STEM image of the interface between an Al:$A$(112) grain and the semiconductor, showing Al(112) to SE(110) lateral matching. Both the image and the model (top right panel) show an undisturbed epitaxial matching. The red line highlights coherent matching over a step on the SE surface. (b) HAADF-STEM image of a grain of the same orientation (Al:$A$), but rotated by 90$^\circ$ around the [111] axis, i.e., showing Al(110) to SE(110) matching (Al:$A$(110)). The Bragg-filtered image (bottom right panel) shows that the lattice mismatch along this direction is relaxed by networks of misfit dislocations with AlxSE 5x3 periodicity.}
\label{fig2}
\end{figure}

\begin{figure}[hb!]
\vspace{0.2cm}
\includegraphics[scale=0.125]{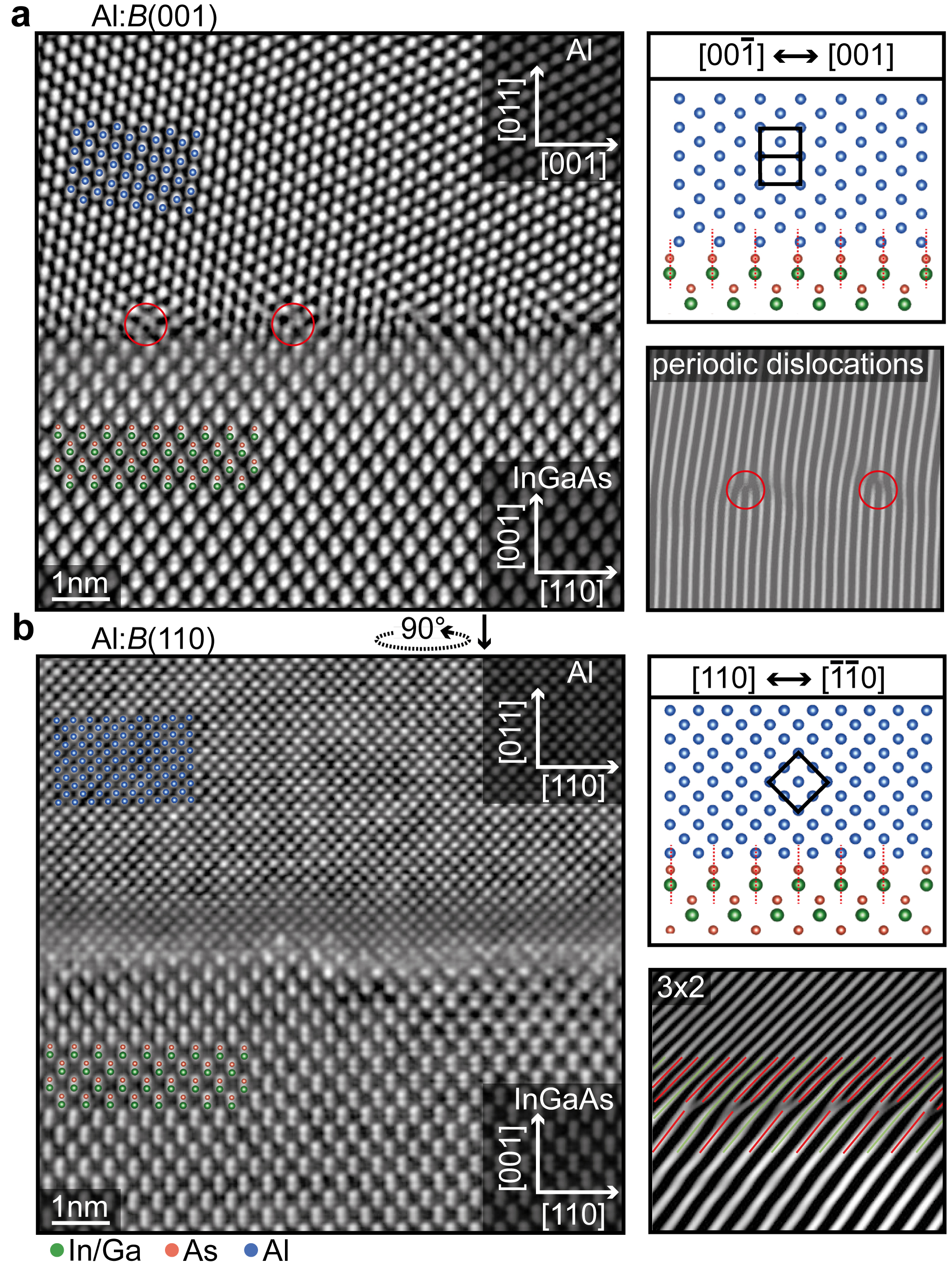}
\vspace{-0.4cm}
\caption{(a) HAADF-STEM image of the interface between an Al:$B$(001) grain and the semiconductor, showing Al(001) to SE(110) lateral matching. The Bragg-filtered image (bottom right panel), shows that the strain along this direction is relaxed by the typical formation of misfit dislocations (highlighted by red circles). (b) HAADF-STEM image of a grain of the same orientation (Al:$B$), but rotated by 90$^\circ$ around the [011] axis, i.e., showing Al(110) to SE(110) matching (Al:$B$(110)). Both the image and the model (top right panel) show significant mismatch. The Bragg-filtered image (bottom right) shows that the lattice mismatch along this direction is relaxed by networks of misfit dislocations with AlxSE 3x2 periodicity.} 
\label{fig3}
\end{figure}

\section{Growth of the semiconductor}

The epitaxial Al films, which are the main focus of this work, were deposited on shallow InAs-based 2DEG structures \cite{shabani2016two}. The InAs QWs were grown on semi-insulating Fe-doped (001) InP substrates by MBE. The native oxide of InP was desorbed at 530$^{\circ}$C for 3 minutes (growth temperature was measured by optical monitoring of the InP band edge \cite{weilmeier1991new}). The first lattice-matched 100 nm thick layer (x = 0.52) was grown at $\sim$ 510$^{\circ}$C, with a V/III ratio of 3.5 and at 1 {\AA}/s. For the purpose of this work, the V/III ratio corresponds to the ratio of As$_{4}$ and group III growth rate (details given in Methods \cite{suppinfo}). 

The lattice mismatch between the InP and InAs was compensated by growth of a step-graded metamorphic In$_{x}$Al$_{1-x}$As buffer layer. The buffer consists of 20 steps with thicknesses of 50 nm, where x increases from \text{0.52 to 0.89} and was grown at $\sim$ 460$^{\circ}$C. We found the lowest semiconductor surface roughness for samples where the growth rate was increased from 1 to 1.5 {\AA}/s and the V/III ratio was reduced from 2.3 to 1.8 when reaching x = 0.705. After the metamorphic buffer, a step-back step with x = 0.82 and a thickness of 8 nm was grown while maintaining the other growth conditions. 

To further smoothen the surface before growth of the QW, a 50 nm thick virtual substrate was grown at 480$^{\circ}$C, using growth rate of 1.5 {\AA}/s and \text{V/III ratio of 3}. The QW region consists of a 6.3 nm thick bottom In$_{0.75}$Ga$_{0.25}$As barrier, grown with V/III ratio of 2.75 while maintaining the growth rate of 1.5 {\AA}/s and growth temperature of 480$^{\circ}$C. The 8.5 nm thick QW is grown at 1.125 {\AA}/s and V/III ratio of 3.25. The In$_{0.75}$Ga$_{0.25}$As top barrier was grown under the same conditions as the bottom barrier. The full structure is illustrated and shown in the  Annular Dark Field (ADF) Scanning Transmission Electron Microscope (STEM) image in Fig. \ref{fig1}a.

For the purpose of this study, we fixed the top barrier thickness to 13.4 nm, as this material configuration provided us with a good ratio of as-grown mobility and strength of the induced proximity effect, as discussed below. In our standard samples, the surface is capped with 2 mono-layers (MLs) of GaAs (at 0.375 {\AA}/s), which provide a barrier for In diffusion into the Al film and improve chemical stability during device fabrication. 

Adjustment of the growth conditions of the metamorphic buffer layer typically results in semiconductor surface without pronounced cross-hatching, isotropic cross-hatching or anisotropic cross-hatching, as shown in the Supplemental Material \cite{suppinfo}. For the optimized samples with strongly anisotropic cross-hatching, grown as described above, we measured electron mobilities around \text{50 000 cm$^{2}$/(V$\cdot$s)} at electron densities around \text{4$\cdot$10$^{11}$ cm$^{-2}$} at 4.2 K (measured in van der Pauw configuration on a 5x5 mm$^{2}$ sample). 

 The structure of the SE is partially adapted from Ref. \cite{shabani2016two}. In our case, we grew a thicker bottom barrier, upper barrier and QW. Yet, the SE/SC coupling in such a structure remained on a scale relevant for our transport experiments \cite{haxell2022large, haxell2022microwave, hinderling2022flip, suppinfo}. We observed that the behaviour of our proximitized devices exceeds the theoretical limits given for specific barrier thicknesses in Ref. \cite{shabani2016two}. This should be thoroughly investigated in future works, as the proximity effect in hybrid materials is expected to be affected by the interplay of strain, material composition of the barrier, QW dimensions, quality of the Al film and the epitaxial interface.  

One of the important aspects of the SE structure is strain, which can strongly affect its transport properties, but also the epitaxial relation to the Al film. Strain fields related to defect formation and lattice relaxation in the In$_{x}$Al$_{1-x}$As metamorphic buffer are clearly visible in the ADF STEM  image in Fig. \ref{fig1}a. Importantly, they do not extend into the QW region, which remains defect free. The most abundant defects which dominate the relaxation and generate the strain fields are shown in High Angle Annular Dark Field (HAADF) STEM images in Fig. \ref{fig1}b and c. The first type, shown in b, are isolated misfit dislocations. The second type are single twin planes extending over tenths of nanometers, shown in c. Both defect types are randomly distributed throughout the strain relaxation region. 

To further understand the strain distribution within the structure, we extracted the lateral and vertical lattice constants from STEM images by measuring the (1$\bar{1}$0) and (001) lattice plane spacings (details are given in the Supplemental Material \cite{suppinfo}) in the center of each layer along the red dashed line in Fig \ref{fig1}a. A percentual deviation from the bulk lattice constant is plotted as a function of distance from the substrate in Fig. \ref{fig1}d. As expected, the (110) lattice spacing is rather stable, while the (001) spacing expands and varies throughout the buffer layer. The (001) spacing shows two local minima, which overlap with the most strained regions visible in Fig. \ref{fig1}a. The measured lattice spacing is shown in Fig. \ref{fig1}d, showing that the lateral (110) lattice spacing remains almost constant in the QW region, as expected for growth on a relaxed virtual substrate. The average spacing and deviation is 6.00 $\pm$ 0.02 \AA $ $ for the bottom barrier, QW and top barrier, which yields 0.9 $\%$ compressive strain compared to InAs. This negligible variation of the lattice constant is important, as the strain can be considered as fixed between the SE and Al. We observed a similar relaxation behaviour in multiple samples.

\begin{figure*}[htb!]
\vspace{0.2cm}
\includegraphics[scale=0.25]{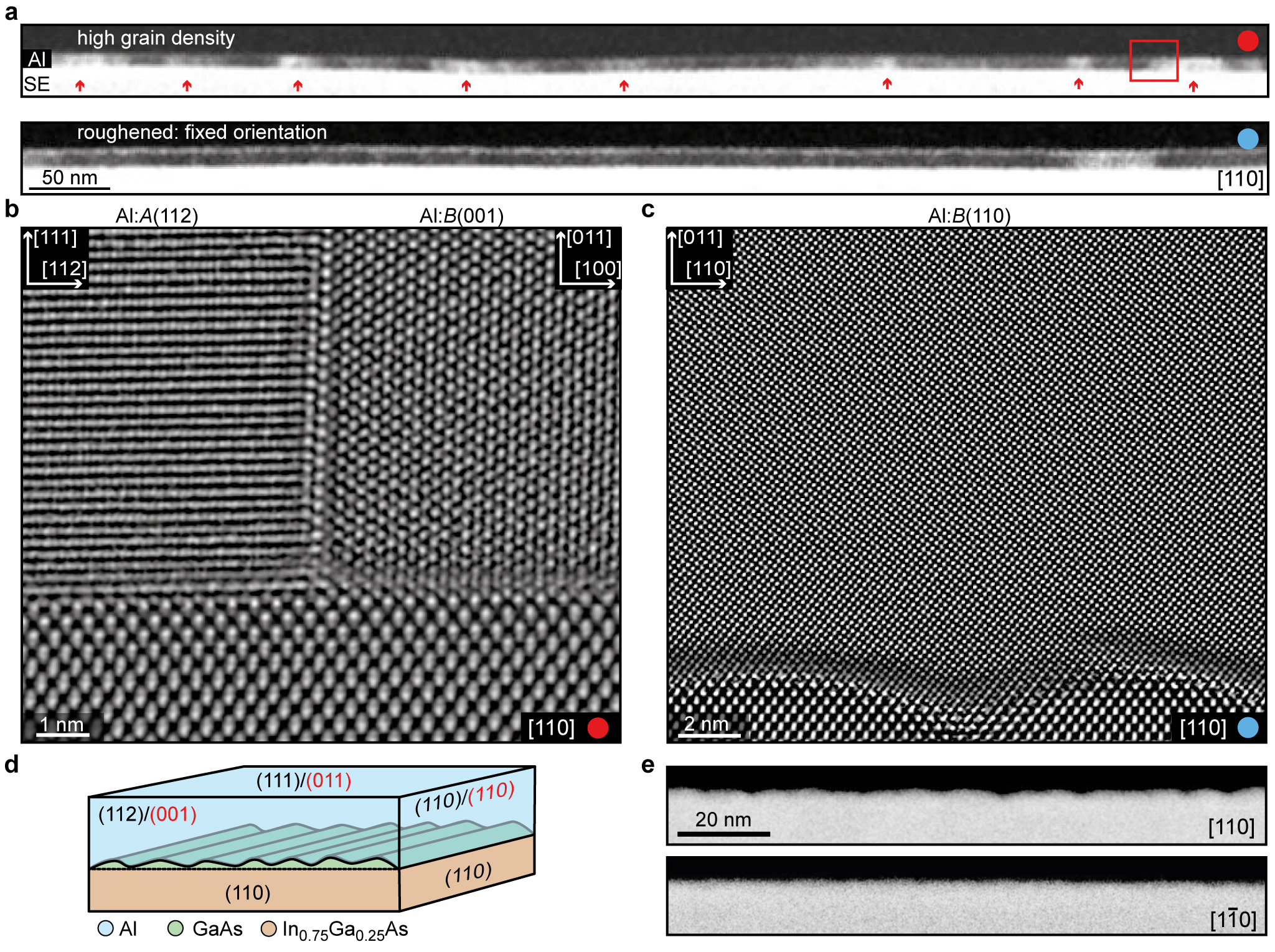}
\vspace{-0.4cm}
\caption{ (a) ADF-STEM images of high grain density (top panel) and a single orientation (bottom panel) Al thin films. The red arrows highlight regions with strong changes in contrast. (b) A higher magnification and contrast normalized HAADF image showing structure of one of the possible grain boundaries, highlighted by red arrows in (a). (c) Contrast normalized HAADF-STEM image of an Al film deposited on a roughened surface. (d) Schematic of the possible orientations of different Al films on a heavily anisotropically roughened surface. (e) HAADF-STEM images of two perpendicular projections ([110] and [1$\bar{1}$0]) of the top interface of the semiconductor, demonstrating the anisotropy and scale of the intentional roughening. } 
\label{fig4}
\end{figure*} 

\section{Al thin films}

For the in-situ Al deposition, the wafer was moved to an ultra-high vacuum buffer chamber (1$\times$10$^{-11}$ mbar) directly after growth of the shallow InAs 2DEG, and re-transferred when the pressure in the growth chamber reached 1$\times$10$^{-10}$ mbar, i.e., once the As background dropped after closing the As valve. The manipulator was rotated towards the liquid nitrogen shroud. No active cooling was involved and therefore the cooling power is only related to the MBE system being set to idle state and the sample holder going towards thermal equilibrium with LN$_2$ cooled cryo-shrouds (in our system for at least 12 hours to reach approximately -30$^{\circ}$C). The Al films (\text{12 nm} thick in samples used in the presented transport experiments) were deposited at a rate of 1\AA/s at a pressure of 3-5$\times$10$^{-11}$ mbar in the growth chamber. We note that the sample surface must remain cold during the deposition. Therefore, the used growth rate (controlled by the cell temperature) needs to be optimized for each specific MBE system geometry, as it determines heat delivered to the surface of the sample during growth. After the deposition, the wafer was moved to the load lock chamber (below 5$\times$10$^{-10}$ mbar). In order to prevent thermal dewetting of the Al film, we transferred the sample as fast as possible to the load lock, i.e., it remained cold until controlled oxidation. In our case, the sample was oxidized and brought to room temperature in the load lock by slowly venting with an Ar/O$_{2}$ (90/10 \%) mixture over the course of 15 minutes (at 25$^{\circ}$C). An AFM image of the oxidized surface for both the controlled oxidation and venting into atmosphere is shown in the Supplemental Material \cite{suppinfo}. The samples that were unloaded via controlled oxidation shown lower surface roughness and reduced formation of large AlO$_{x}$ grains than what we observed in samples directly unloaded into ambient atmosphere.

\begin{figure*}[htb!]
\vspace{0.2cm}
\includegraphics[scale=0.25]{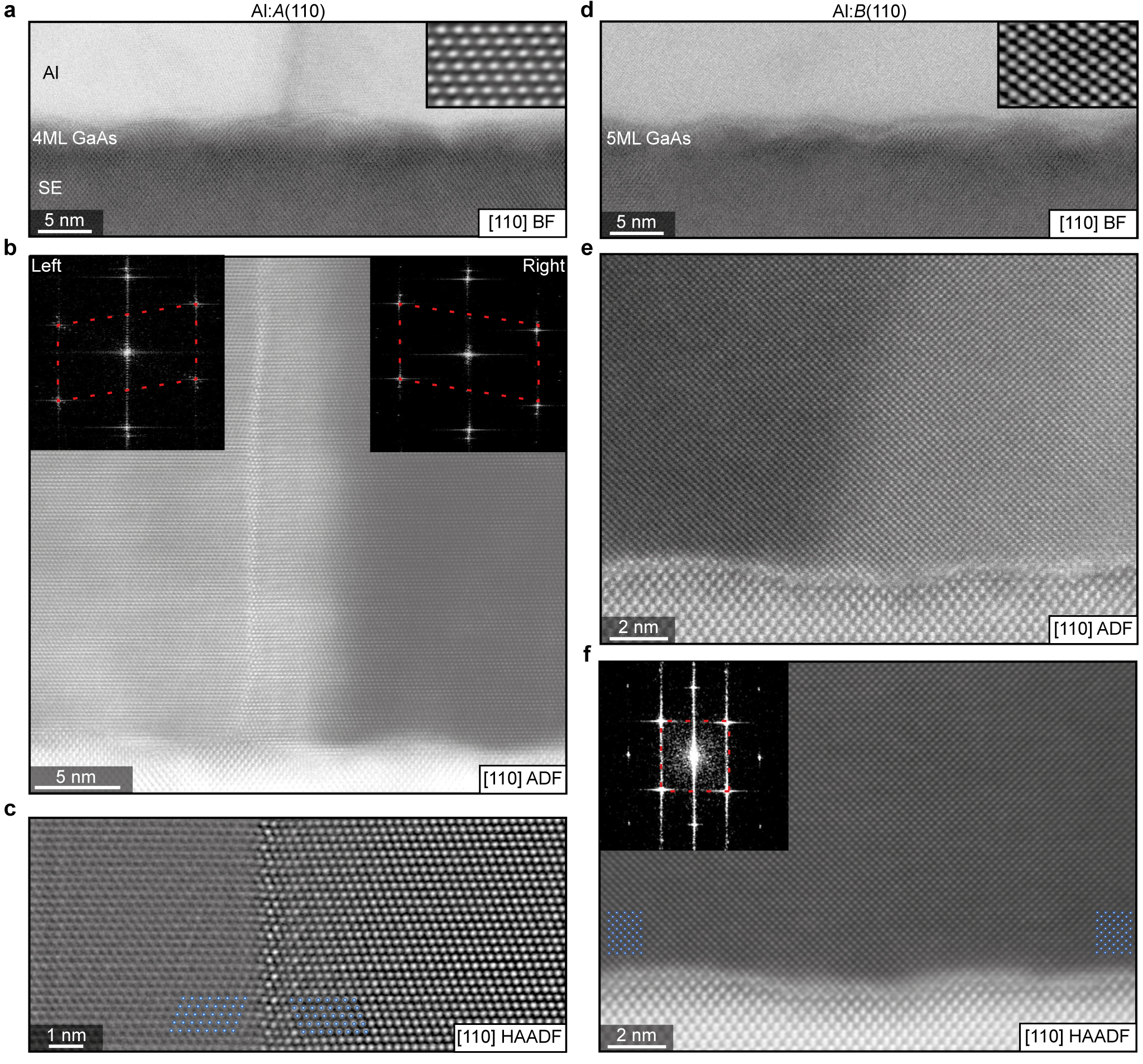}
\vspace{-0.4cm}
\caption{ (a) BF-STEM image of Al films deposited on a surface capped with 4 MLs of GaAs showing a change of contrast in the center. The HAADF image in the inset shows the orientation of the Al film. (b) ADF-STEM image, showing the origin of the contrast in (a). The fast Fourier transform in the inset shows that the Al crystal is mirrored over the boundary. (c) Contrast normalized HAADF image of the crystal around the boundary. (d) BF-STEM image of Al films deposited on a surface capped with 5 MLs of GaAs. The HAADF-STEM image in the inset shows the orientation of the Al film. (e) ADF-STEM image, showing an abrupt change in contrast in the same film as in the lower panel of Fig. \ref{fig4} a. (f) HAADF-STEM image of the same area. The Fast Fourier Transform does not show a detectable change in the crystal structure.} 
\label{fig5}
\end{figure*} 

\begin{figure}[hb!]
\vspace{0.2cm}
\includegraphics[scale=0.12]{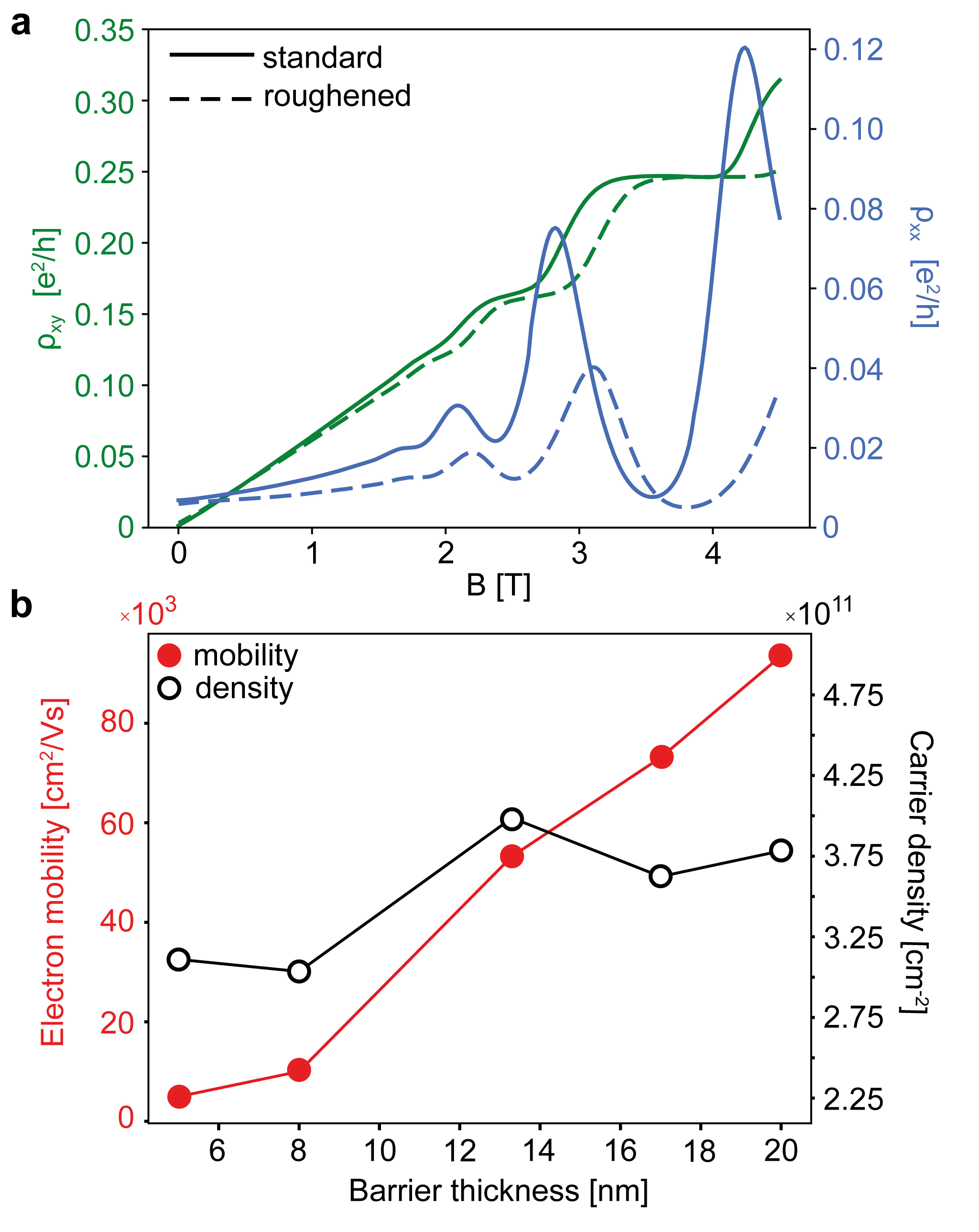}
\vspace{-0.4cm}
\caption{(a) Hall resistance (green) and longitudinal resistance (blue) as a function of magnetic field for the InAs 2DEG on standard (full line) and intentionally roughened (dashed line) SE surface, after removal of the epitaxial Al. (b) Electron mobility and carrier density as a function of top barrier thickness (for standard structures 2 MLs of GaAs cap and after Al wet etching).} 
\label{fig6}
\end{figure}

In our standard samples, where the surface was capped by depositing 2 MLs of GaAs and grown as described above, we mainly observe the presence of two distinctly oriented grains of Al. The first type is shown in Fig. \ref{fig2}. In this case, the Al with (111) out-of-plane orientation adapts either (112) in (a) or (110) in (b) lateral matching to the (110) planes of the semiconductor (here labeled as Al:$A$(112) and Al:$A$(110)). Therefore in STEM, it is possible to observe two distinct projections and epitaxial relations of the Al:$A$ crystal, depending on its alignment to the substrate. For the Al:$A$(112) orientation shown in Fig. \ref{fig2}a, we observe a clean, fully epitaxial interface, with apparent (and ordered) intermixing within the first matching monolayers. For the Al:$A$(110) orientation shown in Fig. \ref{fig2}b, we observe 5x3 matching, i.e., there are two misfit dislocations at the interface per 3 planes in the substrate. 

The second grain type is shown in Fig. \ref{fig3}, where the Al adapts (011) out-of-plane orientation and lateral matching as either (001) in (a) or (110) in (b) to the (110) planes of the semiconductor (here labeled as Al:$B$(001) and Al:$B$(110)). For this grain, we observe formation of misfit dislocations for both alignments to the (110) planes of the semiconductor. For the Al:$B$(001) matching, the film relaxes by the formation of periodic arrays of misfit dislocations with larger spacing. In contrast, there is a single misfit dislocation per 2 planes in the substrate, i.e., a 3x2 match for the Al:$B$(110) orientation. 
We observed that both grain types were equally present in the samples and often adapted small tilts (e.g. the slight tilt visible in Fig. \ref{fig3}a). A larger scale ADF-STEM overview of such a sample is shown in the top panel of Fig. \ref{fig4}a. The changing ADF contrast indicates a high density of grains with different orientation. The contrast can originate from the two different types of grains, mutual tilt of grains with the same orientation and/or from the two possible projections (i.e. 90$^{\circ}$ grain rotation). An example of a boundary between the two different grain types is shown in Fig. \ref{fig4}b. The smoothness of the SE surface in the HAADF STEM image indicates that the presence of the boundary is not associated with neither a surface step nor a crystallographic defect in the semiconductor.

These partially polycrystalline Al films performed consistently with previous reports in literature in our transport experiments which rely on the combined SE/SC system \cite{haxell2022large, haxell2022microwave, hinderling2022flip, suppinfo}. On the other hand, consistent results were achieved only when processing temperatures did not exceed 175$^{\circ}$C. This is due to degradation of both the Al film and the SE/Al interface, which was investigated by in-situ annealing of a lamella prepared from the standard material in a scanning transmission electron microscope. While ramping the temperature up to 225$^{\circ}$C, we observed degradation of individual grains into amorphous Al, diffusion grain boundaries and also intermixing of Al at the interface and even local recrystallization into a zinc-blende structure, as reported in the Supplemental Material \cite{suppinfo}. Reaching such high temperatures during device fabrication had a negative impact on the transport properties of the material and the fabricated devices could not be further utilized in our experiments. The observed degradation is also expected to happen in the smallest features of our devices, since their dimensions are often comparable to the lamella (5 micrometers in length and below 50 nm thick). Importantly, the observed recrystallization was selective to the specific grain type and some of the investigated grains remained crystalline and their interface to the SE was stable even at 225$^{\circ}$C. This difference in thermal energy necessary do dissolve the interface for specific grain types needs to be considered during development of fabrication processes and is one of the motivations to develop growth of single crystalline Al films on InAs 2DEGs.

\section{Effect of roughening on crystallography}

The Al film is locked into a single orientation using intentional roughening of the In$_{0.75}$Ga$_{0.25}$As surface by deposition of more GaAs on the InGaAs top barrier. This is shown for a sample capped with 5 MLs of GaAs in the bottom panel of Fig. \ref{fig4}a. In such films, we did not detect the presence of any grain boundaries, i.e., grains with different orientations over a range of 5 $\mu$m, the typical size of our investigated lamellae. 

We found that nanoscale surface roughening of the SE induced by GaAs deposition is anisotropic along the [110] and [1$\bar{1}$0] directions, as shown in the STEM images in Fig. \ref{fig4}e and as investigated by STEM tomography \cite{nicolai2021application} in the Supplemental Material \cite{suppinfo}. This is likely related to highly anisotropic diffusion of Ga atoms during deposition at elevated temperatures \cite{ohta1989anisotropic} and a large mismatch between the In$_{0.75}$Ga$_{0.25}$As top barrier and GaAs cap \cite{dieguez1997defects}. 

The nanoscale roughening of the SE surface seems to affect the orientation of the Al in two ways, summarized in Fig. \ref{fig4}c. Firstly, it leads to a selectivity of the out-of-plane grain orientation and reduced formation of grain boundaries. Similarly, growth on roughened or nano-patterned substrates, i.e. nano-heteroepitaxy, was previously used to reduce formation of threading dislocations and residual strain in heterostructures with highly mismatched materials \cite{chen2018study,feng2016grouped}. Secondly, the modulation of the SE surface locks the in-plane orientation into either the Al:$A$ or Al:$B$ grains, so that the $\{$110$\}$ planes of the Al align with the (1$\bar{1}$0) planes of the SE. Both effects are likely related to the enhanced 3D character of the roughened surface, which gives the Al layer an additional degree of freedom for strain relaxation. 

We observed this behaviour in samples where the InGaAs top barrier was capped with more than 3 MLs of GaAs. An example is shown in Fig \ref{fig5}, where the Al layer is locked in \ref{fig5}a as Al:$A$(110) after capping with 4 MLs of GaAs and \ref{fig5}b as Al:$B$(110) after capping with 5 MLs of GaAs. For the sample capped with 4 MLs GaAs in Fig. \ref{fig5}a, we observed that the Al only adapted the Al:$A$(110) orientation. Surprisingly, we observed regions with abruptly changing Bright Field (BF) and ADF contrast in the STEM images. Yet, this contrast was not related to a major change in the crystal orientation, but to mirroring with respect to the Al(111) planes, as shown in Fig. \ref{fig5}b. This is apparent from the mirroring of the fast Fourier transform spectra and the atomic arrangement in the HAADF-STEM zoom-in in Fig. \ref{fig5}c. This implies that even when the whole Al film is locked as Al:$A$\{110\}, the Al:$A$($\bar{1}$10) and Al:$A$(1$\bar{1}$0) are not degenerate in this grain orientation. Hence, the Al film in this crystal orientation is naturally prone to twinning and the related formation of incoherent grain boundaries.

This is different for a sample capped with 5 MLs of GaAs, shown in Fig. \ref{fig5}d, where the crystal orientation was locked into Al:$B$(100). Similarly to the previous case, the lateral matching to the substrate was fixed as Al(110) to SE(110) for the whole layer. In this case, we detected only subtle differences in ADF and BF contrast, as shown in Fig. \ref{fig5}e. Such a detail in the acquired image of the Al layer corresponds to the region of the film with a change in contrast shown in the lower panel of Fig. \ref{fig4}a. These changes of contrast were sparse in the films and are likely related to subtle tilts of the crystal, as no apparent crystallographic change was seen in neither the HAADF-STEM image nor the fast Fourier transform in Fig. \ref{fig5}f. Compared to Al:$A$, the symmetry of Al:$B$ type grain with respect to the \{110\} in-plane direction results in the formation of a fully single crystalline film. The subtle observed tilts might be related to relaxation of the strain induced by natural bending of the lamella for samples incorporating metamorphic buffer layers. We note that the differences in the capping layer thickness used in this study are at the experimental limits even with the utilized MBE technique. A consistent study of the capping layer thickness is needed in future works to gain full control over the Al grain growth selectivity. 

In addition, we observed that the grain distribution in the films was affected by various additional factors, such as lamella preparation, strain, oxidation etc., which complicated the capability to perform consistent studies. Also, the Al film degraded and new grains appeared if the lamella was stored in ambient conditions for more over one month, indicating room temperature recrystallization in the films, as shown in the Supplemental Material \cite{suppinfo}.  

\section{Effect of roughening on transport properties}

\begin{figure*}[ht]
\vspace{0.2cm}
\includegraphics[scale=0.25]{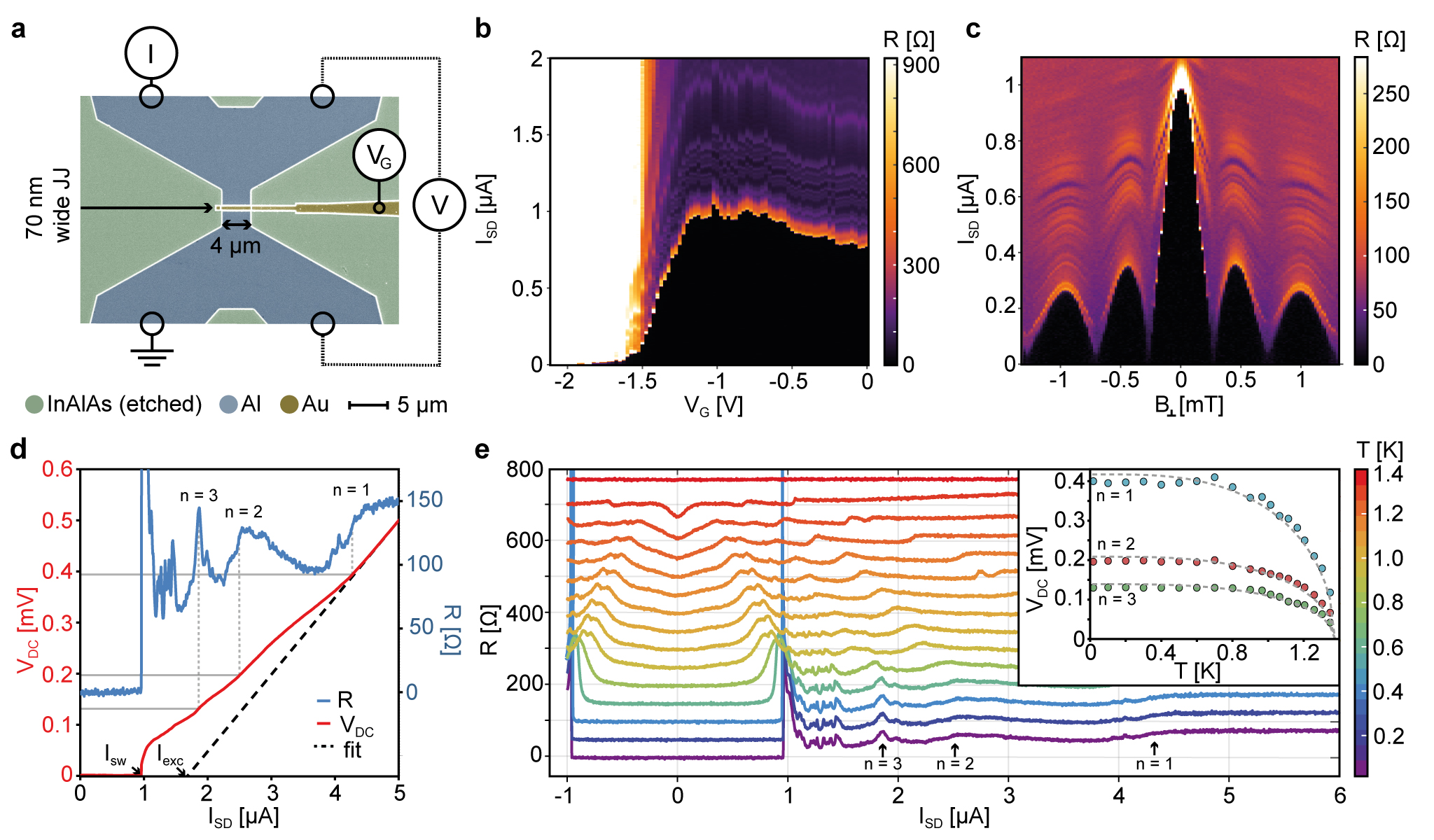}
\vspace{-0.4cm}
\caption{(a) False-colored SEM image of the measured Josephson junction, highlighting the top-gate (gold) and the epitaxial Al leads of the JJ (blue). The terminals for the bias current $I$ = $I_\mathrm{SD}$ + $I_\mathrm{AC}$, measured 4-probe voltage $V_\mathrm{DC}$ and $V_\mathrm{AC}$ and gate voltage $V_\mathrm{G}$ are highlighted. (b) Differential resistance $R$ = $V_\mathrm{AC}$/$I_\mathrm{AC}$  of the JJ as a function of source-drain current $I_\mathrm{SD}$ and gate voltage $V_\mathrm{G}$. (c) Dependence of $R$ of the JJ as a function of $I_\mathrm{SD}$ and out-of-plane magnetic field $B_{\perp}$. (d) Dependence of $V_\mathrm{DC}$ (red, left axis) and differential resistance $R$ (blue, right axis) on $I_\mathrm{SD}$, measured at a base temperature of 18 mK. The black dashed line shows a linear fit to the I-V trace above $\sim$ 0.4 mV. The individual MAR peaks in $R$ are labeled by $n$. They are related to the measured voltage via the I-V trace (grey dotted lines). The black arrows point to the position of the switching and excess current, respectively. (e) Temperature dependence of $R$ as function of bias current $I_\mathrm{SD}$. For clarity, each temperature trace is offset by 50 $\Omega$. Extracted $V_\mathrm{DC}$ positions of the MAR peaks in $R$ for $n$ = 1, 2, 3 as a function of temperature are shown in the inset. The dashed lines correspond to fits to the BCS theory for the nominal Al superconducting gap.} 
\label{fig7}
\end{figure*}

Even though the reduction of grain boundaries by surface roughening offers many benefits in terms of the crystal structure of the Al films, as discussed above, it is crucial to maintain the transport properties of the 2DEG. In Fig. \ref{fig6}a, we compare magneto-transport measurements of the roughened and standard materials. The measurements were performed in a van der Pauw configuration at 4 K on 5x5 mm as-grown samples after the Al film was removed by wet-etching. For the standard non-roughened surface we measured an electron mobility of \text{53 000 cm$^{2}$/(V$\cdot$s)} at an electron density of \text{4.0$\cdot 10^{11}$ cm$^{-2}$}. For the sample with the same structure but intentionally roughened surface (5 MLs of GaAs) we measured \text{51 000 cm$^{2}$/(V$\cdot$s)} at an electron density of \text{3.5$\cdot 10^{11}$ cm$^{-2}$}. The Drude mobility and charge carrier density do not significantly change for the roughened sample. We consider this difference to be negligible, since the properties and chemistry of the surface were reported to have a strong effect on the transport in shallow InAs 2DEGs \cite{pauka2020repairing, hatke2017mobility}. We observed clear quantum Hall and Shubnikov-de Haas transport features at higher magnetic fields in both samples. The variation in quantum Hall behaviour at higher magnetic fields can be related to the difference in electron density and sample geometry, for example. In practical applications of the hybrid material, the rather low critical magnetic field of the superconducting aluminium typically limits the magnetic field range in transport experiments to below 1.5 T in a parallel field and tens of mT in a perpendicular field, where the behaviour of the compared materials is almost identical. 

To further investigate the effect of the near-surface structure on the electron mobility, we grew a series of standard samples (capped with 2 MLs of GaAs) with the In$_{0.75}$Ga$_{0.25}$As top barrier thickness varying from 5 to 20 nm. The electron mobility increased almost linearly from 5 000 up to 100 000 cm$^{2}$/(V$\cdot$s) while the density remained between 3$\cdot 10^{11}$ and 4$\cdot 10^{11}$ cm$^{-2}$, as shown in Fig. \ref{fig6}b. This demonstrates that the effect of the intentional roughening of the surface on the electron mobility is indeed negligible in comparison to the effect of other growth parameters and changes in the structural design. An enhancement in mobility is desirable, but an increase in the top barrier thickness can also affect the strength of the proximitized superconductivity, as will be discussed below and in the Supplemental Material \cite{suppinfo}.

Finally, to assess the influence of the intentional roughening of the SE/SC interface on the superconducting properties of the proximitized system, we investigated the basic characteristics of a top-gated SC-normal-SC (SNS) planar Josephson junction (JJ) device, shown in Fig. \ref{fig7}a. The JJ device was defined by selectively etching Al (blue) to expose the III-V semiconductor below, which was controlled via an electrostatic gate (gold). The differential resistance $R$ = $V_\mathrm{AC}$/$I_\mathrm{AC}$ as a function of bias current $I_\mathrm{SD}$ and top gate voltage $V_\mathrm{G}$, is shown in Fig. \ref{fig7}b. The transition from superconducting to the resistive state (at the switching current $I_\mathrm{sw}$) was tuned by sweeping $V_\mathrm{G}$, where full suppression of the supercurrent appeared around \text{$V_\mathrm{G}$ = -1.8 V}. For all following experiments we set \text{$V_\mathrm{G}$ = -1.1 V} where we expected the SE segment to be tuned to the single sub-band regime.

Next, we investigated the SC-to-normal transition as a function of the out-of-plane magnetic field $B_{\perp}$. We observed a Fraunhofer pattern typical for planar JJs, shown in Fig. \ref{fig7}c. The suppression of the switching current occurs when the flux penetrating the junction area equals the magnetic flux quantum, as further discussed in the Supplemental Material \cite{suppinfo}. For large current bias the junction resistance attained a normal state value \text{$R_\mathrm{n}$ = 150 $\Omega$} (the same as above the critical temperature of \text{$T_\mathrm{c}$ $\approx$ 1.38 K}). 

At zero magnetic field we found a maximum switching current $I_\mathrm{sw}$ $\approx$ 1 $\mu$A, and excess current \text{$I_\mathrm{exc}$ $\approx$ 1.7 $\mu$A}, as shown in Fig. \ref{fig7}d. Both the products $I_\mathrm{sw}R_\mathrm{n}$ and $I_\mathrm{exc}R_\mathrm{n}$ can be related to the interface transparency and the induced superconducting gap energy \cite{flensberg1988subharmonic,kjaergaard2017transparent}. This is consistent with other reports in literature and discussed in more detail in the Supplementary Material \cite{suppinfo}.

The induced gap energy can also be extracted by analysing the modulation of the differential resistance in the resistive state. This originates from multiple Andreev reflection (MAR) and became apparent when the current bias was larger than the switching current for both the gate and magnetic field dependencies. The MAR is a signature of coherent charge transport at finite bias within the induced superconducting gap, $\Delta$*, i.e., at sub-gap voltages $V_\mathrm{DC} < 2 \Delta_\mathrm{Al}$/$e$ ($\Delta_\mathrm{Al}$ = $1.76 \cdot k_\mathrm{B} \cdot T_\mathrm{c}$ $\approx$ 210 $\mu$eV) \cite{flensberg1988subharmonic, kjaergaard2017transparent}. The position in $V_\mathrm{DC}$ of individual MAR peaks is related to the size of the induced gap, as shown in Fig. \ref{fig7}d-e. The induced gap $\Delta$* is given via $e \cdot V_\mathrm{DC}$ = 2 ${\Delta}^*/{n}$, where \text{$n$ = 1, 2, 3, ...} This is highlighted in Fig. \ref{fig7}d for a measurement at 18 mK. By averaging the value extracted for n = 1, 2 and 3, we found \text{$\Delta$* = 197 $\pm$ 2 $\mu$eV} which is close to the nominal $\Delta_\mathrm{Al}$. This value of $\Delta$* remains almost constant up to 700 mK, as apparent from the dependency of $V_\mathrm{DC}$ assigned to different $n$ on temperature shown \ref{fig7}e. Above 700 mK, $\Delta$* starts to change and follows the BCS relation \cite{waldram1996book} for the Al gap energy \text{$\Delta_\mathrm{Al}(T,n)$ = 2/$n\cdot \Delta_\mathrm{Al}(0) \cdot \textrm{tanh} (1.74 \sqrt{T/T_\mathrm{c}-1})$}.

To demonstrate the quality of the intentionally roughened material, we compared the measurements to a nominally identical JJ device fabricated from the standard non-roughened material (2 MLs of GaAs). The same analysis of MAR as introduced above yields \text{$\Delta$*$_\mathrm{r}$ = 197 $\pm$ 1 $\mu$eV} for the roughened material and \text{$\Delta$*$_\mathrm{s}$ =  184 $\pm$ 6 $\mu$eV} for the standard material, both at 500 mK. The similar values of $\Delta$* together with the no apparent change in electron mobility, show that the quality of the hybrid material is not significantly affected by the implemented roughening.

Furthermore, we investigated a JJ fabricated from the standard material where the top barrier thickness was increased from \text{13.4 nm} to 20 nm (20 nm top barrier thickness and 2ML GaAs cap). In this case, we observed a reduced number of clear MAR features which complicated the correct assignment of $n$, as shown in the Supplemental Material. The analysis yields an upper bound of \text{$\Delta$*$_\mathrm{20nm}$ = 170 $\mu$eV} for $n$ = 2 at 550 mK, which is lower but still comparable to the samples with a thinner barrier and unexpected in comparison to theoretical predictions \cite{shabani2016two}. The possibility to increase the electron mobility (i.e. tune the top barrier thickness), while preserving the induced gap size can be advantageous in future device designs. More importantly, it was recently suggested by Awoga et al. \cite{awoga2022mitigating}, that a weaker coupling between the SE and SC can mitigate influence of local disorder, which is expected to be one of the significant factors that can impair device performance. More detailed analysis of the MAR data for all the investigate samples is given in the Supplemental Material \cite{suppinfo}. Our results show that more experimental investigations of growth series with consistent change of parameters are necessary to relate the material properties to current theoretical description of hybrid devices.

Overall, we found that all three material designs exhibited a highly transparent interface, gap size comparable to the BCS theory and other reports in literature, signatures of coherent ballistic transport through a JJ and a good performance in our other experiments focused on InAs/Al hybrids \cite{haxell2022large, haxell2022microwave, hinderling2022flip}. In addition, studying differently designed material structures implemented into hybrid devices is important for further understanding of the influence of individual material parameters on transport properties.

\section{Conclusion}

This work provides a full description of the growth of shallow high-quality InAs 2DEGs with epitaxial Al films by MBE. We have demonstrated that deposition of 5 MLs of GaAs on the top barrier anisotropically roughens the surface. Subsequent in-situ deposition of a thin epitaxial Al film results in the formation of an Al layer with a single crystal orientation, which remains free of grain boundaries on at least a 5 $\mu$m scale. Our results also indicate that the intentional roughening can be further optimized to achieve full control over the crystallographic orientation of the Al film.

We found that the introduced roughening did not impair the electron mobility and carrier density of the shallow 2DEG. In addition, the material showed qualities comparable to state-of-the-art SE/SC hybrid JJ-based devices in transport experiments. This implies that local nanotexturing of the semiconductor surface (likely at scales below the Fermi wavelength) does not negatively impact the transport properties of the proximitized system, and provides new experimental insight into the requirements on quality of the interface in hybrid materials, which were recently subject of extensive debate \cite{cole2016proximity,thomas2022disorder}. 

Apart from the possibility of extending the enhanced epitaxy on roughened surfaces to other materials, control over the Al orientation has implications for interface engineering and chemical stability of the well-established hybrid InAs/Al structures. Most importantly, the removal of grain boundaries from the Al film allows for the possibility to form more complex and still fully epitaxial heterostructures on top of the Al. Finally, we hope that the extensive details of material synthesis provided in this work allows for reproducible growth of the material and make it widely accessible to the scientific community. 

\section{Acknowledgement} 
We thank \text{Christian Reichl}, \text{Stefan Fält} and \text{Mattias Beck} for technical support. We also thank \text{Sjoerd Telkamp}, \text{Clemens Todt} and \text{Tilman Tröster} for fruitful discussions. We thank the IBM Quantum Academic Network for financial support. We acknowledge the support of the Cleanroom Operations Team of the Binnig and Rohrer Nanotechnology Center (BRNC). This work was supported by the Swiss National Center of Competence in Research Quantum Science and Technology, QSIT, together with the European Union and the state of Berlin within the frame of the European Regional Development Fund (ERDF) under project number 2016011843 and by Czech Science Foundation grant number 22-22000M. \text{F. N.} acknowledges support from the European Research Council grant number 804273 and the Swiss National Science Foundation grant number 200021 201082.

\bibliography{ms}

\begin{thebibliography}{44}%
\makeatletter
\providecommand \@ifxundefined [1]{%
 \@ifx{#1\undefined}
}%
\providecommand \@ifnum [1]{%
 \ifnum #1\expandafter \@firstoftwo
 \else \expandafter \@secondoftwo
 \fi
}%
\providecommand \@ifx [1]{%
 \ifx #1\expandafter \@firstoftwo
 \else \expandafter \@secondoftwo
 \fi
}%
\providecommand \natexlab [1]{#1}%
\providecommand \enquote  [1]{``#1''}%
\providecommand \bibnamefont  [1]{#1}%
\providecommand \bibfnamefont [1]{#1}%
\providecommand \citenamefont [1]{#1}%
\providecommand \href@noop [0]{\@secondoftwo}%
\providecommand \href [0]{\begingroup \@sanitize@url \@href}%
\providecommand \@href[1]{\@@startlink{#1}\@@href}%
\providecommand \@@href[1]{\endgroup#1\@@endlink}%
\providecommand \@sanitize@url [0]{\catcode `\\12\catcode `\$12\catcode
  `\&12\catcode `\#12\catcode `\^12\catcode `\_12\catcode `\%12\relax}%
\providecommand \@@startlink[1]{}%
\providecommand \@@endlink[0]{}%
\providecommand \url  [0]{\begingroup\@sanitize@url \@url }%
\providecommand \@url [1]{\endgroup\@href {#1}{\urlprefix }}%
\providecommand \urlprefix  [0]{URL }%
\providecommand \Eprint [0]{\href }%
\providecommand \doibase [0]{http://dx.doi.org/}%
\providecommand \selectlanguage [0]{\@gobble}%
\providecommand \bibinfo  [0]{\@secondoftwo}%
\providecommand \bibfield  [0]{\@secondoftwo}%
\providecommand \translation [1]{[#1]}%
\providecommand \BibitemOpen [0]{}%
\providecommand \bibitemStop [0]{}%
\providecommand \bibitemNoStop [0]{.\EOS\space}%
\providecommand \EOS [0]{\spacefactor3000\relax}%
\providecommand \BibitemShut  [1]{\csname bibitem#1\endcsname}%
\let\auto@bib@innerbib\@empty
\bibitem [{\citenamefont {Hays}\ \emph {et~al.}(2021)\citenamefont {Hays},
  \citenamefont {Fatemi}, \citenamefont {Bouman}, \citenamefont {Cerrillo},
  \citenamefont {Diamond}, \citenamefont {Serniak}, \citenamefont {Connolly},
  \citenamefont {Krogstrup}, \citenamefont {Nyg{\aa}rd}, \citenamefont
  {Levy~Yeyati} \emph {et~al.}}]{hays2021coherent}%
  \BibitemOpen
  \bibfield  {author} {\bibinfo {author} {\bibfnamefont {M.}~\bibnamefont
  {Hays}}, \bibinfo {author} {\bibfnamefont {V.}~\bibnamefont {Fatemi}},
  \bibinfo {author} {\bibfnamefont {D.}~\bibnamefont {Bouman}}, \bibinfo
  {author} {\bibfnamefont {J.}~\bibnamefont {Cerrillo}}, \bibinfo {author}
  {\bibfnamefont {S.}~\bibnamefont {Diamond}}, \bibinfo {author} {\bibfnamefont
  {K.}~\bibnamefont {Serniak}}, \bibinfo {author} {\bibfnamefont
  {T.}~\bibnamefont {Connolly}}, \bibinfo {author} {\bibfnamefont
  {P.}~\bibnamefont {Krogstrup}}, \bibinfo {author} {\bibfnamefont
  {J.}~\bibnamefont {Nyg{\aa}rd}}, \bibinfo {author} {\bibfnamefont
  {A.}~\bibnamefont {Levy~Yeyati}},  \emph {et~al.},\ }\href@noop {} {\bibfield
   {journal} {\bibinfo  {journal} {Science}\ }\textbf {\bibinfo {volume}
  {373}},\ \bibinfo {pages} {430} (\bibinfo {year} {2021})}\BibitemShut
  {NoStop}%
\bibitem [{\citenamefont {Larsen}\ \emph {et~al.}(2015)\citenamefont {Larsen},
  \citenamefont {Petersson}, \citenamefont {Kuemmeth}, \citenamefont
  {Jespersen}, \citenamefont {Krogstrup}, \citenamefont {Nyg{\aa}rd},\ and\
  \citenamefont {Marcus}}]{larsen2015semiconductor}%
  \BibitemOpen
  \bibfield  {author} {\bibinfo {author} {\bibfnamefont {T.~W.}\ \bibnamefont
  {Larsen}}, \bibinfo {author} {\bibfnamefont {K.~D.}\ \bibnamefont
  {Petersson}}, \bibinfo {author} {\bibfnamefont {F.}~\bibnamefont {Kuemmeth}},
  \bibinfo {author} {\bibfnamefont {T.~S.}\ \bibnamefont {Jespersen}}, \bibinfo
  {author} {\bibfnamefont {P.}~\bibnamefont {Krogstrup}}, \bibinfo {author}
  {\bibfnamefont {J.}~\bibnamefont {Nyg{\aa}rd}}, \ and\ \bibinfo {author}
  {\bibfnamefont {C.~M.}\ \bibnamefont {Marcus}},\ }\href@noop {} {\bibfield
  {journal} {\bibinfo  {journal} {Physical Review Letters}\ }\textbf {\bibinfo
  {volume} {115}},\ \bibinfo {pages} {127001} (\bibinfo {year}
  {2015})}\BibitemShut {NoStop}%
\bibitem [{\citenamefont {Casparis}\ \emph {et~al.}(2018)\citenamefont
  {Casparis}, \citenamefont {Connolly}, \citenamefont {Kjaergaard},
  \citenamefont {Pearson}, \citenamefont {Kringh{\o}j}, \citenamefont {Larsen},
  \citenamefont {Kuemmeth}, \citenamefont {Wang}, \citenamefont {Thomas},
  \citenamefont {Gronin} \emph {et~al.}}]{casparis2018superconducting}%
  \BibitemOpen
  \bibfield  {author} {\bibinfo {author} {\bibfnamefont {L.}~\bibnamefont
  {Casparis}}, \bibinfo {author} {\bibfnamefont {M.~R.}\ \bibnamefont
  {Connolly}}, \bibinfo {author} {\bibfnamefont {M.}~\bibnamefont
  {Kjaergaard}}, \bibinfo {author} {\bibfnamefont {N.~J.}\ \bibnamefont
  {Pearson}}, \bibinfo {author} {\bibfnamefont {A.}~\bibnamefont
  {Kringh{\o}j}}, \bibinfo {author} {\bibfnamefont {T.~W.}\ \bibnamefont
  {Larsen}}, \bibinfo {author} {\bibfnamefont {F.}~\bibnamefont {Kuemmeth}},
  \bibinfo {author} {\bibfnamefont {T.}~\bibnamefont {Wang}}, \bibinfo {author}
  {\bibfnamefont {C.}~\bibnamefont {Thomas}}, \bibinfo {author} {\bibfnamefont
  {S.}~\bibnamefont {Gronin}},  \emph {et~al.},\ }\href@noop {} {\bibfield
  {journal} {\bibinfo  {journal} {Nature nanotechnology}\ }\textbf {\bibinfo
  {volume} {13}},\ \bibinfo {pages} {915} (\bibinfo {year} {2018})}\BibitemShut
  {NoStop}%
\bibitem [{\citenamefont {Sarma}\ \emph {et~al.}(2015)\citenamefont {Sarma},
  \citenamefont {Freedman},\ and\ \citenamefont {Nayak}}]{sarma2015majorana}%
  \BibitemOpen
  \bibfield  {author} {\bibinfo {author} {\bibfnamefont {S.~D.}\ \bibnamefont
  {Sarma}}, \bibinfo {author} {\bibfnamefont {M.}~\bibnamefont {Freedman}}, \
  and\ \bibinfo {author} {\bibfnamefont {C.}~\bibnamefont {Nayak}},\
  }\href@noop {} {\bibfield  {journal} {\bibinfo  {journal} {npj Quantum
  Information}\ }\textbf {\bibinfo {volume} {1}},\ \bibinfo {pages} {1}
  (\bibinfo {year} {2015})}\BibitemShut {NoStop}%
\bibitem [{\citenamefont {Krogstrup}\ \emph {et~al.}(2015)\citenamefont
  {Krogstrup}, \citenamefont {Ziino}, \citenamefont {Chang}, \citenamefont
  {Albrecht}, \citenamefont {Madsen}, \citenamefont {Johnson}, \citenamefont
  {Nyg{\aa}rd}, \citenamefont {Marcus},\ and\ \citenamefont
  {Jespersen}}]{krogstrup2015epitaxy}%
  \BibitemOpen
  \bibfield  {author} {\bibinfo {author} {\bibfnamefont {P.}~\bibnamefont
  {Krogstrup}}, \bibinfo {author} {\bibfnamefont {N.}~\bibnamefont {Ziino}},
  \bibinfo {author} {\bibfnamefont {W.}~\bibnamefont {Chang}}, \bibinfo
  {author} {\bibfnamefont {S.}~\bibnamefont {Albrecht}}, \bibinfo {author}
  {\bibfnamefont {M.}~\bibnamefont {Madsen}}, \bibinfo {author} {\bibfnamefont
  {E.}~\bibnamefont {Johnson}}, \bibinfo {author} {\bibfnamefont
  {J.}~\bibnamefont {Nyg{\aa}rd}}, \bibinfo {author} {\bibfnamefont {C.~M.}\
  \bibnamefont {Marcus}}, \ and\ \bibinfo {author} {\bibfnamefont
  {T.}~\bibnamefont {Jespersen}},\ }\href@noop {} {\bibfield  {journal}
  {\bibinfo  {journal} {Nature Materials}\ }\textbf {\bibinfo {volume} {14}},\
  \bibinfo {pages} {400} (\bibinfo {year} {2015})}\BibitemShut {NoStop}%
\bibitem [{\citenamefont {Shabani}\ \emph {et~al.}(2016)\citenamefont
  {Shabani}, \citenamefont {Kj{\ae}rgaard}, \citenamefont {Suominen},
  \citenamefont {Kim}, \citenamefont {Nichele}, \citenamefont {Pakrouski},
  \citenamefont {Stankevic}, \citenamefont {Lutchyn}, \citenamefont
  {Krogstrup}, \citenamefont {Feidenhans} \emph {et~al.}}]{shabani2016two}%
  \BibitemOpen
  \bibfield  {author} {\bibinfo {author} {\bibfnamefont {J.}~\bibnamefont
  {Shabani}}, \bibinfo {author} {\bibfnamefont {M.}~\bibnamefont
  {Kj{\ae}rgaard}}, \bibinfo {author} {\bibfnamefont {H.~J.}\ \bibnamefont
  {Suominen}}, \bibinfo {author} {\bibfnamefont {Y.}~\bibnamefont {Kim}},
  \bibinfo {author} {\bibfnamefont {F.}~\bibnamefont {Nichele}}, \bibinfo
  {author} {\bibfnamefont {K.}~\bibnamefont {Pakrouski}}, \bibinfo {author}
  {\bibfnamefont {T.}~\bibnamefont {Stankevic}}, \bibinfo {author}
  {\bibfnamefont {R.~M.}\ \bibnamefont {Lutchyn}}, \bibinfo {author}
  {\bibfnamefont {P.}~\bibnamefont {Krogstrup}}, \bibinfo {author}
  {\bibfnamefont {R.}~\bibnamefont {Feidenhans}},  \emph {et~al.},\ }\href@noop
  {} {\bibfield  {journal} {\bibinfo  {journal} {Physical Review B}\ }\textbf
  {\bibinfo {volume} {93}},\ \bibinfo {pages} {155402} (\bibinfo {year}
  {2016})}\BibitemShut {NoStop}%
\bibitem [{\citenamefont {Mikkelsen}\ \emph {et~al.}(2018)\citenamefont
  {Mikkelsen}, \citenamefont {Kotetes}, \citenamefont {Krogstrup},\ and\
  \citenamefont {Flensberg}}]{mikkelsen2018hybridization}%
  \BibitemOpen
  \bibfield  {author} {\bibinfo {author} {\bibfnamefont {A.~E.}\ \bibnamefont
  {Mikkelsen}}, \bibinfo {author} {\bibfnamefont {P.}~\bibnamefont {Kotetes}},
  \bibinfo {author} {\bibfnamefont {P.}~\bibnamefont {Krogstrup}}, \ and\
  \bibinfo {author} {\bibfnamefont {K.}~\bibnamefont {Flensberg}},\ }\href@noop
  {} {\bibfield  {journal} {\bibinfo  {journal} {Physical Review X}\ }\textbf
  {\bibinfo {volume} {8}},\ \bibinfo {pages} {031040} (\bibinfo {year}
  {2018})}\BibitemShut {NoStop}%
\bibitem [{\citenamefont {Chang}\ \emph {et~al.}(2015)\citenamefont {Chang},
  \citenamefont {Albrecht}, \citenamefont {Jespersen}, \citenamefont
  {Kuemmeth}, \citenamefont {Krogstrup}, \citenamefont {Nyg{\aa}rd},\ and\
  \citenamefont {Marcus}}]{chang2015hard}%
  \BibitemOpen
  \bibfield  {author} {\bibinfo {author} {\bibfnamefont {W.}~\bibnamefont
  {Chang}}, \bibinfo {author} {\bibfnamefont {S.}~\bibnamefont {Albrecht}},
  \bibinfo {author} {\bibfnamefont {T.}~\bibnamefont {Jespersen}}, \bibinfo
  {author} {\bibfnamefont {F.}~\bibnamefont {Kuemmeth}}, \bibinfo {author}
  {\bibfnamefont {P.}~\bibnamefont {Krogstrup}}, \bibinfo {author}
  {\bibfnamefont {J.}~\bibnamefont {Nyg{\aa}rd}}, \ and\ \bibinfo {author}
  {\bibfnamefont {C.~M.}\ \bibnamefont {Marcus}},\ }\href@noop {} {\bibfield
  {journal} {\bibinfo  {journal} {Nature nanotechnology}\ }\textbf {\bibinfo
  {volume} {10}},\ \bibinfo {pages} {232} (\bibinfo {year} {2015})}\BibitemShut
  {NoStop}%
\bibitem [{\citenamefont {Mourik}\ \emph {et~al.}(2012)\citenamefont {Mourik},
  \citenamefont {Zuo}, \citenamefont {Frolov}, \citenamefont {Plissard},
  \citenamefont {Bakkers},\ and\ \citenamefont
  {Kouwenhoven}}]{mourik2012signatures}%
  \BibitemOpen
  \bibfield  {author} {\bibinfo {author} {\bibfnamefont {V.}~\bibnamefont
  {Mourik}}, \bibinfo {author} {\bibfnamefont {K.}~\bibnamefont {Zuo}},
  \bibinfo {author} {\bibfnamefont {S.~M.}\ \bibnamefont {Frolov}}, \bibinfo
  {author} {\bibfnamefont {S.}~\bibnamefont {Plissard}}, \bibinfo {author}
  {\bibfnamefont {E.~P.}\ \bibnamefont {Bakkers}}, \ and\ \bibinfo {author}
  {\bibfnamefont {L.~P.}\ \bibnamefont {Kouwenhoven}},\ }\href@noop {}
  {\bibfield  {journal} {\bibinfo  {journal} {Science}\ }\textbf {\bibinfo
  {volume} {336}},\ \bibinfo {pages} {1003} (\bibinfo {year}
  {2012})}\BibitemShut {NoStop}%
\bibitem [{\citenamefont {Vaitiek{\.e}nas}\ \emph {et~al.}(2020)\citenamefont
  {Vaitiek{\.e}nas}, \citenamefont {Winkler}, \citenamefont {van Heck},
  \citenamefont {Karzig}, \citenamefont {Deng}, \citenamefont {Flensberg},
  \citenamefont {Glazman}, \citenamefont {Nayak}, \citenamefont {Krogstrup},
  \citenamefont {Lutchyn} \emph {et~al.}}]{vaitiekenas2020flux}%
  \BibitemOpen
  \bibfield  {author} {\bibinfo {author} {\bibfnamefont {S.}~\bibnamefont
  {Vaitiek{\.e}nas}}, \bibinfo {author} {\bibfnamefont {G.}~\bibnamefont
  {Winkler}}, \bibinfo {author} {\bibfnamefont {B.}~\bibnamefont {van Heck}},
  \bibinfo {author} {\bibfnamefont {T.}~\bibnamefont {Karzig}}, \bibinfo
  {author} {\bibfnamefont {M.-T.}\ \bibnamefont {Deng}}, \bibinfo {author}
  {\bibfnamefont {K.}~\bibnamefont {Flensberg}}, \bibinfo {author}
  {\bibfnamefont {L.}~\bibnamefont {Glazman}}, \bibinfo {author} {\bibfnamefont
  {C.}~\bibnamefont {Nayak}}, \bibinfo {author} {\bibfnamefont
  {P.}~\bibnamefont {Krogstrup}}, \bibinfo {author} {\bibfnamefont
  {R.}~\bibnamefont {Lutchyn}},  \emph {et~al.},\ }\href@noop {} {\bibfield
  {journal} {\bibinfo  {journal} {Science}\ }\textbf {\bibinfo {volume}
  {367}},\ \bibinfo {pages} {eaav3392} (\bibinfo {year} {2020})}\BibitemShut
  {NoStop}%
\bibitem [{\citenamefont {Kurtossy}\ \emph {et~al.}(2021)\citenamefont
  {Kurtossy}, \citenamefont {Scherübl}, \citenamefont {Fülöp}, \citenamefont
  {Luk{\'a}cs}, \citenamefont {Kanne}, \citenamefont {Nyg{\aa}rd},
  \citenamefont {Makk},\ and\ \citenamefont {Csonka}}]{kurtossy2021andreev}%
  \BibitemOpen
  \bibfield  {author} {\bibinfo {author} {\bibfnamefont {O.}~\bibnamefont
  {Kurtossy}}, \bibinfo {author} {\bibfnamefont {Z.}~\bibnamefont {Scherübl}},
  \bibinfo {author} {\bibfnamefont {G.}~\bibnamefont {Fülöp}}, \bibinfo
  {author} {\bibfnamefont {I.~E.}\ \bibnamefont {Luk{\'a}cs}}, \bibinfo
  {author} {\bibfnamefont {T.}~\bibnamefont {Kanne}}, \bibinfo {author}
  {\bibfnamefont {J.}~\bibnamefont {Nyg{\aa}rd}}, \bibinfo {author}
  {\bibfnamefont {P.}~\bibnamefont {Makk}}, \ and\ \bibinfo {author}
  {\bibfnamefont {S.}~\bibnamefont {Csonka}},\ }\href@noop {} {\bibfield
  {journal} {\bibinfo  {journal} {Nano Letters}\ }\textbf {\bibinfo {volume}
  {21}},\ \bibinfo {pages} {7929} (\bibinfo {year} {2021})}\BibitemShut
  {NoStop}%
\bibitem [{\citenamefont {Friedl}\ \emph {et~al.}(2018)\citenamefont {Friedl},
  \citenamefont {Cerveny}, \citenamefont {Weigele}, \citenamefont
  {Tütüncüoglu}, \citenamefont {Mart{\'\i}-S{\'a}nchez}, \citenamefont
  {Huang}, \citenamefont {Patlatiuk}, \citenamefont {Potts}, \citenamefont
  {Sun}, \citenamefont {Hill} \emph {et~al.}}]{friedl2018template}%
  \BibitemOpen
  \bibfield  {author} {\bibinfo {author} {\bibfnamefont {M.}~\bibnamefont
  {Friedl}}, \bibinfo {author} {\bibfnamefont {K.}~\bibnamefont {Cerveny}},
  \bibinfo {author} {\bibfnamefont {P.}~\bibnamefont {Weigele}}, \bibinfo
  {author} {\bibfnamefont {G.}~\bibnamefont {Tütüncüoglu}}, \bibinfo
  {author} {\bibfnamefont {S.}~\bibnamefont {Mart{\'\i}-S{\'a}nchez}}, \bibinfo
  {author} {\bibfnamefont {C.}~\bibnamefont {Huang}}, \bibinfo {author}
  {\bibfnamefont {T.}~\bibnamefont {Patlatiuk}}, \bibinfo {author}
  {\bibfnamefont {H.}~\bibnamefont {Potts}}, \bibinfo {author} {\bibfnamefont
  {Z.}~\bibnamefont {Sun}}, \bibinfo {author} {\bibfnamefont {M.~O.}\
  \bibnamefont {Hill}},  \emph {et~al.},\ }\href@noop {} {\bibfield  {journal}
  {\bibinfo  {journal} {Nano Letters}\ }\textbf {\bibinfo {volume} {18}},\
  \bibinfo {pages} {2666} (\bibinfo {year} {2018})}\BibitemShut {NoStop}%
\bibitem [{\citenamefont {Krizek}\ \emph {et~al.}(2018)\citenamefont {Krizek},
  \citenamefont {Sestoft}, \citenamefont {Aseev}, \citenamefont
  {Marti-Sanchez}, \citenamefont {Vaitiek{\.e}nas}, \citenamefont {Casparis},
  \citenamefont {Khan}, \citenamefont {Liu}, \citenamefont {Stankevi{\v{c}}},
  \citenamefont {Whiticar} \emph {et~al.}}]{krizek2018field}%
  \BibitemOpen
  \bibfield  {author} {\bibinfo {author} {\bibfnamefont {F.}~\bibnamefont
  {Krizek}}, \bibinfo {author} {\bibfnamefont {J.~E.}\ \bibnamefont {Sestoft}},
  \bibinfo {author} {\bibfnamefont {P.}~\bibnamefont {Aseev}}, \bibinfo
  {author} {\bibfnamefont {S.}~\bibnamefont {Marti-Sanchez}}, \bibinfo {author}
  {\bibfnamefont {S.}~\bibnamefont {Vaitiek{\.e}nas}}, \bibinfo {author}
  {\bibfnamefont {L.}~\bibnamefont {Casparis}}, \bibinfo {author}
  {\bibfnamefont {S.~A.}\ \bibnamefont {Khan}}, \bibinfo {author}
  {\bibfnamefont {Y.}~\bibnamefont {Liu}}, \bibinfo {author} {\bibfnamefont
  {T.}~\bibnamefont {Stankevi{\v{c}}}}, \bibinfo {author} {\bibfnamefont
  {A.~M.}\ \bibnamefont {Whiticar}},  \emph {et~al.},\ }\href@noop {}
  {\bibfield  {journal} {\bibinfo  {journal} {Physical Review Materials}\
  }\textbf {\bibinfo {volume} {2}},\ \bibinfo {pages} {093401} (\bibinfo {year}
  {2018})}\BibitemShut {NoStop}%
\bibitem [{\citenamefont {Aghaee}\ \emph {et~al.}(2022)\citenamefont {Aghaee},
  \citenamefont {Akkala}, \citenamefont {Alam}, \citenamefont {Ali},
  \citenamefont {Ramirez}, \citenamefont {Andrzejczuk}, \citenamefont
  {Antipov}, \citenamefont {Astafev}, \citenamefont {Bauer}, \citenamefont
  {Becker} \emph {et~al.}}]{aghaee2022inas}%
  \BibitemOpen
  \bibfield  {author} {\bibinfo {author} {\bibfnamefont {M.}~\bibnamefont
  {Aghaee}}, \bibinfo {author} {\bibfnamefont {A.}~\bibnamefont {Akkala}},
  \bibinfo {author} {\bibfnamefont {Z.}~\bibnamefont {Alam}}, \bibinfo {author}
  {\bibfnamefont {R.}~\bibnamefont {Ali}}, \bibinfo {author} {\bibfnamefont
  {A.~A.}\ \bibnamefont {Ramirez}}, \bibinfo {author} {\bibfnamefont
  {M.}~\bibnamefont {Andrzejczuk}}, \bibinfo {author} {\bibfnamefont {A.~E.}\
  \bibnamefont {Antipov}}, \bibinfo {author} {\bibfnamefont {M.}~\bibnamefont
  {Astafev}}, \bibinfo {author} {\bibfnamefont {B.}~\bibnamefont {Bauer}},
  \bibinfo {author} {\bibfnamefont {J.}~\bibnamefont {Becker}},  \emph
  {et~al.},\ }\href@noop {} {\bibfield  {journal} {\bibinfo  {journal} {arXiv
  preprint arXiv:2207.02472}\ } (\bibinfo {year} {2022})}\BibitemShut {NoStop}%
\bibitem [{\citenamefont {Sarney}\ \emph {et~al.}(2018)\citenamefont {Sarney},
  \citenamefont {Svensson}, \citenamefont {Wickramasinghe}, \citenamefont
  {Yuan},\ and\ \citenamefont {Shabani}}]{sarney2018reactivity}%
  \BibitemOpen
  \bibfield  {author} {\bibinfo {author} {\bibfnamefont {W.~L.}\ \bibnamefont
  {Sarney}}, \bibinfo {author} {\bibfnamefont {S.~P.}\ \bibnamefont
  {Svensson}}, \bibinfo {author} {\bibfnamefont {K.~S.}\ \bibnamefont
  {Wickramasinghe}}, \bibinfo {author} {\bibfnamefont {J.}~\bibnamefont
  {Yuan}}, \ and\ \bibinfo {author} {\bibfnamefont {J.}~\bibnamefont
  {Shabani}},\ }\href@noop {} {\bibfield  {journal} {\bibinfo  {journal}
  {Journal of Vacuum Science \& Technology B, Nanotechnology and
  Microelectronics: Materials, Processing, Measurement, and Phenomena}\
  }\textbf {\bibinfo {volume} {36}},\ \bibinfo {pages} {062903} (\bibinfo
  {year} {2018})}\BibitemShut {NoStop}%
\bibitem [{\citenamefont {Sarney}\ \emph {et~al.}(2020)\citenamefont {Sarney},
  \citenamefont {Svensson}, \citenamefont {Leff}, \citenamefont {Schiela},
  \citenamefont {Yuan}, \citenamefont {Dartiailh}, \citenamefont {Mayer},
  \citenamefont {Wickramasinghe},\ and\ \citenamefont
  {Shabani}}]{sarney2020aluminum}%
  \BibitemOpen
  \bibfield  {author} {\bibinfo {author} {\bibfnamefont {W.~L.}\ \bibnamefont
  {Sarney}}, \bibinfo {author} {\bibfnamefont {S.~P.}\ \bibnamefont
  {Svensson}}, \bibinfo {author} {\bibfnamefont {A.~C.}\ \bibnamefont {Leff}},
  \bibinfo {author} {\bibfnamefont {W.~F.}\ \bibnamefont {Schiela}}, \bibinfo
  {author} {\bibfnamefont {J.~O.}\ \bibnamefont {Yuan}}, \bibinfo {author}
  {\bibfnamefont {M.~C.}\ \bibnamefont {Dartiailh}}, \bibinfo {author}
  {\bibfnamefont {W.}~\bibnamefont {Mayer}}, \bibinfo {author} {\bibfnamefont
  {K.~S.}\ \bibnamefont {Wickramasinghe}}, \ and\ \bibinfo {author}
  {\bibfnamefont {J.}~\bibnamefont {Shabani}},\ }\href@noop {} {\bibfield
  {journal} {\bibinfo  {journal} {Journal of Vacuum Science \& Technology B,
  Nanotechnology and Microelectronics: Materials, Processing, Measurement, and
  Phenomena}\ }\textbf {\bibinfo {volume} {38}},\ \bibinfo {pages} {032212}
  (\bibinfo {year} {2020})}\BibitemShut {NoStop}%
\bibitem [{\citenamefont {Wang}\ \emph {et~al.}(2020)\citenamefont {Wang},
  \citenamefont {Thomas}, \citenamefont {Diaz}, \citenamefont {Gronin},
  \citenamefont {Passarello}, \citenamefont {Gardner}, \citenamefont {Capano},\
  and\ \citenamefont {Manfra}}]{wang2020dependence}%
  \BibitemOpen
  \bibfield  {author} {\bibinfo {author} {\bibfnamefont {T.}~\bibnamefont
  {Wang}}, \bibinfo {author} {\bibfnamefont {C.}~\bibnamefont {Thomas}},
  \bibinfo {author} {\bibfnamefont {R.~E.}\ \bibnamefont {Diaz}}, \bibinfo
  {author} {\bibfnamefont {S.}~\bibnamefont {Gronin}}, \bibinfo {author}
  {\bibfnamefont {D.}~\bibnamefont {Passarello}}, \bibinfo {author}
  {\bibfnamefont {G.~C.}\ \bibnamefont {Gardner}}, \bibinfo {author}
  {\bibfnamefont {M.~A.}\ \bibnamefont {Capano}}, \ and\ \bibinfo {author}
  {\bibfnamefont {M.~J.}\ \bibnamefont {Manfra}},\ }\href@noop {} {\bibfield
  {journal} {\bibinfo  {journal} {Journal of Crystal Growth}\ }\textbf
  {\bibinfo {volume} {535}},\ \bibinfo {pages} {125570} (\bibinfo {year}
  {2020})}\BibitemShut {NoStop}%
\bibitem [{\citenamefont {Ferguson}\ \emph {et~al.}(2007)\citenamefont
  {Ferguson}, \citenamefont {Clark} \emph {et~al.}}]{ferguson2007energy}%
  \BibitemOpen
  \bibfield  {author} {\bibinfo {author} {\bibfnamefont {A.}~\bibnamefont
  {Ferguson}}, \bibinfo {author} {\bibfnamefont {R.}~\bibnamefont {Clark}},
  \emph {et~al.},\ }\href@noop {} {\bibfield  {journal} {\bibinfo  {journal}
  {Superconductor Science and Technology}\ }\textbf {\bibinfo {volume} {21}},\
  \bibinfo {pages} {015013} (\bibinfo {year} {2007})}\BibitemShut {NoStop}%
\bibitem [{\citenamefont {Pracht}\ \emph {et~al.}(2016)\citenamefont {Pracht},
  \citenamefont {Bachar}, \citenamefont {Benfatto}, \citenamefont {Deutscher},
  \citenamefont {Farber}, \citenamefont {Dressel},\ and\ \citenamefont
  {Scheffler}}]{pracht2016enhanced}%
  \BibitemOpen
  \bibfield  {author} {\bibinfo {author} {\bibfnamefont {U.~S.}\ \bibnamefont
  {Pracht}}, \bibinfo {author} {\bibfnamefont {N.}~\bibnamefont {Bachar}},
  \bibinfo {author} {\bibfnamefont {L.}~\bibnamefont {Benfatto}}, \bibinfo
  {author} {\bibfnamefont {G.}~\bibnamefont {Deutscher}}, \bibinfo {author}
  {\bibfnamefont {E.}~\bibnamefont {Farber}}, \bibinfo {author} {\bibfnamefont
  {M.}~\bibnamefont {Dressel}}, \ and\ \bibinfo {author} {\bibfnamefont
  {M.}~\bibnamefont {Scheffler}},\ }\href@noop {} {\bibfield  {journal}
  {\bibinfo  {journal} {Physical Review B}\ }\textbf {\bibinfo {volume} {93}},\
  \bibinfo {pages} {100503} (\bibinfo {year} {2016})}\BibitemShut {NoStop}%
\bibitem [{\citenamefont {Nguyen}\ \emph {et~al.}(2018)\citenamefont {Nguyen},
  \citenamefont {Hashimoto}, \citenamefont {Zakharov}, \citenamefont {Stach},
  \citenamefont {Rooney}, \citenamefont {Berkels}, \citenamefont {Thompson},
  \citenamefont {Haigh},\ and\ \citenamefont {Burnett}}]{nguyen2018atomic}%
  \BibitemOpen
  \bibfield  {author} {\bibinfo {author} {\bibfnamefont {L.}~\bibnamefont
  {Nguyen}}, \bibinfo {author} {\bibfnamefont {T.}~\bibnamefont {Hashimoto}},
  \bibinfo {author} {\bibfnamefont {D.~N.}\ \bibnamefont {Zakharov}}, \bibinfo
  {author} {\bibfnamefont {E.~A.}\ \bibnamefont {Stach}}, \bibinfo {author}
  {\bibfnamefont {A.~P.}\ \bibnamefont {Rooney}}, \bibinfo {author}
  {\bibfnamefont {B.}~\bibnamefont {Berkels}}, \bibinfo {author} {\bibfnamefont
  {G.~E.}\ \bibnamefont {Thompson}}, \bibinfo {author} {\bibfnamefont {S.~J.}\
  \bibnamefont {Haigh}}, \ and\ \bibinfo {author} {\bibfnamefont {T.~L.}\
  \bibnamefont {Burnett}},\ }\href@noop {} {\bibfield  {journal} {\bibinfo
  {journal} {ACS Applied Materials \& Interfaces}\ }\textbf {\bibinfo {volume}
  {10}},\ \bibinfo {pages} {2230} (\bibinfo {year} {2018})}\BibitemShut
  {NoStop}%
\bibitem [{\citenamefont {David~Henry}\ \emph {et~al.}(2014)\citenamefont
  {David~Henry}, \citenamefont {Wolfley}, \citenamefont {Monson}, \citenamefont
  {Clark}, \citenamefont {Shaner},\ and\ \citenamefont
  {Jarecki}}]{david2014stress}%
  \BibitemOpen
  \bibfield  {author} {\bibinfo {author} {\bibfnamefont {M.}~\bibnamefont
  {David~Henry}}, \bibinfo {author} {\bibfnamefont {S.}~\bibnamefont
  {Wolfley}}, \bibinfo {author} {\bibfnamefont {T.}~\bibnamefont {Monson}},
  \bibinfo {author} {\bibfnamefont {B.~G.}\ \bibnamefont {Clark}}, \bibinfo
  {author} {\bibfnamefont {E.}~\bibnamefont {Shaner}}, \ and\ \bibinfo {author}
  {\bibfnamefont {R.}~\bibnamefont {Jarecki}},\ }\href@noop {} {\bibfield
  {journal} {\bibinfo  {journal} {Journal of Applied Physics}\ }\textbf
  {\bibinfo {volume} {115}},\ \bibinfo {pages} {083903} (\bibinfo {year}
  {2014})}\BibitemShut {NoStop}%
\bibitem [{\citenamefont {Kj{\ae}rgaard}\ \emph {et~al.}(2016)\citenamefont
  {Kj{\ae}rgaard}, \citenamefont {Nichele}, \citenamefont {Suominen},
  \citenamefont {Nowak}, \citenamefont {Wimmer}, \citenamefont {Akhmerov},
  \citenamefont {Folk}, \citenamefont {Flensberg}, \citenamefont {Shabani},
  \citenamefont {Palmstr{\o}m} \emph {et~al.}}]{kjaergaard2016quantized}%
  \BibitemOpen
  \bibfield  {author} {\bibinfo {author} {\bibfnamefont {M.}~\bibnamefont
  {Kj{\ae}rgaard}}, \bibinfo {author} {\bibfnamefont {F.}~\bibnamefont
  {Nichele}}, \bibinfo {author} {\bibfnamefont {H.}~\bibnamefont {Suominen}},
  \bibinfo {author} {\bibfnamefont {M.}~\bibnamefont {Nowak}}, \bibinfo
  {author} {\bibfnamefont {M.}~\bibnamefont {Wimmer}}, \bibinfo {author}
  {\bibfnamefont {A.}~\bibnamefont {Akhmerov}}, \bibinfo {author}
  {\bibfnamefont {J.}~\bibnamefont {Folk}}, \bibinfo {author} {\bibfnamefont
  {K.}~\bibnamefont {Flensberg}}, \bibinfo {author} {\bibfnamefont
  {J.}~\bibnamefont {Shabani}}, \bibinfo {author} {\bibfnamefont {w.~C.}\
  \bibnamefont {Palmstr{\o}m}},  \emph {et~al.},\ }\href@noop {} {\bibfield
  {journal} {\bibinfo  {journal} {Nature Communications}\ }\textbf {\bibinfo
  {volume} {7}},\ \bibinfo {pages} {1} (\bibinfo {year} {2016})}\BibitemShut
  {NoStop}%
\bibitem [{\citenamefont {Richardson}\ \emph {et~al.}(2016)\citenamefont
  {Richardson}, \citenamefont {Siwak}, \citenamefont {Hackley}, \citenamefont
  {Keane}, \citenamefont {Robinson}, \citenamefont {Arey}, \citenamefont
  {Arslan},\ and\ \citenamefont {Palmer}}]{richardson2016fabrication}%
  \BibitemOpen
  \bibfield  {author} {\bibinfo {author} {\bibfnamefont {C.~J.}\ \bibnamefont
  {Richardson}}, \bibinfo {author} {\bibfnamefont {N.~P.}\ \bibnamefont
  {Siwak}}, \bibinfo {author} {\bibfnamefont {J.}~\bibnamefont {Hackley}},
  \bibinfo {author} {\bibfnamefont {Z.~K.}\ \bibnamefont {Keane}}, \bibinfo
  {author} {\bibfnamefont {J.~E.}\ \bibnamefont {Robinson}}, \bibinfo {author}
  {\bibfnamefont {B.}~\bibnamefont {Arey}}, \bibinfo {author} {\bibfnamefont
  {I.}~\bibnamefont {Arslan}}, \ and\ \bibinfo {author} {\bibfnamefont {B.~S.}\
  \bibnamefont {Palmer}},\ }\href@noop {} {\bibfield  {journal} {\bibinfo
  {journal} {Superconductor Science and Technology}\ }\textbf {\bibinfo
  {volume} {29}},\ \bibinfo {pages} {064003} (\bibinfo {year}
  {2016})}\BibitemShut {NoStop}%
\bibitem [{\citenamefont {de~Leon}\ \emph {et~al.}(2021)\citenamefont
  {de~Leon}, \citenamefont {Itoh}, \citenamefont {Kim}, \citenamefont {Mehta},
  \citenamefont {Northup}, \citenamefont {Paik}, \citenamefont {Palmer},
  \citenamefont {Samarth}, \citenamefont {Sangtawesin},\ and\ \citenamefont
  {Steuerman}}]{de2021materials}%
  \BibitemOpen
  \bibfield  {author} {\bibinfo {author} {\bibfnamefont {N.~P.}\ \bibnamefont
  {de~Leon}}, \bibinfo {author} {\bibfnamefont {K.~M.}\ \bibnamefont {Itoh}},
  \bibinfo {author} {\bibfnamefont {D.}~\bibnamefont {Kim}}, \bibinfo {author}
  {\bibfnamefont {K.~K.}\ \bibnamefont {Mehta}}, \bibinfo {author}
  {\bibfnamefont {T.~E.}\ \bibnamefont {Northup}}, \bibinfo {author}
  {\bibfnamefont {H.}~\bibnamefont {Paik}}, \bibinfo {author} {\bibfnamefont
  {B.}~\bibnamefont {Palmer}}, \bibinfo {author} {\bibfnamefont
  {N.}~\bibnamefont {Samarth}}, \bibinfo {author} {\bibfnamefont
  {S.}~\bibnamefont {Sangtawesin}}, \ and\ \bibinfo {author} {\bibfnamefont
  {D.}~\bibnamefont {Steuerman}},\ }\href@noop {} {\bibfield  {journal}
  {\bibinfo  {journal} {Science}\ }\textbf {\bibinfo {volume} {372}},\ \bibinfo
  {pages} {eabb2823} (\bibinfo {year} {2021})}\BibitemShut {NoStop}%
\bibitem [{\citenamefont {Cole}\ \emph {et~al.}(2016)\citenamefont {Cole},
  \citenamefont {Sau},\ and\ \citenamefont {Sarma}}]{cole2016proximity}%
  \BibitemOpen
  \bibfield  {author} {\bibinfo {author} {\bibfnamefont {W.~S.}\ \bibnamefont
  {Cole}}, \bibinfo {author} {\bibfnamefont {J.~D.}\ \bibnamefont {Sau}}, \
  and\ \bibinfo {author} {\bibfnamefont {S.~D.}\ \bibnamefont {Sarma}},\
  }\href@noop {} {\bibfield  {journal} {\bibinfo  {journal} {Physical Review
  B}\ }\textbf {\bibinfo {volume} {94}},\ \bibinfo {pages} {140505} (\bibinfo
  {year} {2016})}\BibitemShut {NoStop}%
\bibitem [{\citenamefont {Thomas}\ \emph {et~al.}(2022)\citenamefont {Thomas},
  \citenamefont {Sarma},\ and\ \citenamefont {Sau}}]{thomas2022disorder}%
  \BibitemOpen
  \bibfield  {author} {\bibinfo {author} {\bibfnamefont {S.~N.}\ \bibnamefont
  {Thomas}}, \bibinfo {author} {\bibfnamefont {S.~D.}\ \bibnamefont {Sarma}}, \
  and\ \bibinfo {author} {\bibfnamefont {J.~D.}\ \bibnamefont {Sau}},\
  }\href@noop {} {\bibfield  {journal} {\bibinfo  {journal} {Physical Review
  B}\ }\textbf {\bibinfo {volume} {106}},\ \bibinfo {pages} {174501} (\bibinfo
  {year} {2022})}\BibitemShut {NoStop}%
\bibitem [{\citenamefont {Thomas}\ \emph {et~al.}(2019)\citenamefont {Thomas},
  \citenamefont {Diaz}, \citenamefont {Dycus}, \citenamefont {Salmon},
  \citenamefont {Daniel}, \citenamefont {Wang}, \citenamefont {Gardner},\ and\
  \citenamefont {Manfra}}]{thomas2019toward}%
  \BibitemOpen
  \bibfield  {author} {\bibinfo {author} {\bibfnamefont {C.}~\bibnamefont
  {Thomas}}, \bibinfo {author} {\bibfnamefont {R.~E.}\ \bibnamefont {Diaz}},
  \bibinfo {author} {\bibfnamefont {J.~H.}\ \bibnamefont {Dycus}}, \bibinfo
  {author} {\bibfnamefont {M.~E.}\ \bibnamefont {Salmon}}, \bibinfo {author}
  {\bibfnamefont {R.~E.}\ \bibnamefont {Daniel}}, \bibinfo {author}
  {\bibfnamefont {T.}~\bibnamefont {Wang}}, \bibinfo {author} {\bibfnamefont
  {G.~C.}\ \bibnamefont {Gardner}}, \ and\ \bibinfo {author} {\bibfnamefont
  {M.~J.}\ \bibnamefont {Manfra}},\ }\href@noop {} {\bibfield  {journal}
  {\bibinfo  {journal} {Physical Review Materials}\ }\textbf {\bibinfo {volume}
  {3}},\ \bibinfo {pages} {124202} (\bibinfo {year} {2019})}\BibitemShut
  {NoStop}%
\bibitem [{\citenamefont {Moehle}\ \emph {et~al.}(2021)\citenamefont {Moehle},
  \citenamefont {Ke}, \citenamefont {Wang}, \citenamefont {Thomas},
  \citenamefont {Xiao}, \citenamefont {Karwal}, \citenamefont {Lodari},
  \citenamefont {van~de Kerkhof}, \citenamefont {Termaat}, \citenamefont
  {Gardner} \emph {et~al.}}]{moehle2021insbas}%
  \BibitemOpen
  \bibfield  {author} {\bibinfo {author} {\bibfnamefont {C.~M.}\ \bibnamefont
  {Moehle}}, \bibinfo {author} {\bibfnamefont {C.~T.}\ \bibnamefont {Ke}},
  \bibinfo {author} {\bibfnamefont {Q.}~\bibnamefont {Wang}}, \bibinfo {author}
  {\bibfnamefont {C.}~\bibnamefont {Thomas}}, \bibinfo {author} {\bibfnamefont
  {D.}~\bibnamefont {Xiao}}, \bibinfo {author} {\bibfnamefont {S.}~\bibnamefont
  {Karwal}}, \bibinfo {author} {\bibfnamefont {M.}~\bibnamefont {Lodari}},
  \bibinfo {author} {\bibfnamefont {V.}~\bibnamefont {van~de Kerkhof}},
  \bibinfo {author} {\bibfnamefont {R.}~\bibnamefont {Termaat}}, \bibinfo
  {author} {\bibfnamefont {G.~C.}\ \bibnamefont {Gardner}},  \emph {et~al.},\
  }\href@noop {} {\bibfield  {journal} {\bibinfo  {journal} {Nano Letters}\
  }\textbf {\bibinfo {volume} {21}},\ \bibinfo {pages} {9990} (\bibinfo {year}
  {2021})}\BibitemShut {NoStop}%
\bibitem [{\citenamefont {Weilmeier}\ \emph {et~al.}(1991)\citenamefont
  {Weilmeier}, \citenamefont {Colbow}, \citenamefont {Tiedje}, \citenamefont
  {Buuren},\ and\ \citenamefont {Xu}}]{weilmeier1991new}%
  \BibitemOpen
  \bibfield  {author} {\bibinfo {author} {\bibfnamefont {M.}~\bibnamefont
  {Weilmeier}}, \bibinfo {author} {\bibfnamefont {K.}~\bibnamefont {Colbow}},
  \bibinfo {author} {\bibfnamefont {T.}~\bibnamefont {Tiedje}}, \bibinfo
  {author} {\bibfnamefont {T.~v.}\ \bibnamefont {Buuren}}, \ and\ \bibinfo
  {author} {\bibfnamefont {L.}~\bibnamefont {Xu}},\ }\href@noop {} {\bibfield
  {journal} {\bibinfo  {journal} {Canadian Journal of Physics}\ }\textbf
  {\bibinfo {volume} {69}},\ \bibinfo {pages} {422} (\bibinfo {year}
  {1991})}\BibitemShut {NoStop}%
\bibitem [{sup()}]{suppinfo}%
  \BibitemOpen
  \href@noop {} {\bibinfo  {journal} {See Supplemental Material at [URL will be
  inserted by publisher] for following details: Methods; S1: Surface of shallow
  InAs 2DEGs; S2: Extraction of lattice spacings; S3: Controlled oxidation of
  Al; S4: In-situ annealing STEM experiment; S5: Ageing of studied lamellae;
  S6: STEM tomography, S7: Additional analysis of the JJ measurements of the
  roughened sample, S8: JJ measurements of the standard non-roughened
  samples.}\ }\BibitemShut {NoStop}%
\bibitem [{\citenamefont {Haxell}\ \emph
  {et~al.}(2022{\natexlab{a}})\citenamefont {Haxell}, \citenamefont {Cheah},
  \citenamefont {K{\v{r}}{\'\i}{\v{z}}ek}, \citenamefont {Schott},
  \citenamefont {Ritter}, \citenamefont {Hinderling}, \citenamefont {Belzig},
  \citenamefont {Bruder}, \citenamefont {Wegscheider}, \citenamefont {Riel}
  \emph {et~al.}}]{haxell2022large}%
  \BibitemOpen
\bibfield  {journal} {  }\bibfield  {author} {\bibinfo {author} {\bibfnamefont
  {D.}~\bibnamefont {Haxell}}, \bibinfo {author} {\bibfnamefont
  {E.}~\bibnamefont {Cheah}}, \bibinfo {author} {\bibfnamefont
  {F.}~\bibnamefont {K{\v{r}}{\'\i}{\v{z}}ek}}, \bibinfo {author}
  {\bibfnamefont {R.}~\bibnamefont {Schott}}, \bibinfo {author} {\bibfnamefont
  {M.}~\bibnamefont {Ritter}}, \bibinfo {author} {\bibfnamefont
  {M.}~\bibnamefont {Hinderling}}, \bibinfo {author} {\bibfnamefont
  {W.}~\bibnamefont {Belzig}}, \bibinfo {author} {\bibfnamefont
  {C.}~\bibnamefont {Bruder}}, \bibinfo {author} {\bibfnamefont
  {W.}~\bibnamefont {Wegscheider}}, \bibinfo {author} {\bibfnamefont
  {H.}~\bibnamefont {Riel}},  \emph {et~al.},\ }\href@noop {} {\bibfield
  {journal} {\bibinfo  {journal} {arXiv preprint arXiv:2204.05619}\ } (\bibinfo
  {year} {2022}{\natexlab{a}})}\BibitemShut {NoStop}%
\bibitem [{\citenamefont {Haxell}\ \emph
  {et~al.}(2022{\natexlab{b}})\citenamefont {Haxell}, \citenamefont {Coraiola},
  \citenamefont {Sabonis}, \citenamefont {Hinderling}, \citenamefont {ten
  Kate}, \citenamefont {Cheah}, \citenamefont {Krizek}, \citenamefont {Schott},
  \citenamefont {Wegscheider}, \citenamefont {Belzig} \emph
  {et~al.}}]{haxell2022microwave}%
  \BibitemOpen
  \bibfield  {author} {\bibinfo {author} {\bibfnamefont {D.}~\bibnamefont
  {Haxell}}, \bibinfo {author} {\bibfnamefont {M.}~\bibnamefont {Coraiola}},
  \bibinfo {author} {\bibfnamefont {D.}~\bibnamefont {Sabonis}}, \bibinfo
  {author} {\bibfnamefont {M.}~\bibnamefont {Hinderling}}, \bibinfo {author}
  {\bibfnamefont {S.}~\bibnamefont {ten Kate}}, \bibinfo {author}
  {\bibfnamefont {E.}~\bibnamefont {Cheah}}, \bibinfo {author} {\bibfnamefont
  {F.}~\bibnamefont {Krizek}}, \bibinfo {author} {\bibfnamefont
  {R.}~\bibnamefont {Schott}}, \bibinfo {author} {\bibfnamefont
  {W.}~\bibnamefont {Wegscheider}}, \bibinfo {author} {\bibfnamefont
  {W.}~\bibnamefont {Belzig}},  \emph {et~al.},\ }\href@noop {} {\bibfield
  {journal} {\bibinfo  {journal} {arXiv preprint arXiv:2212.03554}\ } (\bibinfo
  {year} {2022}{\natexlab{b}})}\BibitemShut {NoStop}%
\bibitem [{\citenamefont {Hinderling}\ \emph {et~al.}(2022)\citenamefont
  {Hinderling}, \citenamefont {Sabonis}, \citenamefont {Paredes}, \citenamefont
  {Haxell}, \citenamefont {Coraiola}, \citenamefont {ten Kate}, \citenamefont
  {Cheah}, \citenamefont {Krizek}, \citenamefont {Schott}, \citenamefont
  {Wegscheider} \emph {et~al.}}]{hinderling2022flip}%
  \BibitemOpen
  \bibfield  {author} {\bibinfo {author} {\bibfnamefont {M.}~\bibnamefont
  {Hinderling}}, \bibinfo {author} {\bibfnamefont {D.}~\bibnamefont {Sabonis}},
  \bibinfo {author} {\bibfnamefont {S.}~\bibnamefont {Paredes}}, \bibinfo
  {author} {\bibfnamefont {D.}~\bibnamefont {Haxell}}, \bibinfo {author}
  {\bibfnamefont {M.}~\bibnamefont {Coraiola}}, \bibinfo {author}
  {\bibfnamefont {S.}~\bibnamefont {ten Kate}}, \bibinfo {author}
  {\bibfnamefont {E.}~\bibnamefont {Cheah}}, \bibinfo {author} {\bibfnamefont
  {F.}~\bibnamefont {Krizek}}, \bibinfo {author} {\bibfnamefont
  {R.}~\bibnamefont {Schott}}, \bibinfo {author} {\bibfnamefont
  {W.}~\bibnamefont {Wegscheider}},  \emph {et~al.},\ }\href@noop {} {\bibfield
   {journal} {\bibinfo  {journal} {arXiv preprint arXiv:2212.11164}\ }
  (\bibinfo {year} {2022})}\BibitemShut {NoStop}%
\bibitem [{\citenamefont {Nicolai}\ \emph {et~al.}(2021)\citenamefont
  {Nicolai}, \citenamefont {Biermann},\ and\ \citenamefont
  {Trampert}}]{nicolai2021application}%
  \BibitemOpen
  \bibfield  {author} {\bibinfo {author} {\bibfnamefont {L.}~\bibnamefont
  {Nicolai}}, \bibinfo {author} {\bibfnamefont {K.}~\bibnamefont {Biermann}}, \
  and\ \bibinfo {author} {\bibfnamefont {A.}~\bibnamefont {Trampert}},\
  }\href@noop {} {\bibfield  {journal} {\bibinfo  {journal} {Ultramicroscopy}\
  }\textbf {\bibinfo {volume} {224}},\ \bibinfo {pages} {113261} (\bibinfo
  {year} {2021})}\BibitemShut {NoStop}%
\bibitem [{\citenamefont {Ohta}\ \emph {et~al.}(1989)\citenamefont {Ohta},
  \citenamefont {Kojima},\ and\ \citenamefont
  {Nakagawa}}]{ohta1989anisotropic}%
  \BibitemOpen
  \bibfield  {author} {\bibinfo {author} {\bibfnamefont {K.}~\bibnamefont
  {Ohta}}, \bibinfo {author} {\bibfnamefont {T.}~\bibnamefont {Kojima}}, \ and\
  \bibinfo {author} {\bibfnamefont {T.}~\bibnamefont {Nakagawa}},\ }\href@noop
  {} {\bibfield  {journal} {\bibinfo  {journal} {Journal of Crystal Growth}\
  }\textbf {\bibinfo {volume} {95}},\ \bibinfo {pages} {71} (\bibinfo {year}
  {1989})}\BibitemShut {NoStop}%
\bibitem [{\citenamefont {Dieguez}\ \emph {et~al.}(1997)\citenamefont
  {Dieguez}, \citenamefont {Vila}, \citenamefont {Cornet}, \citenamefont
  {Clark}, \citenamefont {Westwood},\ and\ \citenamefont
  {Morante}}]{dieguez1997defects}%
  \BibitemOpen
  \bibfield  {author} {\bibinfo {author} {\bibfnamefont {A.}~\bibnamefont
  {Dieguez}}, \bibinfo {author} {\bibfnamefont {A.}~\bibnamefont {Vila}},
  \bibinfo {author} {\bibfnamefont {A.}~\bibnamefont {Cornet}}, \bibinfo
  {author} {\bibfnamefont {S.}~\bibnamefont {Clark}}, \bibinfo {author}
  {\bibfnamefont {D.}~\bibnamefont {Westwood}}, \ and\ \bibinfo {author}
  {\bibfnamefont {J.}~\bibnamefont {Morante}},\ }\href@noop {} {\bibfield
  {journal} {\bibinfo  {journal} {Journal of Vacuum Science \& Technology B:
  Microelectronics and Nanometer Structures Processing, Measurement, and
  Phenomena}\ }\textbf {\bibinfo {volume} {15}},\ \bibinfo {pages} {687}
  (\bibinfo {year} {1997})}\BibitemShut {NoStop}%
\bibitem [{\citenamefont {Chen}\ \emph {et~al.}(2018)\citenamefont {Chen},
  \citenamefont {Chen}, \citenamefont {Li}, \citenamefont {Chen}, \citenamefont
  {Li}, \citenamefont {Zhan}, \citenamefont {Yu}, \citenamefont {Kang},
  \citenamefont {Jiao}, \citenamefont {Li} \emph {et~al.}}]{chen2018study}%
  \BibitemOpen
  \bibfield  {author} {\bibinfo {author} {\bibfnamefont {Y.}~\bibnamefont
  {Chen}}, \bibinfo {author} {\bibfnamefont {Z.}~\bibnamefont {Chen}}, \bibinfo
  {author} {\bibfnamefont {J.}~\bibnamefont {Li}}, \bibinfo {author}
  {\bibfnamefont {Y.}~\bibnamefont {Chen}}, \bibinfo {author} {\bibfnamefont
  {C.}~\bibnamefont {Li}}, \bibinfo {author} {\bibfnamefont {J.}~\bibnamefont
  {Zhan}}, \bibinfo {author} {\bibfnamefont {T.}~\bibnamefont {Yu}}, \bibinfo
  {author} {\bibfnamefont {X.}~\bibnamefont {Kang}}, \bibinfo {author}
  {\bibfnamefont {F.}~\bibnamefont {Jiao}}, \bibinfo {author} {\bibfnamefont
  {S.}~\bibnamefont {Li}},  \emph {et~al.},\ }\href@noop {} {\bibfield
  {journal} {\bibinfo  {journal} {CrystEngComm}\ }\textbf {\bibinfo {volume}
  {20}},\ \bibinfo {pages} {6811} (\bibinfo {year} {2018})}\BibitemShut
  {NoStop}%
\bibitem [{\citenamefont {Feng}\ \emph {et~al.}(2016)\citenamefont {Feng},
  \citenamefont {Yu}, \citenamefont {Wei}, \citenamefont {Ji}, \citenamefont
  {Cheng}, \citenamefont {Zong}, \citenamefont {Wang}, \citenamefont {Yang},
  \citenamefont {Kang}, \citenamefont {Zhang} \emph
  {et~al.}}]{feng2016grouped}%
  \BibitemOpen
  \bibfield  {author} {\bibinfo {author} {\bibfnamefont {X.}~\bibnamefont
  {Feng}}, \bibinfo {author} {\bibfnamefont {T.}~\bibnamefont {Yu}}, \bibinfo
  {author} {\bibfnamefont {Y.}~\bibnamefont {Wei}}, \bibinfo {author}
  {\bibfnamefont {C.}~\bibnamefont {Ji}}, \bibinfo {author} {\bibfnamefont
  {Y.}~\bibnamefont {Cheng}}, \bibinfo {author} {\bibfnamefont
  {H.}~\bibnamefont {Zong}}, \bibinfo {author} {\bibfnamefont {K.}~\bibnamefont
  {Wang}}, \bibinfo {author} {\bibfnamefont {Z.}~\bibnamefont {Yang}}, \bibinfo
  {author} {\bibfnamefont {X.}~\bibnamefont {Kang}}, \bibinfo {author}
  {\bibfnamefont {G.}~\bibnamefont {Zhang}},  \emph {et~al.},\ }\href@noop {}
  {\bibfield  {journal} {\bibinfo  {journal} {ACS Applied Materials \&
  Interfaces}\ }\textbf {\bibinfo {volume} {8}},\ \bibinfo {pages} {18208}
  (\bibinfo {year} {2016})}\BibitemShut {NoStop}%
\bibitem [{\citenamefont {Pauka}\ \emph {et~al.}(2020)\citenamefont {Pauka},
  \citenamefont {Witt}, \citenamefont {Allen}, \citenamefont {Harlech-Jones},
  \citenamefont {Jouan}, \citenamefont {Gardner}, \citenamefont {Gronin},
  \citenamefont {Wang}, \citenamefont {Thomas}, \citenamefont {Manfra} \emph
  {et~al.}}]{pauka2020repairing}%
  \BibitemOpen
  \bibfield  {author} {\bibinfo {author} {\bibfnamefont {S.}~\bibnamefont
  {Pauka}}, \bibinfo {author} {\bibfnamefont {J.}~\bibnamefont {Witt}},
  \bibinfo {author} {\bibfnamefont {C.}~\bibnamefont {Allen}}, \bibinfo
  {author} {\bibfnamefont {B.}~\bibnamefont {Harlech-Jones}}, \bibinfo {author}
  {\bibfnamefont {A.}~\bibnamefont {Jouan}}, \bibinfo {author} {\bibfnamefont
  {G.}~\bibnamefont {Gardner}}, \bibinfo {author} {\bibfnamefont
  {S.}~\bibnamefont {Gronin}}, \bibinfo {author} {\bibfnamefont
  {T.}~\bibnamefont {Wang}}, \bibinfo {author} {\bibfnamefont {C.}~\bibnamefont
  {Thomas}}, \bibinfo {author} {\bibfnamefont {M.}~\bibnamefont {Manfra}},
  \emph {et~al.},\ }\href@noop {} {\bibfield  {journal} {\bibinfo  {journal}
  {Journal of Applied Physics}\ }\textbf {\bibinfo {volume} {128}},\ \bibinfo
  {pages} {114301} (\bibinfo {year} {2020})}\BibitemShut {NoStop}%
\bibitem [{\citenamefont {Hatke}\ \emph {et~al.}(2017)\citenamefont {Hatke},
  \citenamefont {Wang}, \citenamefont {Thomas}, \citenamefont {Gardner},\ and\
  \citenamefont {Manfra}}]{hatke2017mobility}%
  \BibitemOpen
  \bibfield  {author} {\bibinfo {author} {\bibfnamefont {A.}~\bibnamefont
  {Hatke}}, \bibinfo {author} {\bibfnamefont {T.}~\bibnamefont {Wang}},
  \bibinfo {author} {\bibfnamefont {C.}~\bibnamefont {Thomas}}, \bibinfo
  {author} {\bibfnamefont {G.}~\bibnamefont {Gardner}}, \ and\ \bibinfo
  {author} {\bibfnamefont {M.}~\bibnamefont {Manfra}},\ }\href@noop {}
  {\bibfield  {journal} {\bibinfo  {journal} {Applied Physics Letters}\
  }\textbf {\bibinfo {volume} {111}},\ \bibinfo {pages} {142106} (\bibinfo
  {year} {2017})}\BibitemShut {NoStop}%
\bibitem [{\citenamefont {Flensberg}\ \emph {et~al.}(1988)\citenamefont
  {Flensberg}, \citenamefont {Hansen},\ and\ \citenamefont
  {Octavio}}]{flensberg1988subharmonic}%
  \BibitemOpen
  \bibfield  {author} {\bibinfo {author} {\bibfnamefont {K.}~\bibnamefont
  {Flensberg}}, \bibinfo {author} {\bibfnamefont {J.~B.}\ \bibnamefont
  {Hansen}}, \ and\ \bibinfo {author} {\bibfnamefont {M.}~\bibnamefont
  {Octavio}},\ }\href@noop {} {\bibfield  {journal} {\bibinfo  {journal}
  {Physical Review B}\ }\textbf {\bibinfo {volume} {38}},\ \bibinfo {pages}
  {8707} (\bibinfo {year} {1988})}\BibitemShut {NoStop}%
\bibitem [{\citenamefont {Kj{\ae}rgaard}\ \emph {et~al.}(2017)\citenamefont
  {Kj{\ae}rgaard}, \citenamefont {Suominen}, \citenamefont {Nowak},
  \citenamefont {Akhmerov}, \citenamefont {Shabani}, \citenamefont
  {Palmstr{\o}m}, \citenamefont {Nichele},\ and\ \citenamefont
  {Marcus}}]{kjaergaard2017transparent}%
  \BibitemOpen
  \bibfield  {author} {\bibinfo {author} {\bibfnamefont {M.}~\bibnamefont
  {Kj{\ae}rgaard}}, \bibinfo {author} {\bibfnamefont {H.~J.}\ \bibnamefont
  {Suominen}}, \bibinfo {author} {\bibfnamefont {M.}~\bibnamefont {Nowak}},
  \bibinfo {author} {\bibfnamefont {A.}~\bibnamefont {Akhmerov}}, \bibinfo
  {author} {\bibfnamefont {J.}~\bibnamefont {Shabani}}, \bibinfo {author}
  {\bibfnamefont {C.}~\bibnamefont {Palmstr{\o}m}}, \bibinfo {author}
  {\bibfnamefont {F.}~\bibnamefont {Nichele}}, \ and\ \bibinfo {author}
  {\bibfnamefont {C.~M.}\ \bibnamefont {Marcus}},\ }\href@noop {} {\bibfield
  {journal} {\bibinfo  {journal} {Physical Review Applied}\ }\textbf {\bibinfo
  {volume} {7}},\ \bibinfo {pages} {034029} (\bibinfo {year}
  {2017})}\BibitemShut {NoStop}%
\bibitem [{\citenamefont {Waldram}(1996)}]{waldram1996book}%
  \BibitemOpen
  \bibfield  {author} {\bibinfo {author} {\bibfnamefont {J.~R.}\ \bibnamefont
  {Waldram}},\ }\href@noop {} {\emph {\bibinfo {title} {Superconductivity of
  metals and cuprates}}}\ (\bibinfo  {publisher} {IOP Publishing Ltd},\
  \bibinfo {year} {1996})\BibitemShut {NoStop}%
\bibitem [{\citenamefont {Awoga}\ \emph {et~al.}(2022)\citenamefont {Awoga},
  \citenamefont {Leijnse}, \citenamefont {Black-Schaffer},\ and\ \citenamefont
  {Cayao}}]{awoga2022mitigating}%
  \BibitemOpen
  \bibfield  {author} {\bibinfo {author} {\bibfnamefont {O.~A.}\ \bibnamefont
  {Awoga}}, \bibinfo {author} {\bibfnamefont {M.}~\bibnamefont {Leijnse}},
  \bibinfo {author} {\bibfnamefont {A.~M.}\ \bibnamefont {Black-Schaffer}}, \
  and\ \bibinfo {author} {\bibfnamefont {J.}~\bibnamefont {Cayao}},\
  }\href@noop {} {\bibfield  {journal} {\bibinfo  {journal} {arXiv preprint
  arXiv:2212.06061}\ } (\bibinfo {year} {2022})}\BibitemShut {NoStop}%
\end{thebibliography}%


\begin{thebibliography}{19}%
\makeatletter
\providecommand \@ifxundefined [1]{%
 \@ifx{#1\undefined}
}%
\providecommand \@ifnum [1]{%
 \ifnum #1\expandafter \@firstoftwo
 \else \expandafter \@secondoftwo
 \fi
}%
\providecommand \@ifx [1]{%
 \ifx #1\expandafter \@firstoftwo
 \else \expandafter \@secondoftwo
 \fi
}%
\providecommand \natexlab [1]{#1}%
\providecommand \enquote  [1]{``#1''}%
\providecommand \bibnamefont  [1]{#1}%
\providecommand \bibfnamefont [1]{#1}%
\providecommand \citenamefont [1]{#1}%
\providecommand \href@noop [0]{\@secondoftwo}%
\providecommand \href [0]{\begingroup \@sanitize@url \@href}%
\providecommand \@href[1]{\@@startlink{#1}\@@href}%
\providecommand \@@href[1]{\endgroup#1\@@endlink}%
\providecommand \@sanitize@url [0]{\catcode `\\12\catcode `\$12\catcode
  `\&12\catcode `\#12\catcode `\^12\catcode `\_12\catcode `\%12\relax}%
\providecommand \@@startlink[1]{}%
\providecommand \@@endlink[0]{}%
\providecommand \url  [0]{\begingroup\@sanitize@url \@url }%
\providecommand \@url [1]{\endgroup\@href {#1}{\urlprefix }}%
\providecommand \urlprefix  [0]{URL }%
\providecommand \Eprint [0]{\href }%
\providecommand \doibase [0]{https://doi.org/}%
\providecommand \selectlanguage [0]{\@gobble}%
\providecommand \bibinfo  [0]{\@secondoftwo}%
\providecommand \bibfield  [0]{\@secondoftwo}%
\providecommand \translation [1]{[#1]}%
\providecommand \BibitemOpen [0]{}%
\providecommand \bibitemStop [0]{}%
\providecommand \bibitemNoStop [0]{.\EOS\space}%
\providecommand \EOS [0]{\spacefactor3000\relax}%
\providecommand \BibitemShut  [1]{\csname bibitem#1\endcsname}%
\let\auto@bib@innerbib\@empty
\bibitem [{\citenamefont {Klapetek}\ \emph {et~al.}(2011)\citenamefont
  {Klapetek}, \citenamefont {Ne{\v{c}}as}, \citenamefont {Campbellov{\'a}},
  \citenamefont {Yacoot},\ and\ \citenamefont
  {Koenders}}]{klapetek2011methods}%
  \BibitemOpen
  \bibfield  {author} {\bibinfo {author} {\bibfnamefont {P.}~\bibnamefont
  {Klapetek}}, \bibinfo {author} {\bibfnamefont {D.}~\bibnamefont
  {Ne{\v{c}}as}}, \bibinfo {author} {\bibfnamefont {A.}~\bibnamefont
  {Campbellov{\'a}}}, \bibinfo {author} {\bibfnamefont {A.}~\bibnamefont
  {Yacoot}},\ and\ \bibinfo {author} {\bibfnamefont {L.}~\bibnamefont
  {Koenders}},\ }\href@noop {} {\bibfield  {journal} {\bibinfo  {journal}
  {Measurement Science and Technology}\ }\textbf {\bibinfo {volume} {22}},\
  \bibinfo {pages} {025501} (\bibinfo {year} {2011})}\BibitemShut {NoStop}%
\bibitem [{\citenamefont {Schindelin}\ \emph {et~al.}(2012)\citenamefont
  {Schindelin}, \citenamefont {Arganda-Carreras}, \citenamefont {Frise},
  \citenamefont {Kaynig}, \citenamefont {Longair}, \citenamefont {Pietzsch},
  \citenamefont {Preibisch}, \citenamefont {Rueden}, \citenamefont {Saalfeld},
  \citenamefont {Schmid} \emph {et~al.}}]{schindelin2012fiji}%
  \BibitemOpen
  \bibfield  {author} {\bibinfo {author} {\bibfnamefont {J.}~\bibnamefont
  {Schindelin}}, \bibinfo {author} {\bibfnamefont {I.}~\bibnamefont
  {Arganda-Carreras}}, \bibinfo {author} {\bibfnamefont {E.}~\bibnamefont
  {Frise}}, \bibinfo {author} {\bibfnamefont {V.}~\bibnamefont {Kaynig}},
  \bibinfo {author} {\bibfnamefont {M.}~\bibnamefont {Longair}}, \bibinfo
  {author} {\bibfnamefont {T.}~\bibnamefont {Pietzsch}}, \bibinfo {author}
  {\bibfnamefont {S.}~\bibnamefont {Preibisch}}, \bibinfo {author}
  {\bibfnamefont {C.}~\bibnamefont {Rueden}}, \bibinfo {author} {\bibfnamefont
  {S.}~\bibnamefont {Saalfeld}}, \bibinfo {author} {\bibfnamefont
  {B.}~\bibnamefont {Schmid}}, \emph {et~al.},\ }\href@noop {} {\bibfield
  {journal} {\bibinfo  {journal} {Nature Methods}\ }\textbf {\bibinfo {volume}
  {9}},\ \bibinfo {pages} {676} (\bibinfo {year} {2012})}\BibitemShut {NoStop}%
\bibitem [{\citenamefont {Momma}\ and\ \citenamefont
  {Izumi}(2008)}]{momma2008vesta}%
  \BibitemOpen
  \bibfield  {author} {\bibinfo {author} {\bibfnamefont {K.}~\bibnamefont
  {Momma}}\ and\ \bibinfo {author} {\bibfnamefont {F.}~\bibnamefont {Izumi}},\
  }\href@noop {} {\bibfield  {journal} {\bibinfo  {journal} {Journal of Applied
  Crystallography}\ }\textbf {\bibinfo {volume} {41}},\ \bibinfo {pages} {653}
  (\bibinfo {year} {2008})}\BibitemShut {NoStop}%
\bibitem [{\citenamefont {Nicolai}\ \emph {et~al.}(2021)\citenamefont
  {Nicolai}, \citenamefont {Biermann},\ and\ \citenamefont
  {Trampert}}]{nicolai2021application}%
  \BibitemOpen
  \bibfield  {author} {\bibinfo {author} {\bibfnamefont {L.}~\bibnamefont
  {Nicolai}}, \bibinfo {author} {\bibfnamefont {K.}~\bibnamefont {Biermann}},\
  and\ \bibinfo {author} {\bibfnamefont {A.}~\bibnamefont {Trampert}},\
  }\href@noop {} {\bibfield  {journal} {\bibinfo  {journal} {Ultramicroscopy}\
  }\textbf {\bibinfo {volume} {224}},\ \bibinfo {pages} {113261} (\bibinfo
  {year} {2021})}\BibitemShut {NoStop}%
\bibitem [{\citenamefont {Rosenthal}\ \emph {et~al.}(1998)\citenamefont
  {Rosenthal}, \citenamefont {Beasley}, \citenamefont {Char}, \citenamefont
  {Colclough},\ and\ \citenamefont
  {Zaharchuk}}]{rosenthal1998fluxfocusingtheory}%
  \BibitemOpen
  \bibfield  {author} {\bibinfo {author} {\bibfnamefont {P.~A.}\ \bibnamefont
  {Rosenthal}}, \bibinfo {author} {\bibfnamefont {M.~R.}\ \bibnamefont
  {Beasley}}, \bibinfo {author} {\bibfnamefont {K.}~\bibnamefont {Char}},
  \bibinfo {author} {\bibfnamefont {M.~S.}\ \bibnamefont {Colclough}},\ and\
  \bibinfo {author} {\bibfnamefont {G.}~\bibnamefont {Zaharchuk}},\ }\href@noop
  {} {\bibfield  {journal} {\bibinfo  {journal} {Applied Physics Letters}\
  }\textbf {\bibinfo {volume} {59}},\ \bibinfo {pages} {3482} (\bibinfo {year}
  {1998})}\BibitemShut {NoStop}%
\bibitem [{\citenamefont {Suominen}\ \emph {et~al.}(2017)\citenamefont
  {Suominen}, \citenamefont {Danon}, \citenamefont {Kjaergaard}, \citenamefont
  {Flensberg}, \citenamefont {Shabani}, \citenamefont {Palmstr{\o}m},
  \citenamefont {Nichele},\ and\ \citenamefont
  {Marcus}}]{suominen2017anomalousfraunhofer}%
  \BibitemOpen
  \bibfield  {author} {\bibinfo {author} {\bibfnamefont {H.}~\bibnamefont
  {Suominen}}, \bibinfo {author} {\bibfnamefont {J.}~\bibnamefont {Danon}},
  \bibinfo {author} {\bibfnamefont {M.}~\bibnamefont {Kjaergaard}}, \bibinfo
  {author} {\bibfnamefont {K.}~\bibnamefont {Flensberg}}, \bibinfo {author}
  {\bibfnamefont {J.}~\bibnamefont {Shabani}}, \bibinfo {author} {\bibfnamefont
  {C.}~\bibnamefont {Palmstr{\o}m}}, \bibinfo {author} {\bibfnamefont
  {F.}~\bibnamefont {Nichele}},\ and\ \bibinfo {author} {\bibfnamefont
  {C.}~\bibnamefont {Marcus}},\ }\href@noop {} {\bibfield  {journal} {\bibinfo
  {journal} {Physical Review B}\ }\textbf {\bibinfo {volume} {95}},\ \bibinfo
  {pages} {035307} (\bibinfo {year} {2017})}\BibitemShut {NoStop}%
\bibitem [{\citenamefont {Dartiailh}\ \emph {et~al.}(2021)\citenamefont
  {Dartiailh}, \citenamefont {Cuozzo}, \citenamefont {Elfeky}, \citenamefont
  {Mayer}, \citenamefont {Yuan}, \citenamefont {Wickramasinghe}, \citenamefont
  {Rossi},\ and\ \citenamefont {Shabani}}]{dartiailh2021fluxfocus}%
  \BibitemOpen
  \bibfield  {author} {\bibinfo {author} {\bibfnamefont {M.~C.}\ \bibnamefont
  {Dartiailh}}, \bibinfo {author} {\bibfnamefont {J.~J.}\ \bibnamefont
  {Cuozzo}}, \bibinfo {author} {\bibfnamefont {B.~H.}\ \bibnamefont {Elfeky}},
  \bibinfo {author} {\bibfnamefont {W.}~\bibnamefont {Mayer}}, \bibinfo
  {author} {\bibfnamefont {J.}~\bibnamefont {Yuan}}, \bibinfo {author}
  {\bibfnamefont {K.~S.}\ \bibnamefont {Wickramasinghe}}, \bibinfo {author}
  {\bibfnamefont {E.}~\bibnamefont {Rossi}},\ and\ \bibinfo {author}
  {\bibfnamefont {J.}~\bibnamefont {Shabani}},\ }\href@noop {} {\bibfield
  {journal} {\bibinfo  {journal} {Nature Communications 2021 12:1}\ }\textbf
  {\bibinfo {volume} {12}},\ \bibinfo {pages} {1} (\bibinfo {year}
  {2021})}\BibitemShut {NoStop}%
\bibitem [{\citenamefont {Likharev}(1979)}]{likharev1979superconducting}%
  \BibitemOpen
  \bibfield  {author} {\bibinfo {author} {\bibfnamefont {K.}~\bibnamefont
  {Likharev}},\ }\href@noop {} {\bibfield  {journal} {\bibinfo  {journal}
  {Reviews of Modern Physics}\ }\textbf {\bibinfo {volume} {51}},\ \bibinfo
  {pages} {101} (\bibinfo {year} {1979})}\BibitemShut {NoStop}%
\bibitem [{\citenamefont {Haxell}\ \emph {et~al.}(2022)\citenamefont {Haxell},
  \citenamefont {Cheah}, \citenamefont {K{\v{r}}{\'\i}{\v{z}}ek}, \citenamefont
  {Schott}, \citenamefont {Ritter}, \citenamefont {Hinderling}, \citenamefont
  {Belzig}, \citenamefont {Bruder}, \citenamefont {Wegscheider}, \citenamefont
  {Riel} \emph {et~al.}}]{haxell2022large}%
  \BibitemOpen
  \bibfield  {author} {\bibinfo {author} {\bibfnamefont {D.}~\bibnamefont
  {Haxell}}, \bibinfo {author} {\bibfnamefont {E.}~\bibnamefont {Cheah}},
  \bibinfo {author} {\bibfnamefont {F.}~\bibnamefont
  {K{\v{r}}{\'\i}{\v{z}}ek}}, \bibinfo {author} {\bibfnamefont
  {R.}~\bibnamefont {Schott}}, \bibinfo {author} {\bibfnamefont
  {M.}~\bibnamefont {Ritter}}, \bibinfo {author} {\bibfnamefont
  {M.}~\bibnamefont {Hinderling}}, \bibinfo {author} {\bibfnamefont
  {W.}~\bibnamefont {Belzig}}, \bibinfo {author} {\bibfnamefont
  {C.}~\bibnamefont {Bruder}}, \bibinfo {author} {\bibfnamefont
  {W.}~\bibnamefont {Wegscheider}}, \bibinfo {author} {\bibfnamefont
  {H.}~\bibnamefont {Riel}}, \emph {et~al.},\ }\href@noop {} {\bibfield
  {journal} {\bibinfo  {journal} {arXiv preprint arXiv:2204.05619}\ } (\bibinfo
  {year} {2022})}\BibitemShut {NoStop}%
\bibitem [{\citenamefont {Nikoli{\'c}}\ \emph {et~al.}(2001)\citenamefont
  {Nikoli{\'c}}, \citenamefont {Freericks},\ and\ \citenamefont
  {Miller}}]{nikolic2001intrinsic}%
  \BibitemOpen
  \bibfield  {author} {\bibinfo {author} {\bibfnamefont {B.~K.}\ \bibnamefont
  {Nikoli{\'c}}}, \bibinfo {author} {\bibfnamefont {J.}~\bibnamefont
  {Freericks}},\ and\ \bibinfo {author} {\bibfnamefont {P.}~\bibnamefont
  {Miller}},\ }\href@noop {} {\bibfield  {journal} {\bibinfo  {journal}
  {Physical Review B}\ }\textbf {\bibinfo {volume} {64}},\ \bibinfo {pages}
  {212507} (\bibinfo {year} {2001})}\BibitemShut {NoStop}%
\bibitem [{\citenamefont {Flensberg}\ \emph {et~al.}(1988)\citenamefont
  {Flensberg}, \citenamefont {Hansen},\ and\ \citenamefont
  {Octavio}}]{flensberg1988subharmonic}%
  \BibitemOpen
  \bibfield  {author} {\bibinfo {author} {\bibfnamefont {K.}~\bibnamefont
  {Flensberg}}, \bibinfo {author} {\bibfnamefont {J.~B.}\ \bibnamefont
  {Hansen}},\ and\ \bibinfo {author} {\bibfnamefont {M.}~\bibnamefont
  {Octavio}},\ }\href@noop {} {\bibfield  {journal} {\bibinfo  {journal}
  {Physical Review B}\ }\textbf {\bibinfo {volume} {38}},\ \bibinfo {pages}
  {8707} (\bibinfo {year} {1988})}\BibitemShut {NoStop}%
\bibitem [{\citenamefont {Kjaergaard}\ \emph {et~al.}(2017)\citenamefont
  {Kjaergaard}, \citenamefont {Suominen}, \citenamefont {Nowak}, \citenamefont
  {Akhmerov}, \citenamefont {Shabani}, \citenamefont {Palmstrøm},
  \citenamefont {Nichele},\ and\ \citenamefont {Marcus}}]{kjaergaard2017mar}%
  \BibitemOpen
  \bibfield  {author} {\bibinfo {author} {\bibfnamefont {M.}~\bibnamefont
  {Kjaergaard}}, \bibinfo {author} {\bibfnamefont {H.~J.}\ \bibnamefont
  {Suominen}}, \bibinfo {author} {\bibfnamefont {M.~P.}\ \bibnamefont {Nowak}},
  \bibinfo {author} {\bibfnamefont {A.~R.}\ \bibnamefont {Akhmerov}}, \bibinfo
  {author} {\bibfnamefont {J.}~\bibnamefont {Shabani}}, \bibinfo {author}
  {\bibfnamefont {C.~J.}\ \bibnamefont {Palmstrøm}}, \bibinfo {author}
  {\bibfnamefont {F.}~\bibnamefont {Nichele}},\ and\ \bibinfo {author}
  {\bibfnamefont {C.~M.}\ \bibnamefont {Marcus}},\ }\href@noop {} {\bibfield
  {journal} {\bibinfo  {journal} {Physical Review Applied}\ }\textbf {\bibinfo
  {volume} {7}},\ \bibinfo {pages} {034029} (\bibinfo {year}
  {2017})}\BibitemShut {NoStop}%
\bibitem [{\citenamefont {Naidyuk}\ and\ \citenamefont
  {Gloos}(2018)}]{naidyuk2018excesscurrentreview}%
  \BibitemOpen
  \bibfield  {author} {\bibinfo {author} {\bibfnamefont {Y.~G.}\ \bibnamefont
  {Naidyuk}}\ and\ \bibinfo {author} {\bibfnamefont {K.}~\bibnamefont
  {Gloos}},\ }\href@noop {} {\bibfield  {journal} {\bibinfo  {journal} {Low
  Temperature Physics}\ }\textbf {\bibinfo {volume} {44}},\ \bibinfo {pages}
  {257} (\bibinfo {year} {2018})}\BibitemShut {NoStop}%
\bibitem [{\citenamefont {Artemenko}\ \emph {et~al.}(1979)\citenamefont
  {Artemenko}, \citenamefont {Volkov},\ and\ \citenamefont
  {Zaitsev}}]{artemenko1979excess}%
  \BibitemOpen
  \bibfield  {author} {\bibinfo {author} {\bibfnamefont {S.~N.}\ \bibnamefont
  {Artemenko}}, \bibinfo {author} {\bibfnamefont {A.}~\bibnamefont {Volkov}},\
  and\ \bibinfo {author} {\bibfnamefont {A.}~\bibnamefont {Zaitsev}},\
  }\href@noop {} {\bibfield  {journal} {\bibinfo  {journal} {Solid State
  Communications}\ }\textbf {\bibinfo {volume} {30}},\ \bibinfo {pages} {771}
  (\bibinfo {year} {1979})}\BibitemShut {NoStop}%
\bibitem [{\citenamefont {Lee}\ \emph {et~al.}(2019)\citenamefont {Lee},
  \citenamefont {Shojaei}, \citenamefont {Pendharkar}, \citenamefont
  {McFadden}, \citenamefont {Kim}, \citenamefont {Suominen}, \citenamefont
  {Kjaergaard}, \citenamefont {Nichele}, \citenamefont {Zhang}, \citenamefont
  {Marcus} \emph {et~al.}}]{lee2019transport}%
  \BibitemOpen
  \bibfield  {author} {\bibinfo {author} {\bibfnamefont {J.~S.}\ \bibnamefont
  {Lee}}, \bibinfo {author} {\bibfnamefont {B.}~\bibnamefont {Shojaei}},
  \bibinfo {author} {\bibfnamefont {M.}~\bibnamefont {Pendharkar}}, \bibinfo
  {author} {\bibfnamefont {A.~P.}\ \bibnamefont {McFadden}}, \bibinfo {author}
  {\bibfnamefont {Y.}~\bibnamefont {Kim}}, \bibinfo {author} {\bibfnamefont
  {H.~J.}\ \bibnamefont {Suominen}}, \bibinfo {author} {\bibfnamefont
  {M.}~\bibnamefont {Kjaergaard}}, \bibinfo {author} {\bibfnamefont
  {F.}~\bibnamefont {Nichele}}, \bibinfo {author} {\bibfnamefont
  {H.}~\bibnamefont {Zhang}}, \bibinfo {author} {\bibfnamefont {C.~M.}\
  \bibnamefont {Marcus}}, \emph {et~al.},\ }\href@noop {} {\bibfield  {journal}
  {\bibinfo  {journal} {Nano Letters}\ }\textbf {\bibinfo {volume} {19}},\
  \bibinfo {pages} {3083} (\bibinfo {year} {2019})}\BibitemShut {NoStop}%
\bibitem [{\citenamefont {Chrestin}\ \emph {et~al.}(1997)\citenamefont
  {Chrestin}, \citenamefont {Matsuyama},\ and\ \citenamefont
  {Merkt}}]{chrestin1997implicitexperiment}%
  \BibitemOpen
  \bibfield  {author} {\bibinfo {author} {\bibfnamefont {A.}~\bibnamefont
  {Chrestin}}, \bibinfo {author} {\bibfnamefont {T.}~\bibnamefont
  {Matsuyama}},\ and\ \bibinfo {author} {\bibfnamefont {U.}~\bibnamefont
  {Merkt}},\ }\href@noop {} {\bibfield  {journal} {\bibinfo  {journal}
  {Physical Review B}\ }\textbf {\bibinfo {volume} {55}},\ \bibinfo {pages}
  {8457} (\bibinfo {year} {1997})}\BibitemShut {NoStop}%
\bibitem [{\citenamefont {Aminov}\ \emph {et~al.}(1996)\citenamefont {Aminov},
  \citenamefont {Golubov},\ and\ \citenamefont
  {Kupriyanov}}]{aminov1996implicittheory}%
  \BibitemOpen
  \bibfield  {author} {\bibinfo {author} {\bibfnamefont {B.}~\bibnamefont
  {Aminov}}, \bibinfo {author} {\bibfnamefont {A.}~\bibnamefont {Golubov}},\
  and\ \bibinfo {author} {\bibfnamefont {M.~Y.}\ \bibnamefont {Kupriyanov}},\
  }\href@noop {} {\bibfield  {journal} {\bibinfo  {journal} {Physical Review
  B}\ }\textbf {\bibinfo {volume} {53}},\ \bibinfo {pages} {365} (\bibinfo
  {year} {1996})}\BibitemShut {NoStop}%
\bibitem [{\citenamefont {Schäpers}(2001)}]{tschapers2001}%
  \BibitemOpen
  \bibfield  {author} {\bibinfo {author} {\bibfnamefont {T.}~\bibnamefont
  {Schäpers}},\ }\href@noop {} {\emph {\bibinfo {title}
  {Superconductor/Semiconductor Junctions}}},\ Vol.\ \bibinfo {volume} {174}\
  (\bibinfo  {publisher} {Springer Berlin Heidelberg},\ \bibinfo {year}
  {2001})\BibitemShut {NoStop}%
\bibitem [{\citenamefont {Awoga}\ \emph {et~al.}(2022)\citenamefont {Awoga},
  \citenamefont {Leijnse}, \citenamefont {Black-Schaffer},\ and\ \citenamefont
  {Cayao}}]{awoga2022mitigating}%
  \BibitemOpen
  \bibfield  {author} {\bibinfo {author} {\bibfnamefont {O.~A.}\ \bibnamefont
  {Awoga}}, \bibinfo {author} {\bibfnamefont {M.}~\bibnamefont {Leijnse}},
  \bibinfo {author} {\bibfnamefont {A.~M.}\ \bibnamefont {Black-Schaffer}},\
  and\ \bibinfo {author} {\bibfnamefont {J.}~\bibnamefont {Cayao}},\
  }\href@noop {} {\bibfield  {journal} {\bibinfo  {journal} {arXiv preprint
  arXiv:2212.06061}\ } (\bibinfo {year} {2022})}\BibitemShut {NoStop}%
\end{thebibliography}%

\end{document}


\title{Supplemental Material: Control over epitaxy and the role of the InAs/Al interface in hybrid two-dimensional electron gas systems}

\author{Erik Cheah}
\email{echeah@phys.ethz.ch}
\affiliation{Solid State Physics Laboratory, ETH Zurich, 8093 Zurich, Switzerland}

\author{Daniel Z. Haxell}
\affiliation{IBM Research Europe - Zurich, 8803 Rüschlikon, Switzerland}

\author{R\"udiger Schott}
\affiliation{Solid State Physics Laboratory, ETH Zurich, 8093 Zurich, Switzerland}

\author{Peng Zeng}
\affiliation{ScopeM, ETH Zurich, 8093 Zurich, Switzerland}

\author{Ekaterina Paysen}
\affiliation{Paul-Drude-Institut für Festkörperelektronik, Leibniz-Institut im Forschungsverbund Berlin e. V., 10117 Berlin, Germany}

\author{Sofieke C. ten Kate}
\affiliation{IBM Research Europe - Zurich, 8803 Rüschlikon, Switzerland}

\author{Marco Coraiola}
\affiliation{IBM Research Europe - Zurich, 8803 Rüschlikon, Switzerland}

\author{Max Landstetter}
\affiliation{Solid State Physics Laboratory, ETH Zurich, 8093 Zurich, Switzerland}

\author{Ali B. Zadeh}
\affiliation{ScopeM, ETH Zurich, 8093 Zurich, Switzerland}

\author{Achim Trampert}
\affiliation{Paul-Drude-Institut für Festkörperelektronik, Leibniz-Institut im Forschungsverbund Berlin e. V., 10117 Berlin, Germany}

\author{Marilyne Sousa}
\affiliation{IBM Research Europe - Zurich, 8803 Rüschlikon, Switzerland}

\author{Heike Riel}
\affiliation{IBM Research Europe - Zurich, 8803 Rüschlikon, Switzerland}

\author{Fabrizio Nichele}
\affiliation{IBM Research Europe - Zurich, 8803 Rüschlikon, Switzerland}

\author{Werner Wegscheider}
\affiliation{Solid State Physics Laboratory, ETH Zurich, 8093 Zurich, Switzerland}
\affiliation{Quantum Center, ETH Zurich, 8093 Zurich, Switzerland}

\author{Filip Krizek}
\affiliation{Solid State Physics Laboratory, ETH Zurich, 8093 Zurich, Switzerland}
\affiliation{IBM Research Europe - Zurich, 8803 Rüschlikon, Switzerland}
\affiliation{Institute of Physics, Czech Academy of Sciences, 162 00 Prague, Czech Republic}

\date{\today}

\maketitle

\section{Methods}

\subsection{MBE system and additional growth details}
The samples were grown in a Veeco GEN II MBE system equipped with an As valve cracker cell. The cracker temperature during growth was 400$^{\circ}$C, at which we expected to produce mainly As$_{4}$ molecules. The temperature was measured by tracking the absorption edge of the InP substrate with a commercial kSA BandiT thermometry system. The group III growth rates were determined from RHEED oscillations. The given V/III ratios were determined from the group V and group III growth rates. The As growth rate was determined from As oscillations as follows. We terminated the As flux and deposited roughly 10 MLs of Ga on the surface of GaAs at 0.5 \AA/s and 570$^{\circ}$C. Immediately after, we introduced As at a selected valve setting. This resulted in clear and pronounced RHEED oscillations. The group III growth rates were measured at 570$^{\circ}$C for Ga and Al, and at 450$^{\circ}$C for In. For the deposition of Al, we used a Veeco SUMO effusion cell which is operated with a cold tip. This was to limit the heat delivered to the cold substrate surface during deposition. In our system, the distance from the tip of the Al cell to the sample was 18.26 cm. 

\subsection{AFM imaging}

The surface topography was examined with a BRUKER Dimension FastScan AFM using tapping mode in air. A silicon tip on a silicon nintride cantiliver FASTSCAN-A with the following parameters: T = 0.6 $\mu$m, L = 27$\mu$m, W = 32$\mu$m, f$_0$ = 1.400 kHz and = 18 N/m. These measurements were performed with a scan rate of 3.92 Hz, scan size of 1 $\mu$m and resolution of 512 samples/line. The measured data were analyzed with Gwyddion \cite{klapetek2011methods}, and a planar second order fit was the sole correction applied to the acquired data. 

\subsection{TEM lamellae preparation}
The lamellae of cross-sectional samples were prepared by Focused Ion Beam (FIB) (Helios 5 UX from Thermo Scientific) using AutoTEM 5 software (Thermo Scientific, the Netherlands) at ScopeM, ETH Zurich. A protective carbon layer was deposited on the selected region of interest first by an electron beam (2 kV, 13 nA) and subsequently by an ion beam (30 kV, 1.2 nA). The chuck milling and lamellae thinning were done at 30 kV with FIB current from 9 nA to 90 pA. Finally the lamellae were polished at 5 kV (17 pA) and finished at 2 kV (12 pA). The expected thickness was bellow 50 nm. 

The samples were kept in vacuum between the FIB and TEM measurements. A selected lamella was transferred to a MEMS Fusion Select Heating chip from Protochip and used for the in-situ annealing experiments.

\begin{figure*}[hb!]
\vspace{0.2cm}
\includegraphics[scale=0.25]{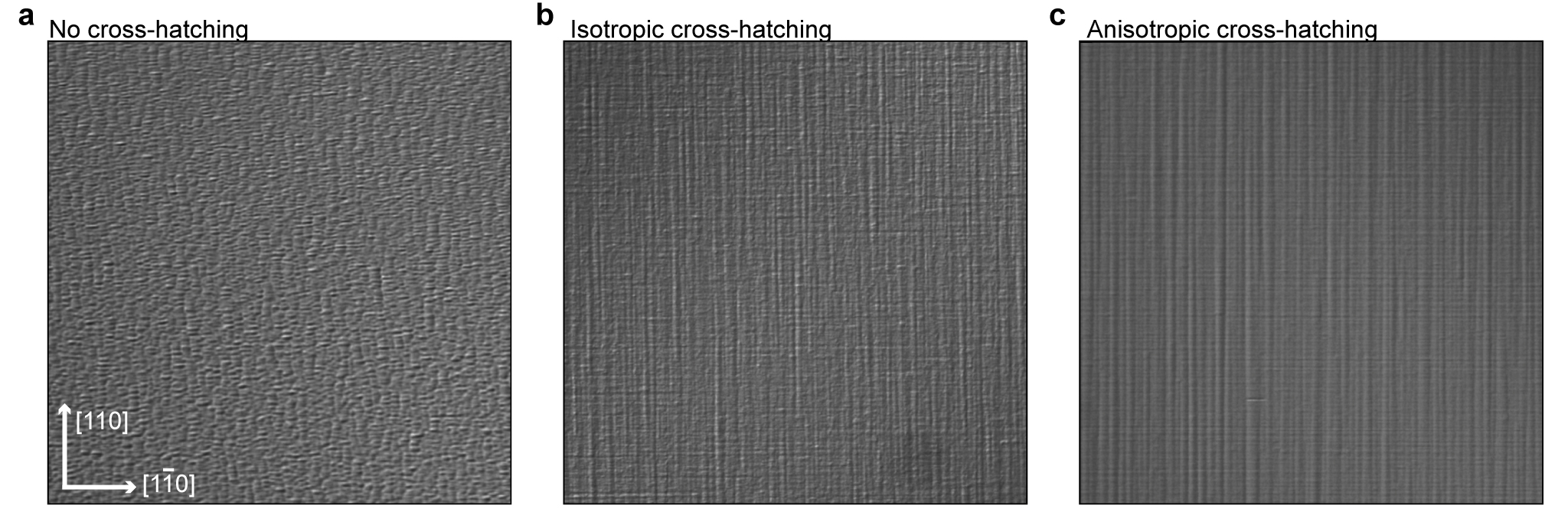}
\vspace{-0.4cm}
\caption{Optical images showing the surface of the shallow InAs 2DEG with the metamorphic buffer layer grown at different conditions with: (a) no cross-hatching, (b) isotropic cross-hatching, (c) anisotropic cross-hatching.}
\label{figS1}
\end{figure*}

\subsection{STEM measurements}
The STEM images were acquired with a JEOL ARM200F cold field emission gun scanning transmission electron microscope located at the Binnig and Rohrer Nanotechnology Center Noise-free laboratories at IBM Research Europe. The images were acquired at 200 kV.

The in-situ heating experiment was performed using a double aberration-corrected JEOL JEM ARM300F GrandARM operated at 200 kV at ScopeM, ETH Zürich. The in-situ heating holder of Protochips Fusion, connected to a KEITHLEY 2614B controller, was used to heat up the lamella. The STEM images were acquired at a camera distance of 8 cm or 12 cm, covering the acceptance semi-angle of 80 mrad to 175 mrad for HAADF detector and 15 mrad to 20 mrad for the ABF detector. 

The tilt series for tomography was recorded on a Cs-corrected JEOL ARM 200F microscope operated at an accelerating voltage of 200 kV in STEM dark-field mode. Annular dark-field detector (ADF) signals were acquired at an acceptance angle of 140-550 mrad to produce HAADF-Z contrast. In addition, a 30 µm diameter condenser aperture was used to achieve sufficient focal depth. The alignment of the HAADF images with respect to each other was performed using the IMOD software package. The 3D reconstruction was then calculated with the simultaneous iterative reconstruction technique (SIRT) algorithm with 55 iterations using ASTRA Toolbox. Amira-AviszoTM software was applied to visualize the 3D data. 

\subsection{STEM image processing and crystal simulations}
The acquired STEM images were processed using FIJI - Fiji is just ImageJ freeware \cite{schindelin2012fiji}. Typically, the FFT pattern of the images was masked by a circular aperture and background noise was removed by subtracting a constant background. In selected images, the contrast was normalized by the contrast normalization plugin, to reduce the large contrast between light Al and relatively heavy III-V material atoms. 
The crystal simulations used for the STEM image overlays were made using the JP-Minerals VESTA visualization program for structural models \cite{momma2008vesta}.

\subsection{Magneto-transport characterization}
The electron mobility and density were extracted on as-grown material after a Transene D wet etch removal of the Al layer. 5x5 mm$^{2}$ chips were contacted by direct soldering of In (without annealing). The data were acquired using standard lock-in techniques and magnetic sweeps up to 250 mT at 4.2 K. 

\subsection{Hybrid device fabrication}
For fabrication of the studied Josephson junctions, the Al film was selectively removed with a Transene D based wet etch after a standard electron beam lithography. Subsequently, the semiconductor mesas were defined by chemical wet etching (220:55:3:3 mixture of H$_2$O:C$_6$H$_8$O$_7$:H$_3$PO$_4$:H$_2$O$_2$) of the V-III structure (depth of approximately 350 nm). The fine features in the remaining Al film were again defined by electron beam lithography and Transene D etching at 50$^\circ$C for 4 s. Subsequently 10 nm Ti and 440 Al were deposited as contacts and gate electrodes. The nominal lead separation of the measured JJs was 50 nm, where from scanning electron microscope imaging of test structures on the same chip, we measured 68 nm after etching. 

\subsection{Hybrid device measurements}

The JJ devices were measured in a dilution refrigerator with a mixing chamber temperature of $\sim$18 mK, and data were obtained through standard lock-in techniques .The JJs were measured in a four-terminal geometry. A current, which consisted of an AC and a DC component, was applied to one contact. The AC current of 2 nA (3 nA) was applied at a frequency of 277 Hz (333 Hz), for device the standard sample (for the intentionally roughened sample). The differential voltage V$_{AC}$ across the device was measured at this frequency, via a differential voltage amplifier with a gain of 1000. For the multiple Andreev reflection (MAR) measurements of the JJs fabricated from the sample with intentionally roughened surface, presented in the main text, the DC voltage $V_\mathrm{DC}$ was recorded in addition to the AC voltage. We note that all the characterized devices had the same critical temperature of $\sim$ 1.37 K.

\section{S1: Surface of InAs 2DEGs}

During optimization of growth steps for the shallow InAs 2DEG metamorphic buffer layers (details given in the main text), we observed the formation of three typical surface morphologies, which are shown in Fig. S\ref{figS1}. The surface texturing depended on multiple growth conditions, where the most prominent was typically the growth temperature. We found that lowering of the growth temperature reduced the surface roughness/strength of cross-hatching. In our case, the samples with pronounced anisotropic cross-hatching yielded the highest as-grown electron mobility. The possibility of surface characterization by optical microscopy provided a fast feedback for tuning of the growth recipe.

\section{S2: Extraction of lattice spacings}

\begin{figure*}[h]
\vspace{0.2cm}
\includegraphics[scale=0.25]{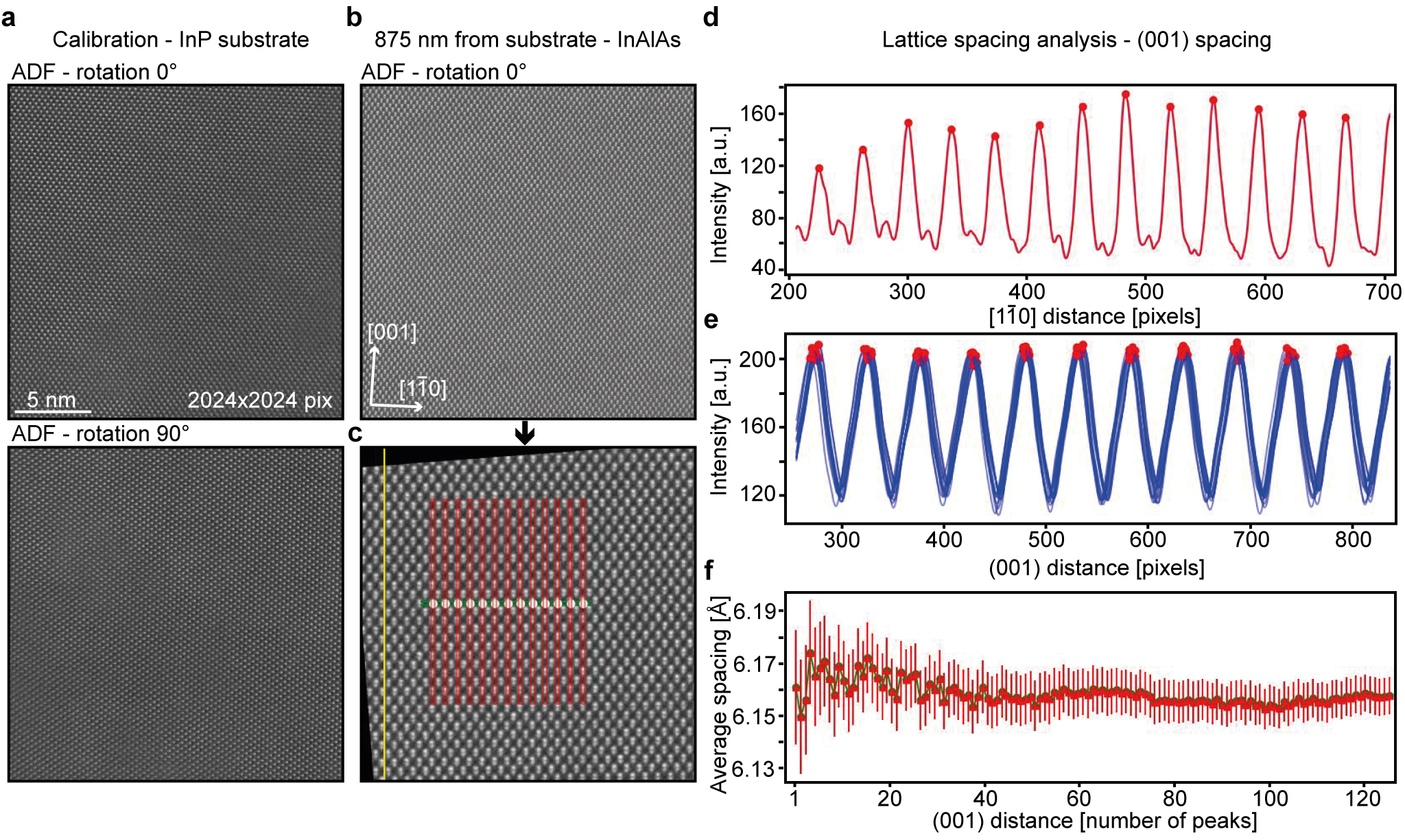}
\vspace{-0.4cm}
\caption{(a) HAADF-STEM micrographs of the InP substrate, with the beam scanning along the [1$\bar{1}$0] (top panel) and [001] directions (bottom panel). (b) HAADF-STEM micrograph taken 875 nm from the interface of the substrate, within the InAlAs region of the metamorphic buffer layer. (c) Zoom-in of b), where the horizontal (1$\bar{1}$0) planes are aligned to the vertical direction (yellow line). (d) Fit to the position of the atomic columns averaged over the area highlighted in c) (green rectangle), where the positions of the peaks (red points) correspond to the white dots in c). (e) Fit of the atomic positions, extracted from the interpolated intensity of the III-V dumbbells averaged over the areas in c) highlighted by red rectangles). (f) Evolution of the average spacing of the (001) planes extracted from the dumbbell positions.}
\label{figS2}
\end{figure*}

In order to monitor the lattice spacing evolution within our shallow InAs 2DEG, a sequence of images was taken roughly in the center of each layer of the structure. The pixel to nm calibration was extracted from images of the InP substrate (5.8687 \AA), which are shown in Fig. S\ref{figS2}a. Separate images were acquired for analysis of the spacing between both (1$\bar{1}$0) and (100) planes (90$^\circ$ rotation of the crystal). The measurements were set up such that the electron beam scans in perpendicular to the measured planes. Individual nm to pixel calibration was extracted for both directions, in order to reduce the influence of artifacts associated with the imaging. In addition, we avoided alignment of the scanning direction with the crystal, in-order to maintain the natural axes of the magnetic lenses. 

An example of the analysis of a HAADF-STEM image acquired at the same acquisition speed, scanning direction, resolution and magnification as the calibration images is shown in Fig. S\ref{figS2}b for part of the metamorphic InAlAs buffer layer at 875 nm from the substrate. First, the image was aligned with the (1$\bar{1}$0) planes of the crystal. A small area of the image was selected for analysis, in order to avoid deviations associated with descanning, Fig. S\ref{figS2}c. The positions of selected atomic columns were fitted as shown in Fig. S\ref{figS2}d. The intensities within selected areas around each selected column were then averaged and interpolated in order to reduce peak shifting due to double peaks induced by the III-V dumbbells. The (001) plane spacing was extracted by fitting to the center of the individual dumbbells, as shown in Figs. S\ref{figS2}e and f. 

This procedure was repeated for all layers within our structure and for both the (1$\bar{1}$0) and (100) plane spacings. Results are shown in Fig. 1 of the main text.

\section{S3: Controlled oxidation of Al}

\begin{figure*}[h]
\vspace{0.2cm}
\includegraphics[scale=0.25]{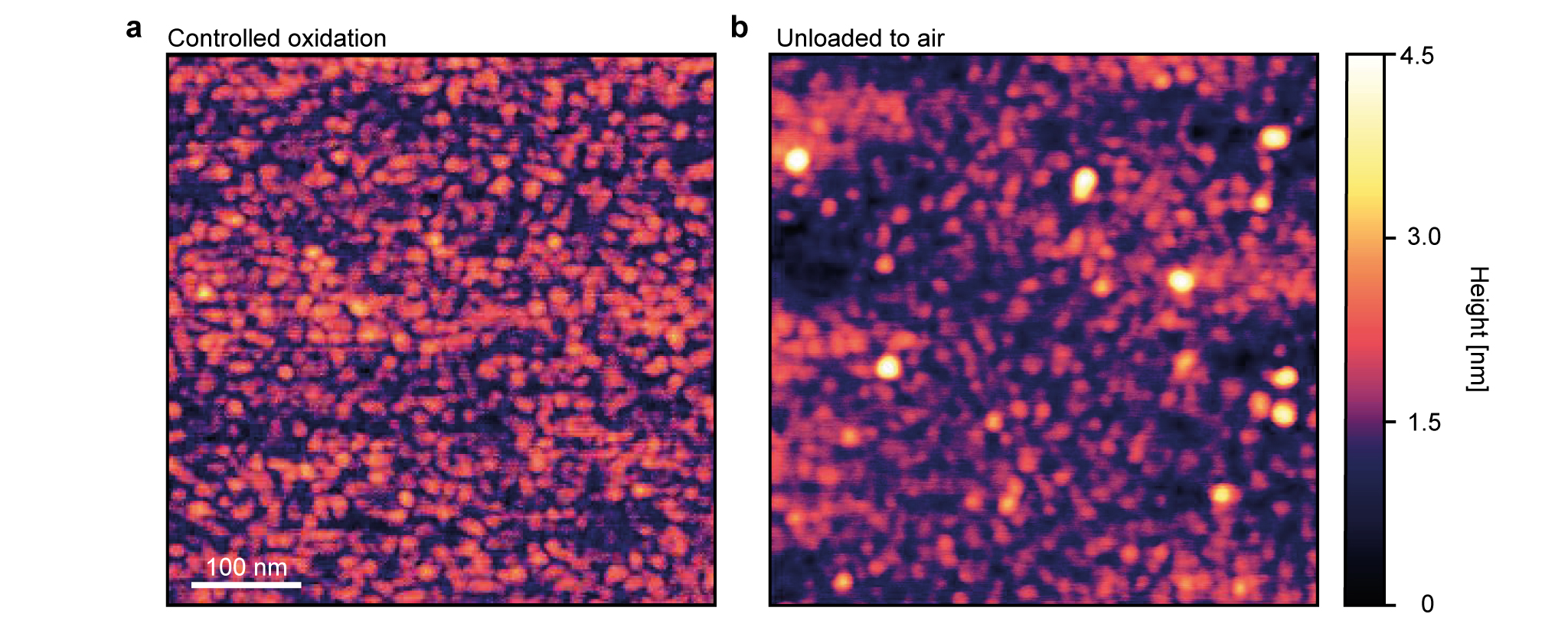}
\vspace{-0.4cm}
\caption{AFM micrographs with 500x500 nm field of view of the surface of the Al films: (a) unloaded through controlled oxidation and (b) directly into air.}
\label{figS3}
\end{figure*}

The quality of the in-situ deposited Al films strongly depends on the unloading procedure after deposition. The sample is slightly heated during deposition and typically removed from the cold sample holder in the growth chamber for transfer. In our case, we turned off all sources of heat that the sample could encounter during the unloading procedure, e.g. the ion gauges in the buffer chamber. Then the sample is transferred as fast as possible from the growth chamber into the load lock (kept at ambient temperature), and we immediately vented the chamber by controlled introduction of a Ar/O$_2$ (90\% /10\%) mixture (described in the main text), in order to warm the cold sample to room temperature and let the Al surface oxidize in a controlled environment. The difference in surface roughness, in comparison to unloading directly into air, is shown in Fig. S\ref{figS3}. The controlled oxidation yielded not only a smoother surface with a more homogeneous distribution of native AlO$_{x}$ grains, but also seemed to limit diffusion of Al and the formation of larger AlO$_{x}$ particles. In our transport experiments, we found that the controlled oxidation allowed for more consistent device fabrication results and transport characteristics of JJ  devices.

\section{S4: In-situ annealing STEM experiment}
\begin{figure*}[h]
\vspace{0.2cm}
\includegraphics[scale=0.25]{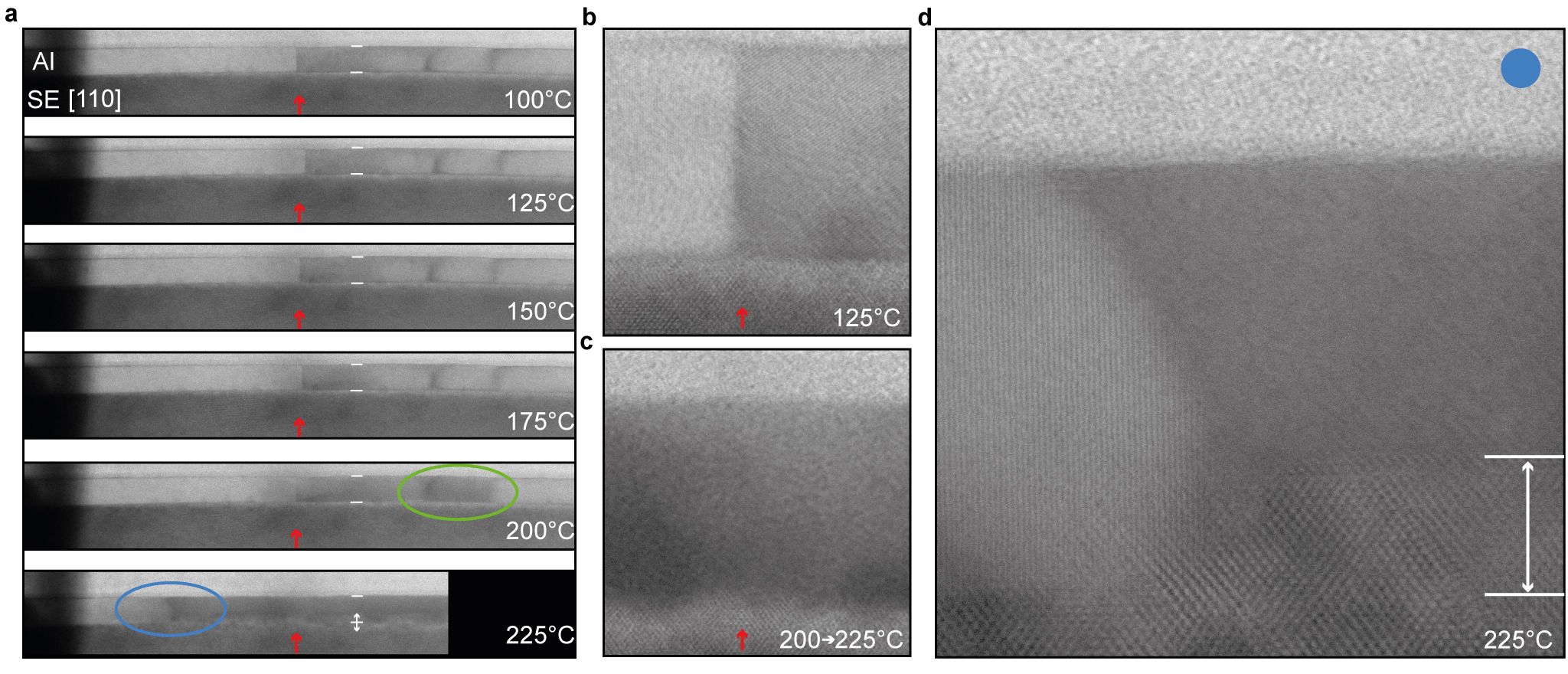}
\vspace{-0.4cm}
\caption{(a) Sequence of BF STEM images acquired at different temperatures. The structural changes that occurred above 175$^{\circ}$C are highlighted in the by green and blue circle. The red arrow highlight the location of a grain boundary, which disappeared, as shown in more detail in (b) and (c). (d) Detailed image of the blue circled area in a), where a significant portion of the Al layer alloyed with the substrate.}
\label{figS4}
\end{figure*}

In-situ annealing experiments up to 225$^{\circ}$C were performed in order to investigate the thermal stability of the Al films. This was partially motivated by the necessity of optimization of annealing steps during the fabrication process, but also to investigate the thermal stability of the individual Al grains. The results are shown in Fig. S\ref{figS4}a. A lamella prepared from one of our standard samples without the roughened surface was annealed in-situ from 100$^{\circ}$C to 225$^{\circ}$C in 5 steps, staying roughly 4 minutes at each step, and with a ramp speed of 25$^{\circ}$C per 1.5 min (the annealing time was similar to typical electron-beam lithography resist bake-out times, but much less than typical ALD gate oxide deposition). The film remained stable up to 175$^{\circ}$C, where only subtle changes in contrast in BF STEM images appeared. At 200$^{\circ}$C, stronger changes in contrast appeared around the observed grain boundaries, indicating slow local changes in the crystal structure. While ramping from 200$^{\circ}$C to 225$^{\circ}$C, the film started to recrystallize around the grain boundaries and at the interface, as shown in Figs. S\ref{figS4}a and b. Interestingly, the recrystallization of part of the of the Al/SE interface into a zinc-blende structure showed selectivity to the Al grains. This indicates that the grains have a reduced chemical and thermal stability. In future experiments higher resolution needs to be achieved to reliably distinguish the individual Al grain orientations. The results could hint towards a preferred Al orientation for growth of stable single crystalline films.

\section{S5: Ageing of studied lamellae}
\begin{figure*}[h]
\vspace{0.2cm}
\includegraphics[scale=0.25]{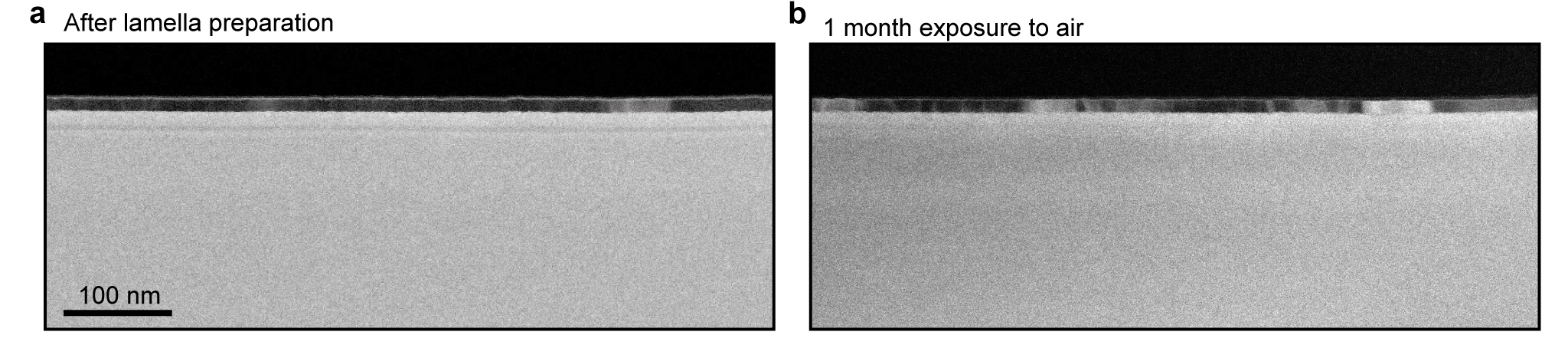}
\vspace{-0.4cm}
\caption{ADF STEM images of the same Al/SE hybrid sample imaged under the same conditions (a) after growth and (b) 1 month after first measurement.}
\label{figS5}
\end{figure*}

During our experiments, we observed degradation of the Al film when the prepared lamellae were exposed to air. A measurement before and after 1 month of exposure to air are shown in Fig. S\ref{figS5}. Even though the 'before' image seems to be acquired with higher quality and the InAs QW is clearly visible, there is a significant amount of new grain boundaries visible in the Al film after the exposure to air. This made consistent investigations of the film crystallinity problematic and samples needed to be stored in vacuum whenever possible, as the results could vary depending on the age of the material. We did not observe similar degradation for non-processed as-grown wafers, where a high-quality lamellas could be prepared months after growth.

\section{S6: STEM tomography}

\begin{figure*}[h]
\vspace{0.2cm}
\includegraphics[scale=0.25]{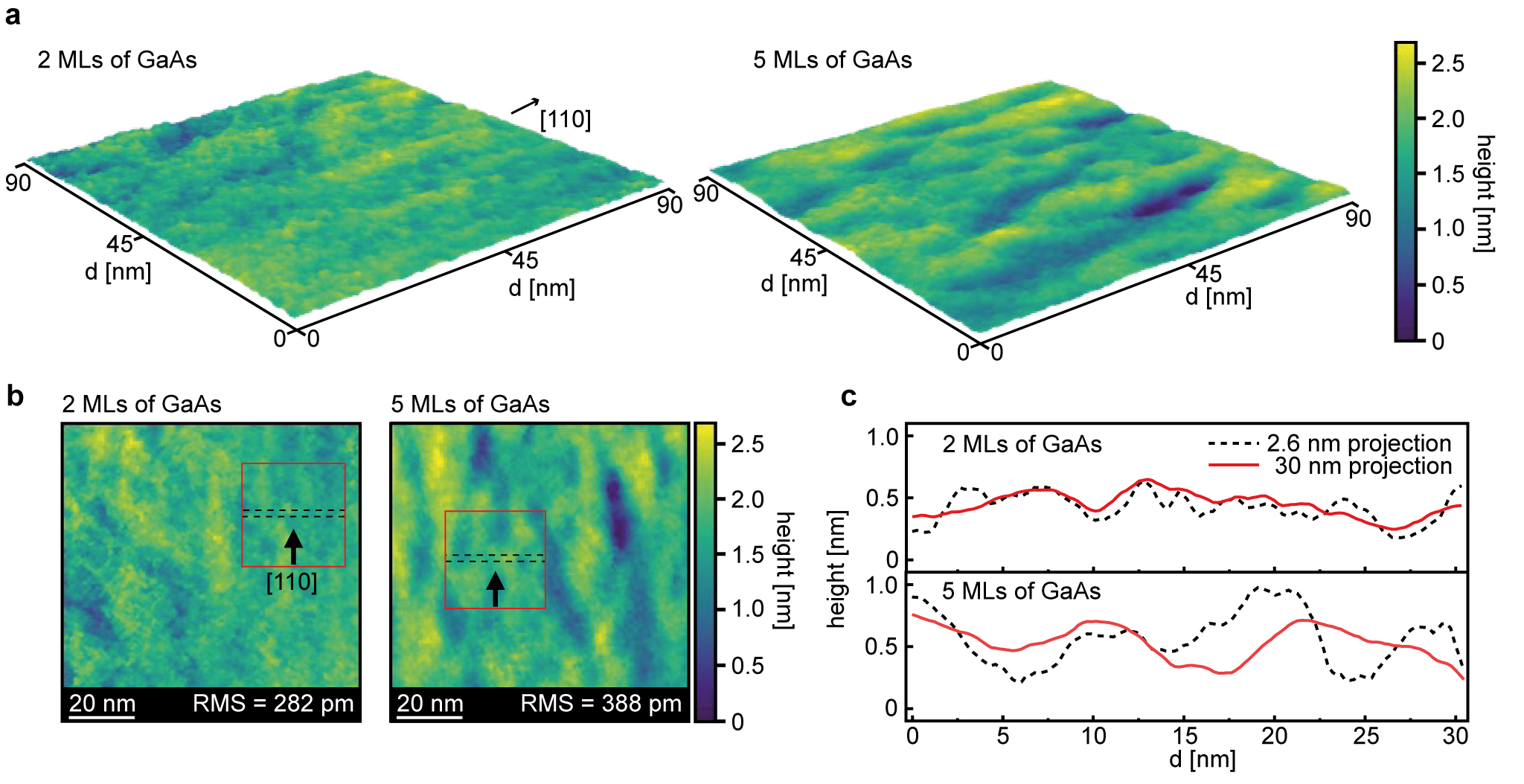}
\vspace{-0.4cm}
\caption{(a) 3D reconstruction acquired by HAADF-STEM tomography at the SE/Al interface of samples capped with 2 MLs and 5 MLs of GaAs. (b) 2D projection of the same data. (c) Projections of the height profiles averaged along the [110] direction (with 2.6 nm and 30 nm depth) for both samples, taken from areas highlighted in b).}
\label{figS6}
\end{figure*}

The interfaces between SE and Al film were characterized by HAADF-STEM tomography for both the standard sample (2 MLs of GaAs) and the intentionally roughened sample (5 MLs of GaAs), in order to characterize them without the necessity to remove the Al layer. The tomography was performed on a cylindrical needle with roughly 250 nm diameter prepared with FIB and where the needle axis coincides with the direction of growth. The preparation was similar to Ref. \citen{nicolai2021application} and the results reflect the surface morphology without affecting the strain as much as in the standard 30 nm thin lamella. In this experiment, the needle was successively rotated around its axis in the microscope and a series of 86 HAADF-STEM projections were recorded in an angular range of 170°. From these projections, a 3D image (tomogram) of the interface region is reconstructed with a sub-nanometer voxel resolution. The tomogram's voxel intensities are linked to the HAADF contrast of the projections and are sensitive to the chemical composition. The strong contrast between the SE and Al allows the extraction of all voxels with intermediate intensity, which in their entirety form the interface. Such a 3D interface is converted into a 2D topography map to quantitatively analyse its morphology with Gwyddion \cite{klapetek2011methods}

The results are summarized in Fig. S\ref{figS6}. The difference in roughness between the standard and roughened interfaces is apparent in the 3D height maps in Fig. S\ref{figS6}a. A 2D projection of the these maps is shown in Fig. S\ref{figS6}b, where the extracted root mean square (RMS) roughness value is 282 pm for the standard interface and 388 pm for the roughened one. The anisotropy along the [110] and [1$\bar{1}$0] directions is apparent at both interfaces, yet it is strongly enhanced on the roughened one. This is highlighted in Fig. S\ref{figS6}c, where an averaged height profile is shown for projection along the [110] direction averaged over 2.6 nm and 30 nm (where 30 nm was selected to resemble the real lamella thickness). The height oscillations related to the anisotropy are more pronounced at the roughened interface and they also correspond well to the $\sim$15 nm period observed in standard STEM in Fig. 4c in the Main Text. We note that the absolute numbers for height can be affected by definition of the location of the interface in the tomogram, but this should not affect the observed differences between the interface morphologies.

\section{S7: Additional analysis of the JJ measurements of the roughened sample}

\begin{figure*}[h]
\vspace{0.2cm}
\includegraphics[scale=0.25]{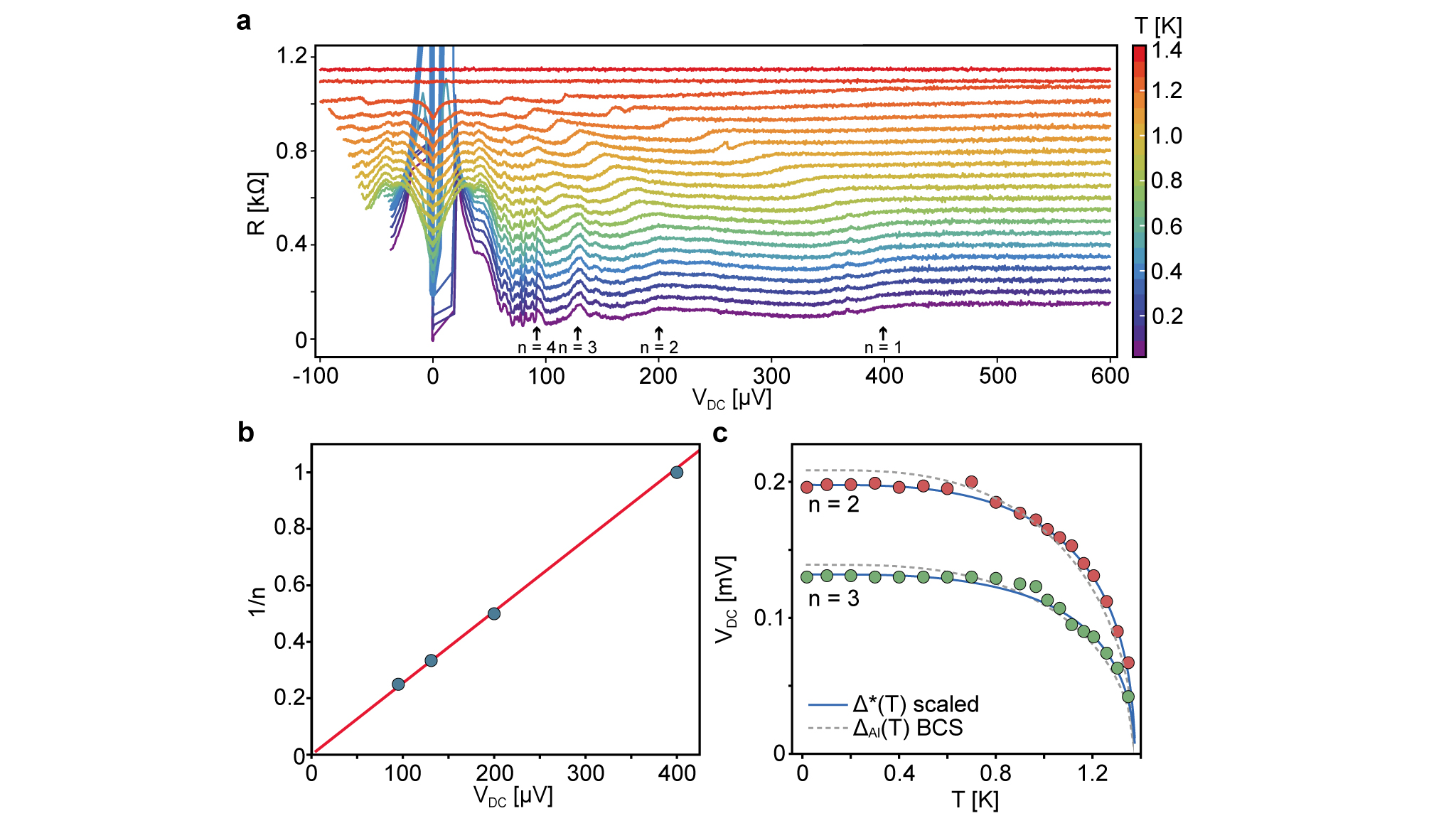}
\vspace{-0.4cm}
\caption{Intentionally roughened sample with 13.4 nm top barrier: (a) Temperature dependence of $R$ as function of $V_\mathrm{DC}$. For clarity, the traces are offset by 50 $\Omega$. The identified MAR peaks are labeled with $n$ = 1, 2, 3, 4 accordingly. (b) The extracted $V_\mathrm{DC}$ values for a measurement at base temperature plotted vs. 1/$n$. The red line corresponds to $V_\mathrm{DC}$ = 2$\Delta$*/$ne$, with $\Delta$* the induced gap energy of 197 $\mu$eV. (c) Extracted $V_\mathrm{DC}$ positions of the MAR peaks in $R$ for $n$ = 2, 3 as a function of temperature. 
The solid lines are the fits of $\Delta$*($T$) with $\gamma_\mathrm{B}$ = 0.78 and scaled by a factor of 1.2. The fit for $n$ = 3 is $\Delta$*$(T)\cdot 2/3$ accordingly. The dashed lines correspond to fits of the BCS theory for the gap of Al.} 
\label{figS9}
\end{figure*}

Here, we provide additional details related to the analysis of the magneto-transport and MAR data presented in Fig. 7 of the Main Text.

Analysis of the Fraunhofer pattern in Fig. 7c in the Main Text shows a node to node separation of $\delta B$ $\approx$ 0.5 mT. This can be described by
\begin{equation} 
I_\mathrm{sw}(B_{\perp}) = I_\mathrm{sw}(0) \cdot \bigg| \displaystyle \frac{\textrm{sin}(\pi B_{\perp} L_\mathrm{eff} W / \Phi_0)}{\pi B_{\perp} L_\mathrm{eff} W / \Phi_0} \bigg|,
\label{franhofer}
\end{equation}
where $L_\mathrm{eff} W$ is the effective area of the normal region where magnetic flux can penetrate. Using $W$ = 4 $\mu$m we estimated $L_\mathrm{eff}$ $\approx$ 1 $\mu$m, approximately 15 times larger than the lithographic lead separation. This discrepancy with respect to the device dimensions is attributed to the Meissner effect in the SC leads, which results in flux focusing in the JJ \cite{rosenthal1998fluxfocusingtheory}, and has been found in previous studies on hybrid JJs \cite{suominen2017anomalousfraunhofer, dartiailh2021fluxfocus}.

Another figure of merit of a JJ, not mentioned in the Main Text, is the product of the critical current and normal state resistance $I_\mathrm{c} R_\mathrm{n}$. Here, $I_\mathrm{c}$ was substituted by a single value of $I_\mathrm{sw}$ = 0.96 $\mu$A extracted at base temperature of 18 mK, yielding $I_\mathrm{c} R_\mathrm{n}$ = 144 $\mu$eV, where $R_\mathrm{n}$ = 150 $\Omega$. This was compared to the energy gap of our bare Al thin film $\Delta_\mathrm{Al}$ = $1.76 \cdot k_\mathrm{B} \cdot T_\mathrm{c}$ = 209 $\mu$eV, where we used $T_\mathrm{c}$ = 1.375 K. The $I_\mathrm{c} R_\mathrm{n}$ value extracted from the JJ is roughly 22\% of the clean junction/KO-limit $\pi \cdot \Delta_\mathrm{Al}/e$ \cite{likharev1979superconducting}. 
This value is likely underestimated as the precise extraction of $I_\mathrm{c}$ is difficult due to the stochastic nature of the superconducting transition \cite{haxell2022large}. Moreover, a variety of different material and junction properties can influence $I_\mathrm{c}$, for example Fermi-level mismatch or non-ideal interface transparency \cite{nikolic2001intrinsic}.

Furthermore, we extracted the excess current $I_\mathrm{exc}$ which occurs due to Andreev reflection at highly transparent SE/SC interfaces \cite{flensberg1988subharmonic}. It is defined as the current $I_\mathrm{SD}$ at which $V_\mathrm{DC}=0$ when linearly extrapolated from the high-bias region where $e \cdot V_\mathrm{DC}$ is larger than twice the energy gap of the Al thin film $\Delta_\mathrm{Al}$. We extracted $I_\mathrm{exc}$ = 1.69 $\mu$A and this value can be related to an induced gap via $I_\mathrm{exc} R_\mathrm{n}$ = 254 $\mu$eV = $\alpha\cdot\Delta$* \cite{flensberg1988subharmonic, kjaergaard2017mar}. The parameter $\alpha$ incorporates elastic scattering events at the SE/SC interface \cite{flensberg1988subharmonic, naidyuk2018excesscurrentreview} and can be used as a measure of the junction transparency. The theoretical upper bounds for a fully transparent S-N interface (i.e., where $\Delta$* = $\Delta_\mathrm{Al}$) were defined as $8/3\cdot \Delta_\mathrm{Al}$ = 556 $\mu$eV in the ballistic case and ($\pi^2$/4)-1 $\cdot \Delta_\mathrm{Al}$ = 306 $\mu$eV in the diffusive case \cite{flensberg1988subharmonic, artemenko1979excess}.
In our device the $I_\mathrm{exc} R_\mathrm{n}$ was 46\% of the ballistic and 83 \% of the diffusive value, which indicates a high SE/SC interface transparency and corresponds well with values found in previous studies \cite{kjaergaard2017mar, lee2019transport}. 

To show the MAR analysis of the roughened material in more detail, we highlight the MAR peaks assigned to $n$ (used in Fig. 7 of the Main Text) in the dependency of $R$ on $V_\mathrm{DC}$ in Fig. S\ref{figS9}a. The linear relationship in 1/$n$, given as $e \cdot V_\mathrm{DC}$ = 2$\cdot$$\Delta$*/$n$ with $\Delta$* = $\Delta$*$_\mathrm{r}$ = 197 $\mu$eV is in agreement with the extracted $V_\mathrm{DC}$ values for $n$ = 1 to 4, as is shown in Fig. S\ref{figS9}b. The temperature dependence of the MAR peaks was found to almost follow the BCS temperature dependence of the nominal energy gap of the Al film given by the 

\begin{equation} %
\Delta_\mathrm{Al}(T) = \Delta_\mathrm{Al}(0) \textrm{ tanh} \left( 1.74 \sqrt{\dfrac{T}{T_\mathrm{c}}-1}\right).
\label{eqDeltabcs}
\end{equation}

In addition, an estimate of the interface transparency can be obtained from a closer look at $\Delta$*($T$) by using the following implicit equation to fit the temperature dependence of the measured MAR peaks \cite{kjaergaard2017mar, chrestin1997implicitexperiment, aminov1996implicittheory}: 

\begin{equation}
\Delta \text{*}(T) = \displaystyle \frac{\Delta_\mathrm{Al}(T)}{1+\gamma_\mathrm{B} \sqrt{\Delta_\mathrm{Al}^2(T) - \Delta^{*2}(T)} / (\pi k_\mathrm{B} T_\mathrm{c})},
\label{eqDeltastar}
\end{equation}
where $\gamma_\mathrm{B}$ is related to the transparency of the SE/SC interface ($\gamma_\mathrm{B}$ = 0 is perfect transmission) \cite{tschapers2001}.

Comparisons of the BCS relation and the fits using a scaled Eq. (\ref{eqDeltastar}) are shown in Figure S\ref{figS9}c. We numerically optimized $\Delta$*$(T)\cdot f$ to fit the MAR peaks of n = 2 as a function of temperature, inserting relation (\ref{eqDeltabcs}), $T_\mathrm{c}$ = 1.375 K and the free parameter $\gamma_\mathrm{B}$. The parameter $f$ is a constant scaling factor. A good agreement is found for $\gamma_\mathrm{B}$ between 0.75 and 0.8 and scaling $f$ = 1.2. As can be seen for $n$ = 2 in Fig. S\ref{figS9}c, the fit follows the temperature evolution of the peaks better than the bare BCS dependence for the gap of Al. An improved agreement is also found for $n$ = 3, for which the curve fitted to $n$ = 2 is multiplied by 2/3. The value for $\gamma_\mathrm{B}$ $\approx$ 0.78 signifies a high interface transparency \cite{kjaergaard2017mar}.

\section{S8: JJ measurements of the standard, non-roughened samples}

\begin{figure*}[h]
\vspace{0.2cm}
\includegraphics[scale=0.25]{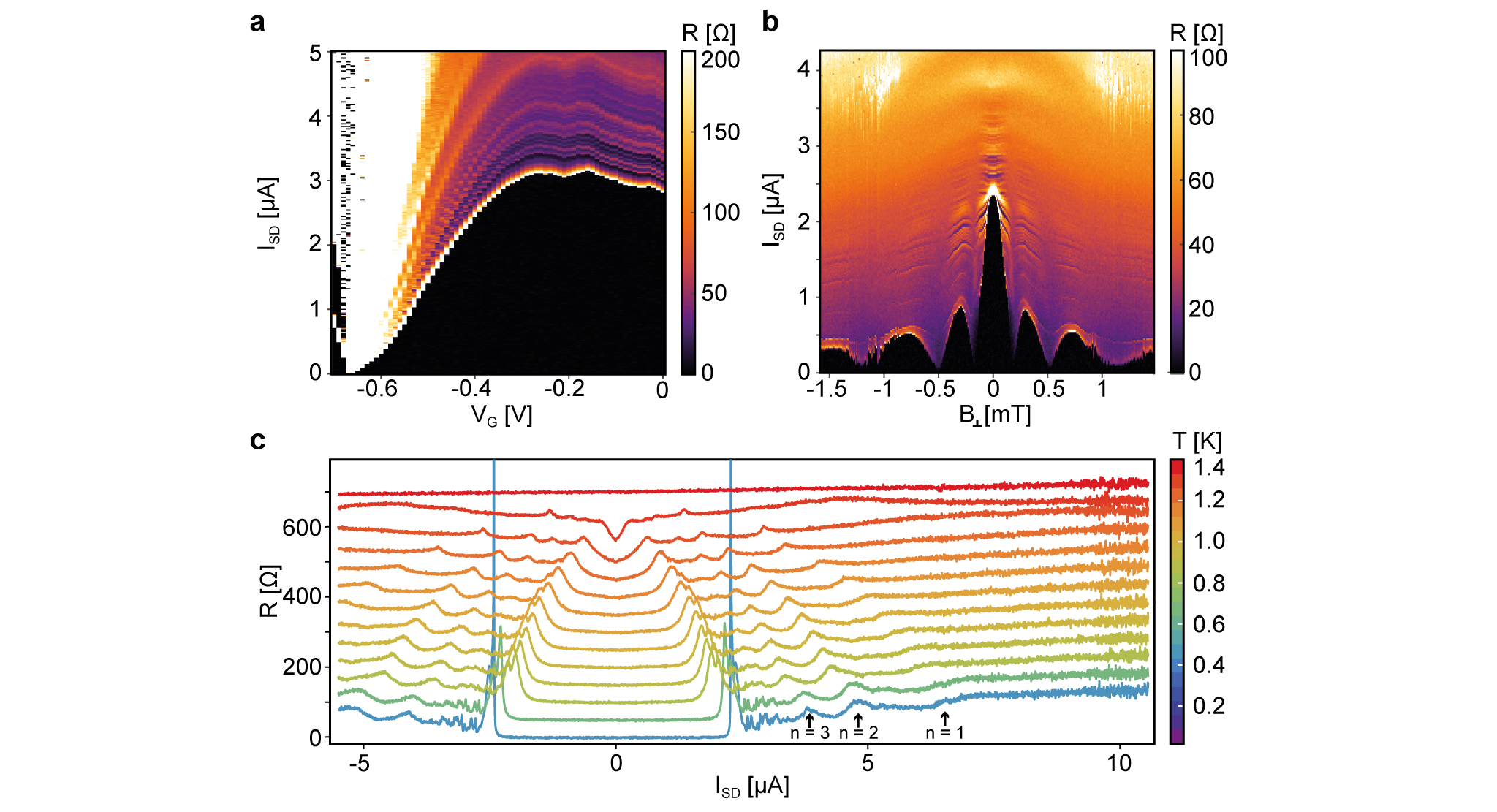}
\vspace{-0.4cm}
\caption{Non-roughened sample with 13.4 nm top barrier: (a) Differential resistance of the Josephson junction as a function of source-drain current $I_\mathrm{SD}$ and gate voltage $V_\mathrm{G}$. The black spots at low gate voltages arise from an overload of the lock-in amplifier. (b) Differential resistance of the Josephson junction as a function of $I_\mathrm{SD}$ and perpendicular magnetic field $B_{\perp}$ measured at $V_\mathrm{G}$ = -0.41 V. (c) Temperature dependence of the differential resistance traces as function of bias current $I_\mathrm{SD}$ at $V_\mathrm{G}$ = -0.41 V. For clarity, each temperature trace is offset by 50 $\Omega$.}
\label{figS7}
\end{figure*}

\begin{figure*}[h]
\vspace{0.2cm}
\includegraphics[scale=0.25]{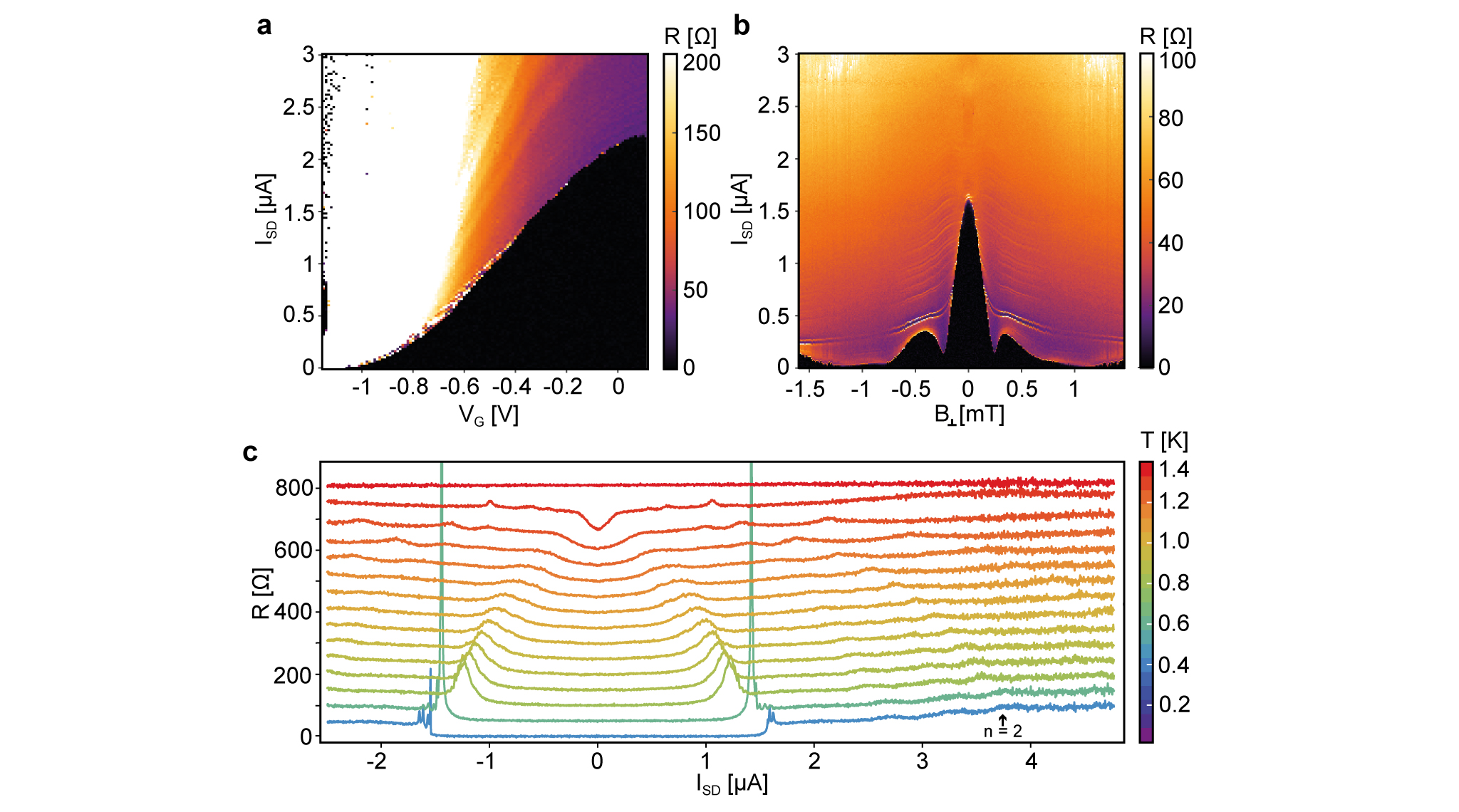}
\vspace{-0.4cm}
\caption{Non-roughened sample with 20 nm top barrier: (a) Differential resistance of the Josephson junction as a function of source-drain current $I_\mathrm{SD}$ and gate voltage $V_\mathrm{G}$. The black spots at low gate voltages arise from an overload of the lock-in amplifier. (b) Differential resistance of the Josephson junction as a function of $I_\mathrm{SD}$ and perpendicular magnetic field $B_{\perp}$ measured at $V_\mathrm{G}$ = -0.23 V. (c) Temperature dependence of the differential resistance traces as function of bias current $I_\mathrm{SD}$ at $V_\mathrm{G}$ = -0.23 V. For clarity, each temperature trace is offset by 50 $\Omega$.}
\label{figS8}
\end{figure*}

For direct comparison with the data acquired for the intentionally roughened sample shown in the Main Text, we present results of measurements on hybrid InAs/Al JJs fabricated from the standard samples with a non-roughened interface (2ML GaAs) both with 13.4 nm and 20 nm top barrier thicknesses. The dependence of the differential resistance $R$ on the top gate voltage $V_\mathrm{G}$ and out-of-plane magnetic field $B_{\perp}$ are shown in Fig. S\ref{figS7}a,b and Fig. S\ref{figS8}a,b respectively for the thinner and thicker top barrier. The temperature dependence of the current bias sweeps at a specific gate-voltage are summarized in Fig. S\ref{figS7}c and Fig. S\ref{figS8}c. 

For the standard sample with a 13.4 nm top-barrier, the MAR peaks followed the relation $e \cdot V_\mathrm{DC}$ = 2 ${\Delta}$*$_\mathrm{s}$$/{n}$ for $n$ = 1, 2, 3 (located at 383, 183, 118 $\mu$V, at 500 mK). The peak assigned to $n$ = 1, was identified from the onset of an almost constant $R$ for large bias current. Around n$\geq$ 4 (at 96 $\mu$V) the peaks started to bunch together and a clear assignment was difficult, as shown in Fig. S\ref{figS7}c. The induced gap value $\Delta$*$_\mathrm{s}$ =  184 $\pm$ 6 $\mu$eV was extracted from the average of the $V_\mathrm{DC}$ values of $n$ = 1 to 3.

The assignment of the MAR peaks was less clear in the standard sample with a 20 nm top-barrier. The peak assigned to $n$=1 (outermost peak) was present at temperatures from $T_\mathrm{c}$ down to 0.9 K, below which it disappeared beyond the used bias range. The peaks, that corresponded to $n$ = 2, 3, 4 (170, 129, 88 $\mu$V at 550 mK) show larger deviation from the 1/$n$ relation in comparison to both 13.4 nm barrier samples. Since the $n = 2$ peak was clearly identified, an upper bound was extracted from $n =2$ with $\Delta$*$_\mathrm{20nm}$ = 170 $\mu$eV. The deviation from the predicted peak positions at $\Delta$*$_\mathrm{20nm}$/$n$ and the absence of smaller peak features for $n$ > 4 is likely due to a weaker coupling that originates from a thicker top barrier. 

The results of the MAR analysis of the roughened and non-roughened JJs (presented here and in S7), together with the $I_\mathrm{sw}$, $I_\mathrm{exc}$ and $R_\mathrm{n}$ values, are summarized Tab. S1. 
In the Main Text we focused on comparing the induced gap energy obtained from the MAR peak positions, while an estimation of the induced gap energy from the $I_\mathrm{c} R_\mathrm{n}$ or $I_\mathrm{exc} R_\mathrm{n}$ was less clear. 

On the one hand, in hybrid planar JJs, $I_\mathrm{c}$ needs to be replaced by $I_\mathrm{sw}$, which likely leads to a underestimation of the induced gap. On the other hand, an apparent difference was found between the roughened and standard sample with respect to the $I_\mathrm{exc} R_\mathrm{n}$ value. We extracted $I_\mathrm{exc} R_\mathrm{n}$ = 466 $\mu$eV for the standard material and 254 $\mu$eV for the roughened material. As the induced gap size extracted from the MAR analysis was found to be similar for these materials, this significant difference in $I_\mathrm{exc} R_\mathrm{n}$  = $\alpha\cdot\Delta$* is likely governed by the pre-factor $\alpha$. For the 70 nm wide junctions, the mobility was sufficiently large for the normal region to be in the ballistic regime. Consequently, the difference in the $\alpha$ values was expected to stem from multiple  parameters such as SE/SC interface transparency, scattering events and local device fluctuations.

For the standard sample with a 20 nm barrier we found a value $I_\mathrm{exc} R_\mathrm{n}$ = 194 $\mu$eV, which is smaller than for the roughened sample. In this case, the difference between the $\Delta$*$_\mathrm{r}$ and ${\Delta}$*$_\mathrm{20nm}$ almost accounts for the difference in the $I_\mathrm{exc} R_\mathrm{n}$ products. This implies that the implemented thicker barriers mainly reduced the induced gap energy value $\Delta$*, with less effect on $\alpha$, despite nearly doubling the electron mobility (shown in Fig. 6b in the Main Text). Together with assuming that the SE/SC coupling is weaker for a thicker top barrier, this indicated that a the thicker barrier could mitigate the influence of local disorder, as recently suggested by Awoga et al. \cite{awoga2022mitigating}. Yet, depending on how exactly the transport regime, scattering events and the interface transparency affect $\alpha$, a clear interpretation remains challenging. In conclusion, a more complex description and analysis of $\alpha$ would be needed in future experimental and theoretical studies.

\begin{table}[h]
\begin{center}
\begin{tabular}{||c | c | c | c | c | c | c||} 
 \hline
 Sample & $I_\mathrm{sw}$ [$\mu$A] & $R_\mathrm{n}$ [$\Omega$] & $I_\mathrm{exc}$ [$\mu A$]& $I_\mathrm{c} R_\mathrm{n}$ [$\mu$eV] & $I_\mathrm{exc} R_\mathrm{n}$ [$\mu$eV] & $\Delta$*$_\mathrm{MAR}$ [$\mu$eV] \\ [0.5ex] 
 \hline\hline
Roughened 13.4 nm barrier (18 mK) & 0.96  & 150 & 1.69 & 144 & 254 & 197 $\pm2$ \\ 
 \hline
Roughened 13.4 nm barrier (500 mK) & 0.9  & 150 & 1.69 & 135 & 254 & 197 $\pm1$\\ 
 \hline
Standard 13.4 nm barrier (500 mK) & 2.34 & 126 & 3.7  & 295 & 510 & 184 $\pm6$ \\ 
 \hline 
Standard 20 nm barrier (550 mK) & 1.69 & 97 & 2 & 164 &194 & 170  \\
 \hline
\end{tabular}
\vspace{0.2cm}
\caption{TAB. S1: Summary of the basic properties of the characterized JJs for the roughened and standard 13.4 nm and 20nm top barriers shown in S7 and S8. The $\Delta$*$_\mathrm{MAR}$ denotes the extracted induced energy gap values from the analysis of the MAR peaks.} 
\label{tab1}
\end{center}
\end{table}

\bibliography{supplement}